\documentclass[twocolumn,showpacs,aps,prd]{revtex4}
\usepackage{dcolumn}
\usepackage{amsmath}
\usepackage{float}

\usepackage{graphicx}

\newcommand{\BABARPubYear}    {16}
\newcommand{\BABARPubNumber}  {006}
\newcommand{\SLACPubNumber} {16855}
\input babarsym

\long\def\inst#1{\par\nobreak\kern 4pt\nobreak
    {\it #1}\par\vskip 10pt plus 3pt minus 3pt}

\begin{document}

{

\begin{flushleft}
\babar-PUB-\BABARPubYear/\BABARPubNumber \\
SLAC-PUB-\SLACPubNumber \\
\end{flushleft}

\vspace{-1cm}

% Title of the paper
\title{
\Large \bf Measurement of the inclusive electron spectrum from
{\boldmath $B$} meson decays and determination of 
{\boldmath $|V_{ub}|$}
}
\bigskip \bigskip \bigskip

% Abstract
\begin{abstract}

Based on the full \babar\ data sample of 466.5 million \BB pairs, we present measurements of the electron spectrum 
from semileptonic $B$ meson decays. We fit the inclusive electron spectrum to distinguish 
Cabibbo-Kobayashi-Maskawa (CKM) suppressed $B \to X_u e \nu$ decays 
from the CKM-favored $B \to X_c e \nu$ decays, and from various other backgrounds, and determine the total semileptonic branching 
fraction ${\cal B}(B\to X e\nu)$ = $(10.34 \pm 0.04_{\textrm{stat}} \pm 0.26_{\textrm{syst}})$\%, averaged over $B^{\pm}$ and 
$B^0$ mesons.   
We determine the spectrum and branching fraction for charmless $B \to X_u e \nu$ decays and extract the CKM element 
$|V_{ub}|$, by relying on four different QCD calculations based on the heavy quark expansion. 
While experimentally, the electron momentum region above 2.1\gevc is favored, because the background is relatively low, 
the uncertainties for the theoretical predictions are largest in the region near the kinematic endpoint. 
Detailed studies to assess the impact of these four predictions on the measurements of the electron spectrum, 
the branching fraction, and the extraction of the CKM matrix element $|V_{ub}|$ are presented, with the lower limit on the 
electron momentum varied from 0.8\gevc to the kinematic endpoint. 
We determine $|V_{ub}|$ using each of these different calculations and find, 
$|V_{ub}|$ =
$(3.794 \pm 0.107_{\textrm{exp}} \,^{+0.292}_{-0.219\,{\textrm{SF}}} \,^{+0.078}_{-0.068\,{\textrm{theory}}}) \times 10^{-3}$ 
(De Fazio and Neubert),
$(4.563 \pm 0.126_{\textrm{exp}} \,^{+0.230}_{-0.208\,{\textrm{SF}}} \,^{+0.162}_{-0.163\,{\textrm{theory}}}) \times 10^{-3}$
(Bosh, Lange, Neubert, and Paz),
$(3.959 \pm 0.104_{\textrm{exp}} \,^{+0.164}_{-0.154\,{\textrm{SF}}} \,^{+0.042}_{-0.079\,{\textrm{theory}}}) \times 10^{-3}$
(Gambino, Giordano, Ossola, and Uraltsev),
$(3.848 \pm 0.108_{\textrm{exp}} \,^{+0.084}_{-0.070\,{\textrm{theory}}}) \times 10^{-3}$
(dressed gluon exponentiation), 
where the stated uncertainties refer to the experimental uncertainties of the partial branching fraction measurement, 
the shape function parameters, and the theoretical calculations.

\vfill

\end{abstract}

% Input author list file
% NOTES
%
% 04-AUG-2016 Add ``deceased'' footnote for Giancarlo Piredda          J.W. Gary
% 02-JUN-2016 Move Marcello Rotondo from Padova to Frascati            J.W. Gary
% 20-FEB-2016 Add footnote for Liang Sun                               J.W. Gary
% 21-DEC-2015 Add Bologna alternative address for Claudia Patrignani   J.W. Gary
% 
\author{J.~P.~Lees}
\author{V.~Poireau}
\author{V.~Tisserand}
\affiliation{Laboratoire d'Annecy-le-Vieux de Physique des Particules (LAPP), Universit\'e de Savoie, CNRS/IN2P3,  F-74941 Annecy-Le-Vieux, France}
\author{E.~Grauges}
\affiliation{Universitat de Barcelona, Facultat de Fisica, Departament ECM, E-08028 Barcelona, Spain }
\author{A.~Palano}
\affiliation{INFN Sezione di Bari and Dipartimento di Fisica, Universit\`a di Bari, I-70126 Bari, Italy }
\author{G.~Eigen}
%\author{B.~Stugu}
\affiliation{University of Bergen, Institute of Physics, N-5007 Bergen, Norway }
\author{D.~N.~Brown}
%\author{L.~T.~Kerth}
\author{Yu.~G.~Kolomensky}
%\author{M.~J.~Lee}
%\author{G.~Lynch}
\affiliation{Lawrence Berkeley National Laboratory and University of California, Berkeley, California 94720, USA }
\author{H.~Koch}
\author{T.~Schroeder}
\affiliation{Ruhr Universit\"at Bochum, Institut f\"ur Experimentalphysik 1, D-44780 Bochum, Germany }
% ...
\collaboration{The \babar\ Collaboration}
\noaffiliation

\pacs{13.20.He,                 % semileptonic bottom meson decays
      12.15.Hh,                 % CKM elements determination
      12.38.Qk,                 % Experimental tests of QCD calculations
      14.40.Nd}                 % properties of bottom mesons

\maketitle

\section{Introduction}
\label{sec:introduction}

Semileptonic decays of $B$ mesons proceed via leading order weak interactions. They are expected to be free of non-Standard-Model 
contributions and therefore play a critical role in the determination of the Cabibbo-Kobayashi-Maskawa (CKM) quark-mixing matrix~\cite{sm} 
elements  $|V_{cb}|$ and  $|V_{ub}|$. In the Standard Model (SM), the CKM elements satisfy unitarity relations that can be illustrated 
geometrically as triangles in the complex plane. For one of these triangles, $CP$ asymmetries determine the angles, $|V_{cb}|$ normalizes 
the length of the sides, and the ratio $|V_{ub}|/|V_{cb}|$ determines the side opposite the well-measured angle $\beta$. 
Thus, precise measurements of  $|V_{cb}|$ and $|V_{ub}|$ are crucial to studies of flavor physics and $CP$ violation in the quark sector. 

There are two methods to determine $|V_{cb}|$ and $|V_{ub}|$, one based on exclusive semileptonic $B$ decays, where the hadron in the 
final state is a $D, D^*, D^{**}$ or $\pi, \rho, \omega, \eta, \eta^{\prime}$ meson, the other based on inclusive decays 
$B \to X e \nu$, where $X$ refers to either $X_c$ or $X_u$, {\it i.e.}, to any hadronic state with or without charm, respectively. 

The extractions of  $|V_{cb}|$ and $|V_{ub}|$ from measured inclusive or exclusive semileptonic $B$ meson decays rely on different 
experimental techniques to isolate the signal and on different theoretical descriptions of QCD contributions to the underlying weak 
decay processes. Thus, they have largely independent uncertainties, and provide important cross-checks of the methods and our 
understanding of these decays in general. At present, these two methods result in values for $|V_{cb}|$ and $|V_{ub}|$ 
that each differ by approximately 3 standard deviations~\cite{pdg2014}.  

In this paper, we present a measurement of the inclusive electron momentum spectrum and branching fraction (BF) for the sum of all 
semileptonic $B \to X e \nu$ decays, as well as measurements of the spectrum and partial BF for charmless semileptonic 
$B \to X_u e \nu$ decays. The total rate for the $B \to X_u e \nu$ decays is suppressed by about a factor 50 compared to the 
$B \to X_c e \nu$ decays. This background dominates the signal spectrum except near the high-momentum endpoint. 
In the rest frame of the $B$ meson, the electron spectrum for $B \to X_u e \nu$ signal extends to 
$\sim 2.6~\gevc$, while for the background $B \to X_c e \nu$ decays the kinematic endpoint is at $\sim 2.3~\gevc$.
In the $\Upsilon(4S)$ rest frame, the two $B$ mesons are produced with momenta of $300~\mevc$ which extends the electron endpoint 
by about $200~\mevc$. The endpoint region above $2.3~\gevc$, which covers only about $10\%$ of the total electron spectrum, 
is more suited for the experimental isolation of the charmless decays.

To distinguish contributions of the CKM suppressed $B \to X_u e \nu$ decays from those of CKM-favored $B \to X_c e \nu$ decays, 
and from various other backgrounds, we fit the inclusive electron momentum spectrum, averaged over $B^{\pm}$ and $B^0$ mesons 
produced in the $\Upsilon(4S)$ decays~\cite{pdg2014, hfag14}. 
For this fit, we need predictions for the shape of the $B \to X_u e \nu$  spectrum. We have employed and studied four 
different QCD calculations based on the heavy quark expansion (HQE)~\cite{ope}. 
The upper limit of the fitted  range of the momentum spectrum is fixed at 3.5 \gevc, 
while the lower limit extends down to 0.8 \gevc, covering up to $90\%$ of the total signal electron spectrum. 
From the fitted spectrum we derive the partial BF for charmless $B \to X_u e \nu$ decays and extract the 
CKM element $|V_{ub}|$. While the experimental sensitivity to the $B \to X_u e \nu$ spectrum and to $|V_{ub}|$ is primarily 
determined from the spectrum above $2.1~\gevc$, due to very large backgrounds at lower momenta,
the uncertainties for the theoretical predictions are largest in the region near the kinematic endpoint. 
Studies of the impact of various theoretical predictions on the measurements are presented. 

Measurements of the total inclusive lepton spectrum in $B \to X e \nu$ decays have been performed by several experiments 
operating at the $\FourS$ resonance~\cite{pdg2014}. The best estimate of this BF has been derived by HFAG~\cite{hfag14}, 
based on a global fit to moments of the lepton momentum and hadron mass spectra in $B \to X e \nu$ decays 
(corrected for $B \to X_u e \nu$ decays) either with 
a constraint on the $c$-quark mass or by including photon energy moments in $B\to X_s \gamma$ decays in the fit.
Inclusive measurements of $|V_{ub}|$ have been performed at the $\FourS$ resonance, by ARGUS~\cite{argus},
CLEO~\cite{cleo,cleo2}, \babar~\cite{babarVub} and Belle~\cite{belle-vub}, and experiments at LEP operating at the
$Z^0$ resonance, L3~\cite{l3-vub}, ALEPH~\cite{aleph-vub}, DELPHI~\cite{delphi-vub}, and OPAL~\cite{opal-vub}. 
Among the $|V_{ub}|$ measurements based on exclusive semileptonic decays~\cite{pdg2014}, 
the most recent by the LHCb experiment at the LHC  
is based on the baryon decay $\Lambda_b \to p \mu \nu$ ~\cite{lhcb-vub}.  

This analysis is based on methods similar to the one used in previous measurements
of the lepton spectrum near the kinematic endpoint at the $\FourS$ resonance~\cite{argus,cleo}.
The results presented here supersede the earlier \babar\ publication~\cite{babarVub},
based on a partial data sample.

\section{Data Sample}
\label{sec:Data}

The data used in this analysis were recorded with the
\babar\ detector~\cite{detector} at the \pep2 energy-asymmetric $e^+e^-$ collider.
A sample of 466.5 million \BB events,
corresponding to an integrated luminosity
of 424.9~$\textrm{fb}^{-1}$~\cite{lumi}, was collected at the \FourS\ resonance. 
An additional sample of 44.4~$\textrm{fb}^{-1}$ was recorded
at a center-of-mass (c.m.) energy 40 \mev below the \FourS\ resonance, {\it i.e.}, just
below the threshold for \BB production.
This off-resonance data sample is used to subtract the non-\BB background
at the \FourS\ resonance.  
The relative normalization of the two data samples
has been derived from luminosity
measurements, which are based on the number of detected $\mu^+\mu^-$
pairs and the QED cross section for
$e^+e^-\to \mu^+\mu^-$
production, adjusted for the small difference in center-of-mass energy.

\section{Detector}
\label{sec:Detector}

The \babar\ detector has been described in detail elsewhere \cite{detector}.
The most important components for this study are the charged-particle
tracking system, consisting of a five-layer silicon vertex tracker and a 40-layer 
cylindrical drift chamber, and the electromagnetic calorimeter consisting of 
6580 CsI(Tl) crystals. These detector components operated in a
1.5 T magnetic field parallel to the beam. Electron candidates are
selected on the basis of the ratio of the energy deposited in the calorimeter
to the track momentum, the shower shape, the energy loss in the drift chamber, 
and the angle of signals in a ring-imaging Cerenkov detector. 
Showers in the electromagnetic calorimeter with energies below 50~\mev\ which 
are dominated by beam background are not used in this analysis.

\section{Simulation}
\label{sec:Simulation}

We use Monte Carlo (MC) techniques to simulate the production
and decay of $B$ mesons and the detector response~\cite{geant4},
to estimate signal and background efficiencies, and to extract 
the observed signal and background distributions. The sample of 
simulated generic \BB events exceeds the \BB data 
sample by about a factor of 3.

The MC simulations include radiative effects such as bremsstrahlung
in the detector material and QED initial and final state radiation~\cite{photos}. 
Information from studies of selected control data samples on efficiencies 
and resolutions is used to adjust and thereby improve the accuracy of the simulation.
Adjustments for small variations of the beam energy over time have been included.

In the MC simulations, the BFs for hadronic $B$ and $D$ meson decays are based on values reported in the
Review of Particle Physics~\cite{pdg2014}. The simulation of inclusive charmless semileptonic decays, $B \to X_u e \nu$, 
is based on calculations by De Fazio and Neubert (DN)~\cite{dFN}. This simulation produces a continuous mass spectrum of 
hadronic states $X_u$. To reproduce and test predictions by other authors this spectrum is reweighted in the course of the 
analysis. Three-body $B \to X_u e \nu$ decays with low-mass hadrons, $X_u = \pi, \rho, \omega, \eta, \eta^{\prime}$, 
make up about $20\%$ of the total charmless rate. They are simulated separately using the ISGW2 model~\cite{isgw2} and added 
to samples of decays to nonresonant and higher-mass resonant states $X_u^{\textrm nr}$, 
so that the cumulative distributions of the hadron mass, the momentum transfer squared, and the electron momentum reproduce 
the inclusive calculation as closely as possible. The hadronization of $X_u$ with masses above $2 m_{\pi}$
is performed according to JETSET~\cite{jetset}. 

The MC-generated electron momentum distributions for $B \to X_u e \nu$ decays are shown in Fig.~\ref{f:buSpectra}
for individual decay modes and for their sum. Here and throughout the paper, the electron momentum and all other kinematic 
variables are measured in the \FourS\ rest frame, unless stated otherwise. Above 2~GeV/c, 
the significant signal contributions are from decays involving the light mesons $\pi$, $\rho$, $\omega$, $\eta$, 
and $\eta^{\prime}$, in addition to some lower mass nonresonant states $X_u^{\textrm nr}$.

\begin{figure}[htbp]
\begin{center}
\includegraphics[height=6.cm]{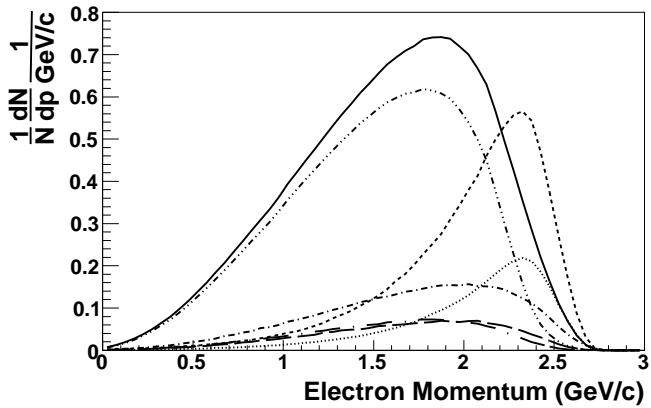}
\caption{
MC-generated electron momentum spectra in the \FourS\ rest frame for charmless semileptonic $B$ decays. 
The full spectrum (solid line) is normalized to 1.0. The largest contribution is from decays
involving higher-mass resonances and nonresonant states ($X_u^{\textrm nr}$) (dash-three-dotted).
The exclusive decays (scaled by a factor of five) are: 
$B \rightarrow \pi e \nu$ (dash-dotted), $B \rightarrow \rho e \nu$ (dashed),
$B \rightarrow \omega e \nu$ (dotted), $B \rightarrow \eta e \nu$ (long-dashed),
$B \rightarrow \eta^{\prime} e \nu$ (long-dash-dotted). 
}
\label{f:buSpectra}
\end{center}
\end{figure}

The simulation of the dominant $B \to X_c e \nu$ decays is based on a variety of theoretical prescriptions.
For $B\to D e \nu$ and $B \to D^* e \nu$ decays we use form factor parametrizations~\cite{hqet,CLN,GL}, 
based on heavy quark effective theory. Decays to pseudoscalar mesons are described in terms of one form factor, 
with a single parameter $\rho_D^2$. The differential decay rate for $B\to D^*e \nu$ is described by three amplitudes, 
with decay rates depending on three parameters: $\rho_{D^*}^2$, $R_1$, and $R_2$. 
These parameters have been measured by many experiments; we use the average values presented in Table \ref{t:parD}.

For the simulation of decays to higher-mass $L=1$ resonances, $D^{**}$, i.e., two wide states
$D^*_0(2400)$, $D_1^{\prime}(2430)$, and two narrow states $D_1(2420)$, $D_2^*(2460)$, 
we have adopted the parametrizations by Leibovich {\it et al.}~\cite{llsw_pr} and the HFAG averages~\cite{hfag14} for the BFs. 
For decays to nonresonant charm states $B \to D^{(*)} \pi e \nu$, we rely on the prescription by
Goity and Roberts~\cite{gr} and the \babar\ and Belle measurements of the BFs~\cite{hfag14}.
The simulations of these decays include the full angular dependence of the rate.

\begin{table}[tbp]
\caption{
Average measured values~\cite{hfag14} of the form factor parameters for $B\to D e \nu$ and $B\to D^* e \nu$ decays,
as defined by Caprini, Lellouch, and Neubert~\cite{CLN}.
}
\label{t:parD}
{\small
\begin{tabular}{ccc}
\hline \hline 
                & $B\to D e \nu$     &  $B\to D^* e \nu$ \\ \hline
$\rho_D^2$      & $1.185 \pm 0.054$  &                 \\

$\rho_{D^*}^2$ 	&                    &  $1.207 \pm 0.026$ \\

$R_1$ 		& 		     &  $1.406 \pm 0.033$ \\

$R_2$ 		& 		     &  $0.853 \pm 0.020$ \\
\hline \hline
\end{tabular}
}
\end{table}

The shapes of the MC-generated electron spectra for individual $B \to X_c e \nu$ decays are shown in Fig.~\ref{f:semilepSpectra}.
Above 2 \gevc the dominant contributions are from semileptonic decays involving the lower-mass charm mesons, 
$D$ and $D^*$.  Higher-mass and nonresonant charm states are expected to contribute at lower electron momenta.
The relative contributions of the individual $B \to X_c e \nu$ decay modes have been adjusted to the results 
of the fit to the observed spectrum (see Sec. \ref{sec:bbfit}). 

\begin{figure}[htbp]
\begin{center}
\includegraphics[height=6.cm]{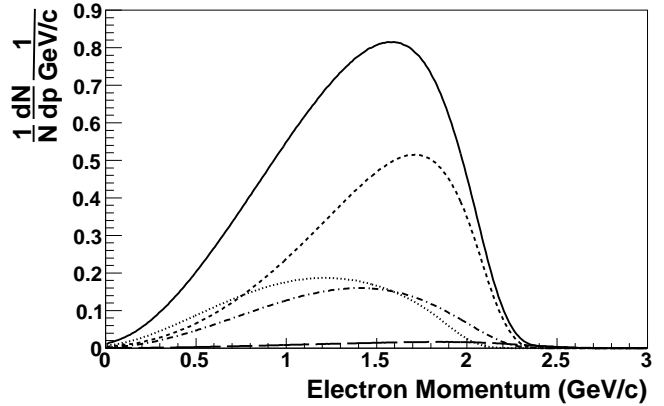}
\caption{
MC-generated electron momentum spectra for semileptonic decays to charm mesons, $B \to X_c e \nu$ with 
the total rate (solid line) normalized to 1.0. The individual components are: $B \to D e\nu$ (dash-dotted),
$B \to D^* e \nu$ (dashed), $B \to D^{**}e \nu$ +
$B \to D^{(*)}\pi e \nu$ (dotted).  The highly suppressed 
$B \to X_u e \nu$ signal spectrum (long dashed) is shown for comparison.
}
\label{f:semilepSpectra}
\end{center}
\end{figure}

The difference between the measured exclusive decays $B \to (D^{(*)},
D^{**}, D^{(*)}\pi) \ell\nu$ and the inclusive rate for semileptonic
$B$ decays to charm final states is $(1.40 \pm 0.28)$\%~\cite{BZT}. 
The decay rate for $\bar{B} \to D^{(*)}\pi^+\pi^-\ell^-\bar{\nu}$ was measured by \babar\ \cite{Dpipi}. 
Based on these results it was estimated that $\bar{B} \to D^{(*)}\pi\pi\ell^-\bar{\nu}$ decays account for 
up to half the difference between measured inclusive and the sum of previously measured exclusive branching fractions. 
Beyond these observed decays, there are missing decay modes, such as $B \to D^{\prime}(2550) e \nu$ 
and $B \to D^{\prime *}(2600) e \nu$. Candidates for the 2S radial excitations were first observed by 
\babar~\cite{babarDprime} and recently confirmed by LHCb~\cite{lhcbDprime}. We have adopted the masses and widths 
($130 \pm 18\mevcc$ and $93\pm 14 \mevcc$) measured by \babar~\cite{babarDprime}, and have simulated these decays 
using the form factor predictions~\cite{BZT}. Both $D^{**}$ and $D^{\prime (*)}$ may contribute by their decays 
to $D^{(*)}\pi\pi$ to $\bar{B} \to D^{(*)}\pi\pi\ell^-\bar{\nu}$ decays. 
The decay rate for $D_1 \to D\pi\pi$ was measured by Belle~\cite{Belle_Dpipi} and LHCb~\cite{LHCbDpipi},  
LHCb also measured the decay rate for $D_2^* \to D\pi\pi$. 
We account for contributions from $\bar{B} \to D^{**} e^-\bar{\nu}$,
$\bar{B} \to D^{\prime (*)} e^-\bar{\nu}$, and 
$\bar{B} \to D^{(*)} \pi e^-\bar{\nu}$ decays to $\bar{B} \to D^{(*)}\pi\pi e^-\bar{\nu}$ final states.

The main sources of secondary electrons are semileptonic charm meson decays and $J/\psi \to e^+e^-$ decays. 
The $J/\psi$ momentum distribution was determined from this data set and the MC simulation was adjusted 
to reproduce these measured spectra. 
The momentum spectra of $D$ and $D_s$ mesons produced in \BB decays were measured earlier by \babar~ \cite{Dspectra}
and the MC simulated spectra were adjusted to reproduce these measurements.

\section{CALCULATIONS OF $B \to X_u \ell \nu$ DECAY RATE}
\label{sec:models}

While at the parton level the rate for $b \to u \ell\nu$  decays can be reliably calculated, 
the theoretical description of inclusive semileptonic $B \to X_u \ell \nu$ decays is more challenging.
Based on HQE the total inclusive rate can be predicted with an uncertainty of about $5\%$, however, 
this rate is very difficult to measure due to very large background from the CKM-favored $B \to X_c \ell \nu$ decays. 
On the other hand, in the endpoint region where the signal to background ratio is much more favorable, calculations of 
the differential decay rates are much more complicated. 
They require the inclusion of additional perturbative and nonperturbative effects. These calculations rely on HQE and 
QCD factorization~\cite{neubert94} and separate perturbative and nonperturbative effects 
by using an expansion in powers of $1/m_b$ and a nonperturbative shape function (SF) which is {\it a priori} unknown. 
This function accounts for the motion of the $b$ quark inside the $B$ meson, and to leading order, it should be universal 
for all transitions of a $b$ quark to a light quark~\cite{kolya94,neubert94a}. It is modeled using arbitrary functions 
for which low-order moments are constrained by measurable parameters.

For the extraction of $|V_{ub}|$, we rely on $\Delta {\cal B}(\Delta p)$, the partial BF for $B \to X_u e \nu$ decays 
measured in the momentum interval $\Delta p$, and $\Delta \zeta (\Delta p)=\Gamma_{\textrm{theory}} \times f_u(\Delta p)/|V_{ub}|^2$, 
the theoretical predictions for partial decay rate normalized by $|V_{ub}|^2$, measured in units of ps$^{-1}$: 

\begin{equation}
\label{f:vub}
   |V_{ub}| = \sqrt{\frac {\Delta {\cal B}(\Delta p)}{\tau_b \, \Delta \zeta (\Delta p)}}. 
\end{equation}
Here $\tau_b=(1.580\pm 0.005)$ ps is the average of the $B^0$ and $B^{+}$ lifetimes~\cite{pdg2014}. 
$\Gamma_{\textrm{theory}}$ is the total predicted decay rate and $f_u(\Delta p)$ refers to the fraction 
of the predicted decay rate for the momentum interval $\Delta p$.

In the following, we briefly describe four different theoretical methods to derive predictions for the partial 
and total BFs. In the original work by De Fazio and Neubert~\cite{dFN} and
Kagan and Neubert~\cite{kagan_neubert} the determination of $|V_{ub}|$ relies on the measurement 
of the electron spectrum for $B \to X_u e \nu$ and on the radiative decays $B \to X_s \gamma$ 
to derive the parameters of the leading SF. More comprehensive calculations were performed by Bosch, Lange,
Neubert, and Paz (BLNP)~\cite{blnp1,blnp2,blnp3,blnp4,neubert_loops,neubert_extract,blnp_nnlo}. 
Calculations in the kinetic scheme were introduced by Gambino, Giordano, Ossola, Uraltsev (GGOU)~\cite{ggou1,ggou2}. 
BLNP and GGOU use $B \to X_c \ell \nu$ and $B \to X_s \gamma$ decays to derive the parameters of the leading SF. 
Inclusive spectra for $B \to X_u e \nu$  decays based on a calculation of nonperturbative functions 
using Sudakov resummation are presented  in the dressed gluon exponentiation (DGE) by Andersen and 
Gardi~\cite{gardi,dge1,dge2,dge3}.

We assess individual contributions to the uncertainty of the predictions of the decay rates 
by the different theoretical approaches. For this purpose, the authors of these calculations have provided software to compute 
the differential rates and to provide guidance for the assessment of the uncertainties on the rate and thereby $|V_{ub}|$. 
We differentiate uncertainties  originating from the SF parametrization, including the sensitivity to $m_b$, the $b$-quark mass, 
from the impact of the other purely theoretical uncertainties.
The uncertainty on $m_b$, the $b$-quark mass, has a large impact. Weak annihilation could contribute significantly 
at high-momentum transfers ($q^2$). 
The impact of weak annihilation is generally assumed to be asymmetric, specifically, it is estimated to decrease $|V_{ub}|$ 
by $O(1-2)\%$~\cite{V3}.

\subsection{DN calculations}

While the calculations by BLNP are to supersede the earlier work by DN, we use DN predictions for comparisons 
with previous measurements based on these predictions and also for comparisons with other calculations.

The early DN calculations~\cite{dFN} predict the differential spectrum with $O(\alpha_s)$ corrections to leading order in HQE. 
This approach is based on a parametrization of the leading-power nonperturbative SF. The long-distance interaction is described 
by a single light-cone distribution. In the region close to phase-space boundaries these nonperturbative corrections to the 
spectrum are large. The prediction for the decay distribution is obtained by a convolution of the parton model spectrum with the SF.
The SF is described by two parameters $\bar{\Lambda}^{\textrm{SF}}=M_B - m_b$ and ${\lambda}_1^{\textrm{SF}}$ which were determined 
from the measured photon energy moments in $B\to X_s \gamma$ decays~\cite{kagan_neubert}. 
We use \babar\ measurements~\cite{babar_photons} of the SF parameters, $m_b^{\textrm{SF}}= (4.79^{+0.06}_{-0.10}) \gev$ 
and ${\lambda}_1^{\textrm{SF}}=-0.24^{+0.09}_{-0.18} \gev^2$ with $94\%$ correlation. 

DN predict the shape of the differential electron spectrum, but they do not provide a normalization. 
Thus to determine the partial rates $\Delta \zeta (\Delta p)$, we rely on the DN predictions for $f_u(\Delta p)$, 
the fractions of $B \to X_u e \nu$ decays in the interval $\Delta p$, and an independent prediction for the normalized 
total decay rate $\zeta = (65.7^{+2.4}_{-2.7})~$ps$^{-1}$~\cite{dge1} 
(the current value of $m_b^{\overline {MS}}=(4.18 \pm 0.03)\gev$~\cite{pdg2014} is used to calculate $\zeta$).
Earlier determinations of $\zeta$ can be found in \cite{V1,V2,V3,V4,V5,V6,V7}. 

The uncertainty on $|V_{ub}|$ due to the application of the shape function 
is derived from $10\%$ variations of ${\bar{\Lambda}}^{\textrm{SF}}$ and ${\lambda}_1^{\textrm{SF}}$,
as prescribed by the authors. The estimated total theoretical uncertainty on $|V_{ub}|$ is about $2.1\%$ 
(for $p_{e} > 0.8 \gevc$).

\subsection{BLNP predictions}

The BLNP calculations incorporate all known perturbative and power corrections and
interpolation between the HQE and SF regions~\cite{blnp1,blnp2,blnp3}. 
The differential and partially integrated spectra for the inclusive 
$B \to X_u l \nu$ decay are calculated in perturbative theory at next-to-leading order (NLO) in renormalization-group, 
and at the leading power in the heavy quark expansion. 
Formulas for the triple differential rate of $B \to X_u l \nu$ and for the $B \to X_s \gamma$ 
photon spectrum are convolution integrals of weight functions with the shape function renormalized
at the intermediate scale $\mu_i$. The ansatz for the leading SF depends on two parameters, 
$m_b$ and $\mu^2_{\pi}$; subleading SFs are treated separately. 

The SF parameters in the kinetic scheme are determined by fits to moments 
of the hadron mass and lepton energy spectra from inclusive $B \to X_c \ell \nu$ decays 
and either additional photon energy moments in $B\to X_s \gamma$ decays or
by applying a constraint on the $c$-quark mass, $m_c^{\overline{MS}}(3\gev)=0.998 \pm 0.029 \gevcc$. 
These parameters are translated from the kinetic to the SF mass scheme \cite{neubert_loops}. 

The impact of the uncertainties in these SFs are estimated by varying the scale parameters $\mu_i$ 
and choices of different subleading SF. The next-to-next-to-leading order (NNLO) corrections were studied 
in detail~\cite{blnp_nnlo}. In extractions of $|V_{ub}|$, the choice $\mu_i = 1.5$~\gev introduces
for the NNLO corrections significant shifts to lower values of the partial decay rates, by $\sim 15\%-20\%$, 
while at the same time reducing the perturbative uncertainty on the scale $\mu_h$.
At NLO, small changes of the value of $\mu_i$ impact the agreement between the NLO and NNLO results. 
We adopt the authors' recommendation and use values $\mu_i = 2.0$~\gev and $\mu_h = 4.25$~\gev, as the default. 
The results obtained in the SF mass scheme with the $m_c$ constraint and $\mu_i = 2.0$~\gev are
$m_b^{\textrm{SF}}=(4.561 \pm 0.023) \gev$ and $\mu_{\pi}^{2\textrm{ SF}} = (0.149\pm 0.040) \gev^2$~\cite{hfag12}.
The $1\sigma$ contours for different choices of these parameters are presented in Fig.~\ref{f:mbmupi2}.

\begin{figure}[htbp]
\begin{center}
\includegraphics[width=9.5cm]{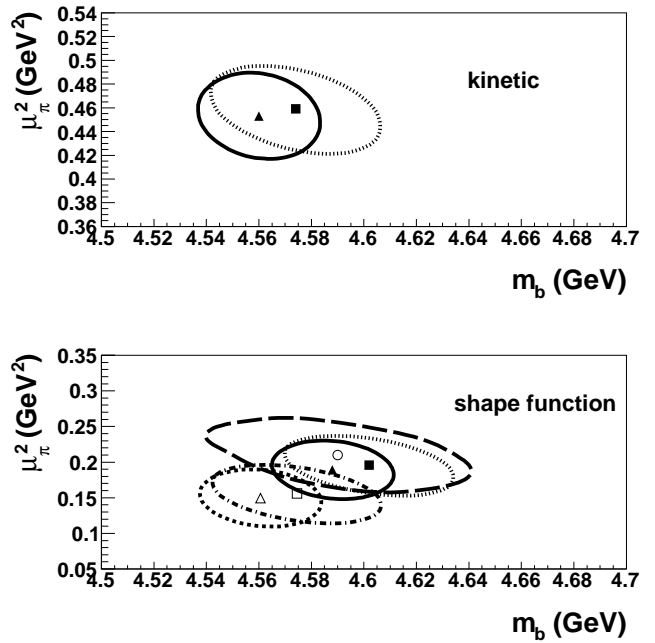}  
\caption{
The shape function parameters $m_b$ and $\mu_{\pi}^2$ in the kinetic scheme (HFAG 2014):
fit to $X_c$ data with constraint on the $c$-quark mass (solid line, solid triangle);
fit to $X_c+X_s\gamma$ data ($\mu_i=1.5$~GeV, $\mu=\mu_i$) (dotted line, solid square).
Translation of fit to $X_c$ data with constraint on the $c$-quark mass (short dashed line, open triangle);
translation of fit to $X_c+X_s\gamma$ data with $\mu_i=2.0$~GeV, $\mu=\mu_i$  (dash-dotted line, open square).
The previous \babar\ endpoint analysis~\cite{babarVub} was based on a $X_s+X_c$ fit (long dashed line, open circle). 
The contours represent $\Delta\chi^2=1$. 
}
\label{f:mbmupi2}
\end{center}   
\end{figure}

In the BLNP framework, the extraction of $|V_{ub}|$ is based
on the predicted partial rate $\zeta(\Delta p)$~\cite{neubert_extract} for
$B \to X_u e \nu$ decays and the measurement of $\Delta {\cal B}$. 
The predictions for total decay rate are: 
\begin{eqnarray}
\zeta = (73.5 \pm 1.9_{\,{\textrm{SF}}} \,^{+5.5}_{-4.9\,{\textrm{theory}}})~\textrm{ps}^{-1} & & \\
m_c \textrm{\ constraint,\ } \mu_i=2.0 \gev, & & \nonumber 
\end{eqnarray} 
\begin{eqnarray}
\zeta = (70.4 \pm 1.9_{\,{\textrm{SF}}} \,^{+6.4}_{-5.2\,{\textrm{theory}}})~\textrm{ps}^{-1} & & \\
m_c \textrm{\ constraint,\ } \mu_i=1.5 \gev, & & \nonumber 
\end{eqnarray}
\begin{eqnarray}
\zeta = (74.5 \pm 2.7_{\,{\textrm{SF}}} \,^{+5.5}_{-4.9\,{\textrm{theory}}})~\textrm{ps}^{-1} & & \\ 
X_s\gamma \textrm{\ constraint,\ } \mu_i=2.0 \gev, & & \nonumber
\end{eqnarray}
\begin{eqnarray}
\zeta = (71.4 \pm 2.7_{\,{\textrm{SF}}} \,^{+6.5}_{-5.3\,{\textrm{theory}}})~\textrm{ps}^{-1} & & \\ 
X_s\gamma \textrm{\ constraint,\ } \mu_i=1.5 \gev. & & \nonumber 
\end{eqnarray}
The estimated SF uncertainty and total theoretical uncertainty on $|V_{ub}|$ 
are about $5.0\%$ and $3.6\%$, respectively (for $p_{e} > 0.8$~\gevc).

\subsection{GGOU predictions}

The GGOU calculations~\cite{ggou1, ggou2} of the triple differential decay rate include all perturbative and 
nonperturbative effects through $O(\alpha_s^2 \beta_0)$ and $O(1/m_b^3)$. The Fermi motion is parametrized in terms of 
a single light-cone function for each structure function and for any value of $q^2$, accounting for all subleading effects. 
The calculations are based on the kinetic mass scheme, with a hard cutoff at $\mu = 1\gev$.

The SF parameters are determined by fits to moments of the hadron mass and lepton energy spectra from inclusive 
$B \to X_c \ell \nu$ decays, and either including photon energy moments in $B\to X_s \gamma$ decays or
by applying a constraint on the $c$-quark mass. The results obtained in the kinetic scheme with the $m_c$ constraint 
are $m_b^{\textrm{kin}}(1.0 \gev)=(4.560 \pm 0.023) \gev$ and    
$\mu_{\pi}^{2\textrm{ kin}}(1.0 \gev) = (0.453\pm 0.036) \gev^2$~\cite{hfag12}.  
The $1\sigma$ contours for the resulting SF parameters are presented in Fig.~\ref{f:mbmupi2}.

The uncertainties are estimated as prescribed in ~\cite{ggou2}. 
To estimate the uncertainties of the higher order perturbative corrections, the
hard cutoff is varied in the range $0.7 < \mu < 1.3$~\gev.
Combined with an estimate of $40\%$ of the uncertainty in $\alpha_s^2 \beta_0$ corrections, 
this is taken as the overall uncertainty of these higher order perturbative and nonperturbative calculations.
The uncertainty due to weak annihilation is assumed to be asymmetric, {\it i.e.}, it tends to decrease $|V_{ub}|$. 
The uncertainty in the modeling of the tail of the $q^2$ distribution is estimated by comparing 
two different assumptions for the range $(8.5-13.5)\gev^2$. 

The extraction of $|V_{ub}|$ is based on the measured partial BF $\Delta {\cal B}(\Delta p)$, 
and the GGOU prediction for the partial normalized rate 
$\zeta(\Delta p)$. The predictions for the total decay rate are,
\begin{eqnarray}
\zeta = (67.2 \pm 1.6_{\,{\textrm{SF}}} \,^{+2.5}_{-1.3\,{\textrm{theory}}})~\textrm{ps}^{-1} & & \\ 
m_c \textrm{\ constraint,} & & \nonumber
\end{eqnarray}  
\begin{eqnarray}
\zeta = (67.9 \pm 2.3_{\,{\textrm{SF}}} \,^{+2.8}_{-5.1\,{\textrm{theory}}})~\textrm{ps}^{-1} & & \\
X_s\gamma \textrm{\ constraint.} & & \nonumber
\end{eqnarray}
The estimated uncertainties on $|V_{ub}|$ for the SF and the total theoretical uncertainty are 
about $4.1\%$ and $2.0\%$, respectively (for $p_{e} > 0.8$~GeV/c).

\subsection{DGE predictions} 

The DGE~\cite{gardi} is a general formalism for inclusive distributions near the kinematic boundaries.
In this approach, the on-shell calculation, converted to hadronic variables, is directly used as an approximation to the decay 
spectrum without the use of a leading-power nonperturbative function. The perturbative expansion includes NNLO resummation in 
momentum space as well as full $O(\alpha_s)$ and  $O(\alpha^2_s\beta_0)$ corrections. 
The triple differential rate of $B \to X_u l \nu$ was calculated~\cite{dge1,dge3}. 
The DGE calculations rely on the $\overline{MS}$ renormalization scheme. 

Based on the prescriptions by the authors~\cite{dge3}, we have estimated the  uncertainties in these calculations and their 
impact on $|V_{ub}|$. The theoretical uncertainty is obtained by accounting for the uncertainty in $\alpha_s = 0.1184 \pm 0.0007$ 
and $m_b^{\overline {MS}}=(4.18 \pm 0.03)\gev$~\cite{pdg2014}. 
The renormalization scale factor $\mu/m_b$=1.0 is varied between 0.5 and 2.0, and the default values 
of $(C_{3/2},f^{pv})=(1.0,0.0)$ are changed to $(C_{3/2},f^{pv})=(6.2,0.4)$ to assess the uncertainties in the 
nonperturbative effects.  

DGE predict the shape of differential electron spectrum, but do not provide a normalization.
Thus we rely on the DGE predictions for $f_u(\Delta p)$, the fraction of $B \to X_u e \nu$ decays in the interval $\Delta p$,  
and an independent prediction for the normalized total decay rate, 
$\zeta = (65.7^{+2.4}_{-2.7})~$ps$^{-1}$~\cite{dge1} to derive $\Delta \zeta (\Delta p)$ 
(the current value of $m_b^{\overline {MS}}=(4.18 \pm 0.03)\gev$~\cite{pdg2014} is used to calculate $\zeta$).

The estimated total theoretical uncertainty on $|V_{ub}|$ for DGE calculations is about $2.2\%$ (for $p_{e} > 0.8$~\gevc). 

\section{Analysis}
\label{sec:analysis}

\subsection{Event Selection}
\label{sec:Selection}

To select \BB events with a candidate electron from a semileptonic $B$ meson decay, we apply the following criteria:
\begin{itemize}
\item[]{\it Electron selection:}
We select events with at least one electron candidate in the c.m. momentum range 
$0.8 < p_{\textrm{cms}} < 5.0\,\gevc$ and within the polar angle acceptance in the laboratory frame 
of $-0.71 < \cos\theta_e < 0.90$. Within these constraints the identification efficiency for electrons exceeds $94\%$. 
The average hadron misidentification rate is about $0.1\%$. 
\item[] {\it Track multiplicity:}
To suppress background from non-\BB events, primarily low-multiplicity QED processes, 
including $\tau^+ \tau^-$ pair production and $e^+e^- \to q \bar{q} (\gamma)$ annihilation 
($q$ represents a $u, d, s$ or $c$ quark), we reject events with fewer than four charged tracks.
\item[] {\it $J/\psi$ suppression:} 
To reject electrons from the decay $J/\psi \to e^+e^-$, we combine the selected electron with other 
electron candidates of opposite charge and reject the event if the invariant mass of any pair is consistent 
with a $J/\psi$ decay, $3.00<m_{e^+e^-}<3.15\,\gevcc$. 
\end{itemize}
\noindent  If an event in the remaining sample has more than one electron that passes this selection, 
the one with the highest momentum is chosen as the signal candidate. 

To further suppress non-\BB events we build a neural network (NN) with the following input 
variables which rely on the momenta of all charged particles and energies of photons above 50~\mev\ detected 
in the event:
\begin{itemize}
\item[(i)] ${\cal R}_2$, the ratio of the second to the zeroth Fox-Wolfram moments \cite{foxw}, calculated
from all detected particles in the event [Fig.~\ref{fig:par_net}(a)]. 
\item[(ii)] $l_2=\sum_i p_i \cos^2\theta_i / 2E_{\textrm{beam}}$, where the sum includes all detected particles except the electron,  
and $\theta_i$ is the angle between the momentum of particle $i$ and the direction of the electron momentum [Fig.~\ref{fig:par_net}(b)]. 
\item[(iii)] $\cos\theta_{e-\textrm{roe}}$, the cosine of the angle between the electron momentum and the axis of the thrust 
of the rest of the event [Fig.~\ref{fig:par_net}(c)].
\end{itemize}

\begin{figure}[htbp]
\begin{center}
\includegraphics[height=5.7cm]{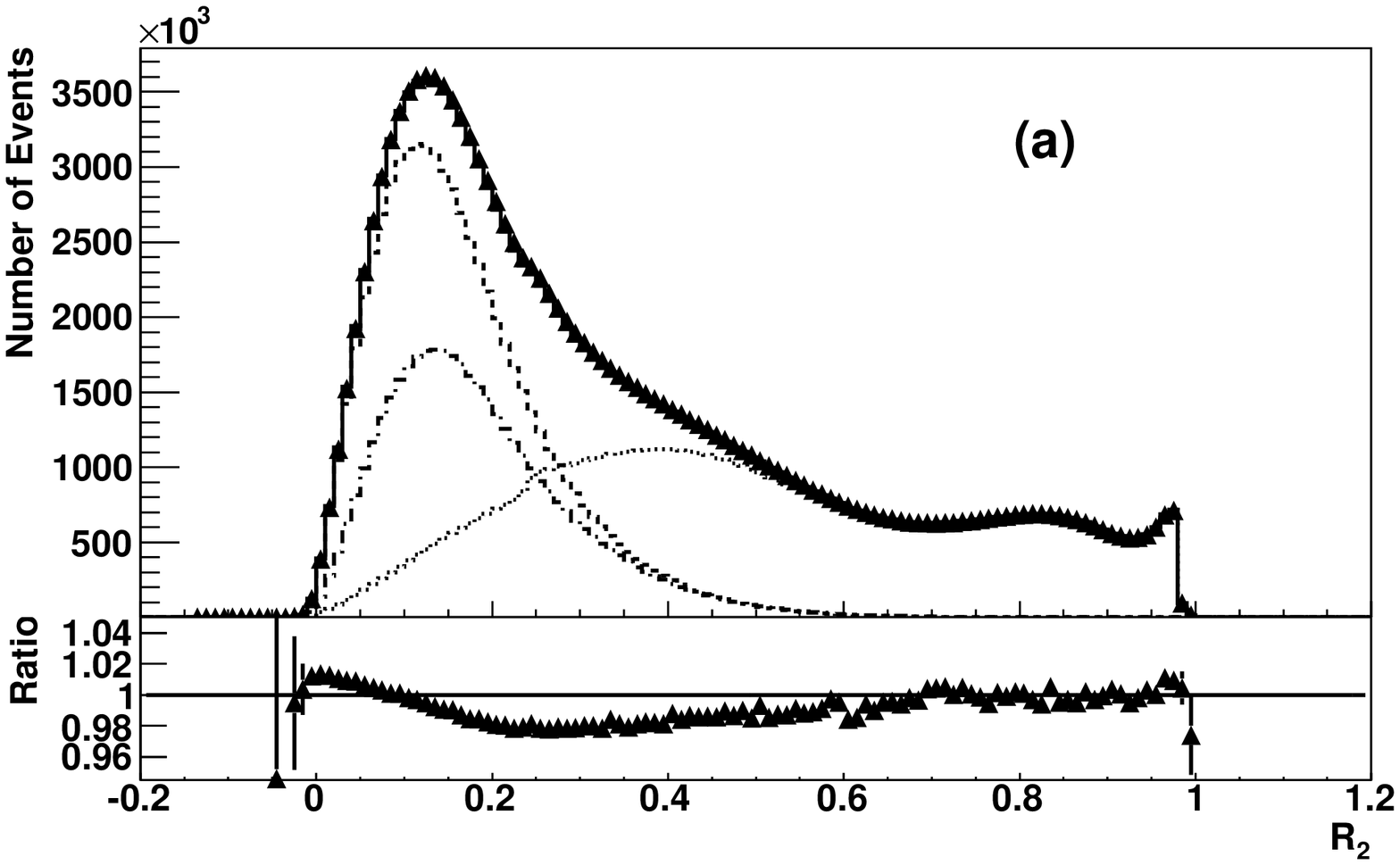}
\includegraphics[height=5.7cm]{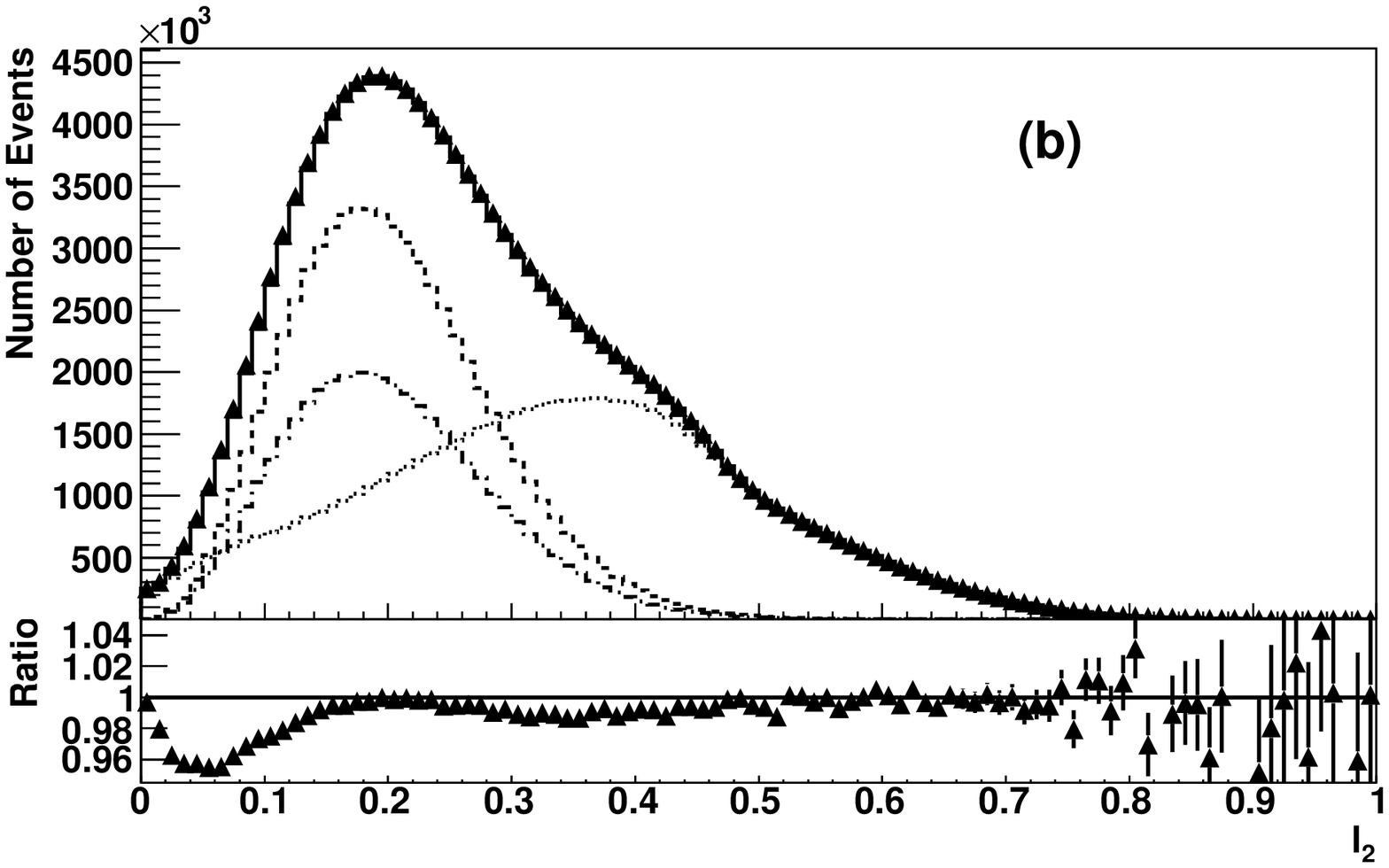}
\includegraphics[height=5.7cm]{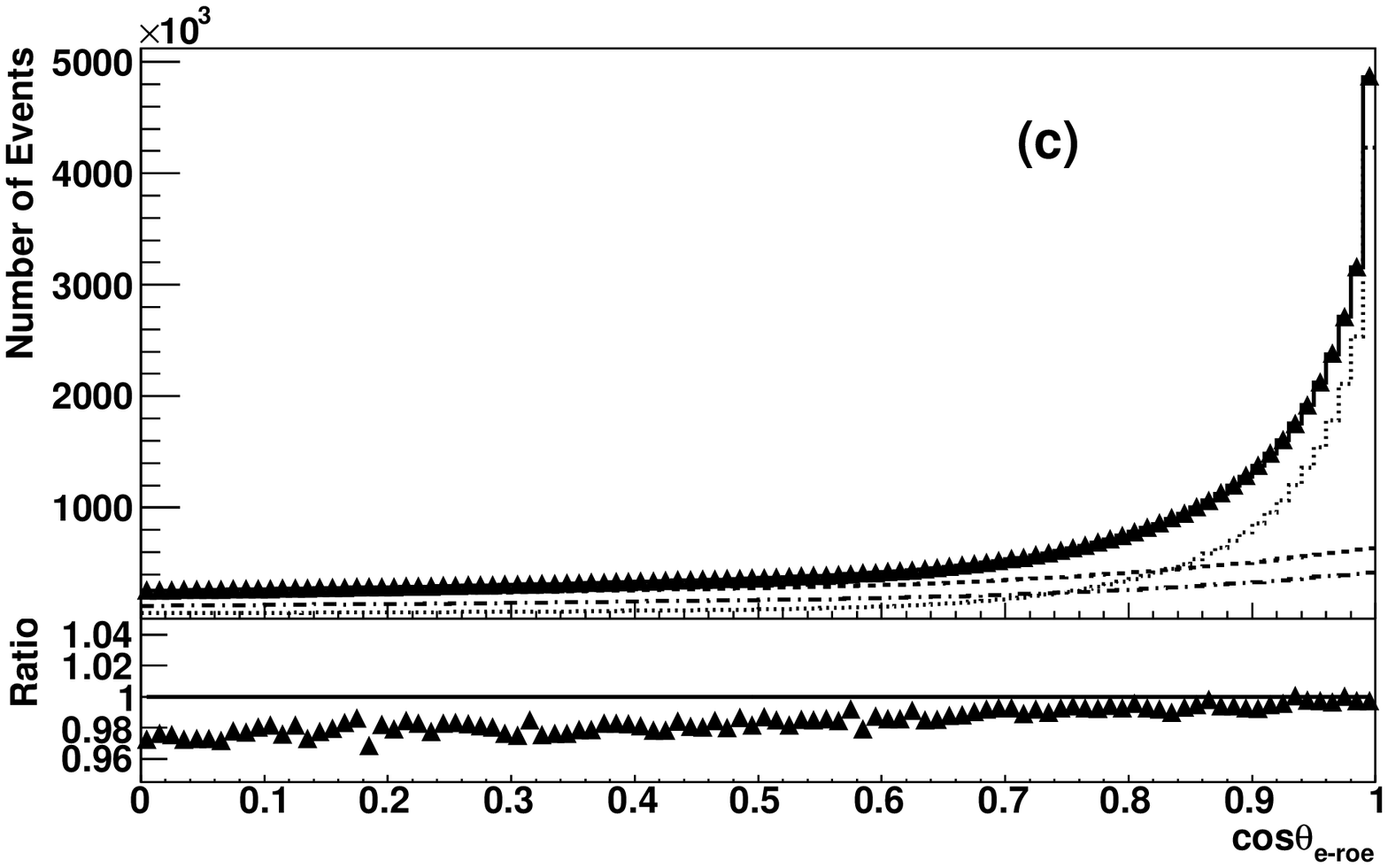}
\caption{
The number of events before the NN selection, as a function of (a) ${\cal R}_2$, (b) $l_2$, (c) $cos\theta_{e-\textrm{roe}}$: 
on-resonance data (triangles), the sum of simulated \BB events and off-resonance data (solid histogram),
MC simulated \BB events (dashed histogram), and off-resonance data (dotted histogram). 
For comparison, the distributions for \BB events with a signal $B\to X_u e \nu$ decay (dash-dotted histogram) 
are shown (scaled by a factor of 50). Ratio = [\BB(MC) + off-resonance]/on-resonance. 
}
\label{fig:par_net}
\end{center}
\end{figure}
%t 

\noindent The distribution of the NN output is shown in Fig.~\ref{fig:net}. 
Only events with positive output values are retained, this selects $\sim 90\%$ of $B \to X_u e \nu$ and $\sim 20\%$ non-\BB events. 
The positive output corresponds the selection with maximum significance level. 

This selection results in an efficiency of $50\%-60$\% 
for $B \to X_u e \nu$ decays; the dependence on the electron momentum is shown in Fig.~\ref{fig:eff}. 

\begin{figure}[htbp]
\begin{center}
\includegraphics[height=5.5cm]{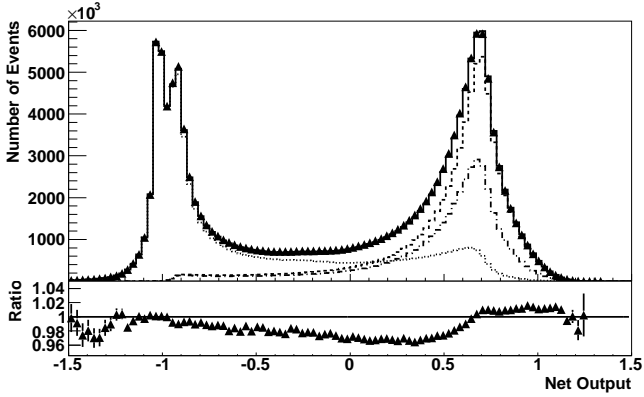}
\caption{
The distribution of events as a function of the NN output: 
On-resonance data (triangles), the sum of MC simulated \BB and off-resonance data (solid histogram),
MC simulated \BB events (dashed histogram), off-resonance data (dotted histogram). 
For comparison, the distributions for \BB events with a signal $B\to X_u e \nu$ decay (dash-dotted histogram),
are shown (scaled by a factor of 50). Ratio = [\BB(MC) + off-resonance]/on-resonance.}
\label{fig:net}
\end{center}
\end{figure}
\begin{figure}[htbp]
\begin{center}
\includegraphics[height=6.cm]{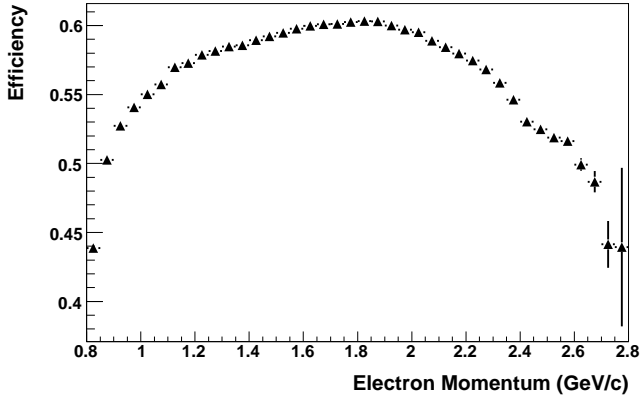}
\caption{
Selection efficiency for \BB events with a $B \rightarrow X_u e \nu$ decay as a
function of the MC-generated electron momentum. The error bars represent the statistical uncertainties only. 
}
\label{fig:eff}
\end{center}
\end{figure}

\subsection{Background subtraction}
\label{sec:background}

The selected sample of events from the on-resonance data contains considerable 
background from \BB events and non-\BB events. The \BB background is dominated by primary electrons 
from semileptonic $B$ decays and secondary electrons from decays of charm mesons and $J/\psi$ mesons. 
Hadronic $B$ decays contribute mostly via the misidentification of charged particles.
Non-\BB events originate from $e^+e^- \to \qqbar(\gamma)$ annihilation and lepton pair production, 
especially $e^+ e^- \to \tau^+\tau^-$.

\subsubsection{Non-\BB background}
\label{sec:nonbb}

To determine the momentum-dependent shape of the non-\BB background, we perform a binned $\chi^2$ fit to
the off-resonance data in the momentum interval 0.8 to 3.5~\gevc, combined with on-resonance data in 
the momentum interval 2.8 to 3.5~\gevc, {\it i.e.,} above the endpoint for electrons from $B$ decays. 
Since the c.m. energy for the off-resonance data is $0.4\%$ lower than for the on-resonance data, 
we scale the electron momenta for the off-resonance data by the ratio of the c.m.\ energies. 

The relative normalization for the two data samples \cite{lumi, detector} is 
$$ r_{L}^{(0)} = \frac{s_{\textrm{OFF}}} 
{s_{\textrm{ON}}} \frac{\int\! L_{\textrm{ON}}\,dt} 
{\int\! L_{\textrm{OFF}}\,dt} = 9.560 \pm 0.003_{\textrm{stat}} \pm 0.006_{\textrm{syst}}, $$
where $s$ and $\int\! L\,dt$ refer to the c.m.\ energy squared and integrated luminosity of the on- and off-resonance data.
The statistical uncertainty of $r_L$ of $0.03\%$ is based on the number of detected $\mu^+\mu^-$ pairs used for 
the measurement of the integrated luminosity; the relative systematic uncertainty on the ratio is estimated to be $0.06\%$.

The binned $\chi^2$ for the fit to the electron spectrum for selected non-\BB events is defined as  

\begin{eqnarray}
\chi^2_c & = & \sum_{p_i>0.8\gevc} \frac{(f(\vec{a},p_i)-r_{L}  n_i)^2}{r_{L}^{2}  n_i} \nonumber \\
&+& \sum_{p_j>2.8\gevc} \frac{(f(\vec{a},p_j)-N_j)^2}{ N_j}  \nonumber \\
&+& \frac{(r_L-r_L^{(0)})^2}{\sigma_{r_L}^2}.
\end{eqnarray}

\noindent Here $n_i$ and $N_j$ refer to the number of selected events in the off- and on-resonance samples 
for momentum bins $i$ and $j$, respectively, $\vec{a}$ represents the set of free parameters of the fit, and
$\sigma_{r_L}$ is the uncertainty on $r^{(0)}_L$. To fit the momentum spectrum, we have chosen an exponential 
expression of the form, 

\begin{eqnarray}
\label{function:cf}
f(\vec{a},p) & = & a_0 ( \exp(a_1 p + a_2 p^2 + a_3 p^3)  \nonumber \\ 
&+& \exp(a_4 p + a_5 p^2) ).
\end{eqnarray}

\noindent We perform three different fits and they all describe the data well, see Table {\ref{t:nonBB}}.
The results of the fit to both on- and off-resonance data are shown in Fig.~\ref{f:cfit}.
In the fit to the full on-resonance data spectrum, the constraint on the ratio  
$r_L/r_L^{(0)}=1.0000 \pm 0.0007$ is applied.

\begin{table}[tbp]
\caption{
Results of the fit to the non-\BB background
}
\label{t:nonBB}
{\small
\begin{ruledtabular}
\begin{tabular}{lcc}
Data                   & $ \chi^2/N_{d.o.f.}$     &  $r_L/r_L^{(0)}$ \\ \hline
Off-resonance only     & $  65.0/48 $       &                 \\

Off- and on-resonance  & $  70.6/61 $       &  $1.007 \pm 0.004$ \\

Off- and on-resonance  &        &   \\

with $r_L/r_L^{(0)}$ constrained  & $  73.0/62 $  &  $1.0002 \pm 0.0007$       \\

\end{tabular}
\end{ruledtabular}
}
\end{table}

\begin{figure}[htbp]
\begin{center}
\includegraphics[height=6.cm]{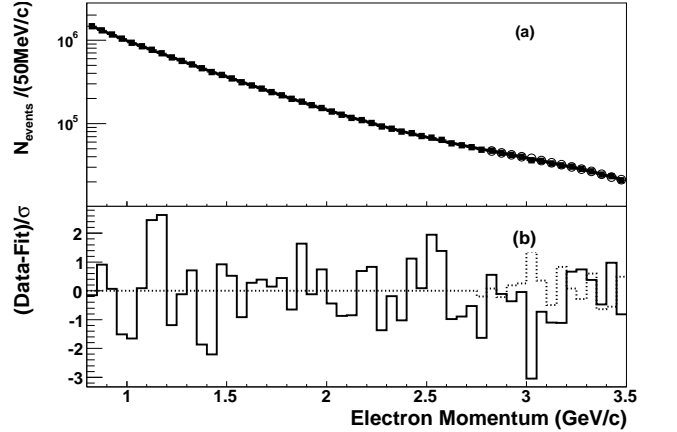}
\caption{
The combined fit to off-resonance data in the momentum interval of
0.8 to 3.5 \gevc and to on-resonance data in the momentum
interval of 2.8 to 3.5 \gevc, with the constraint on $r_L/r_L^{(0)}$;
(a) comparison of off-resonance data (solid squares), on-resonance data (open circles) and fitted function;
(b) $\textrm{(Data-Fit)}/\sigma$: off-resonance data (solid histogram),  
on-resonance data (dotted histogram) 
}
\label{f:cfit}
\end{center}
\end{figure}

In Fig.~\ref{fig:pe_f}(a) the data and the result of this fit to the non-\BB background are compared to the 
full spectrum of the highest momentum electron in selected events observed in the on-resonance data sample. 
By subtracting the fitted non-\BB background we obtain the inclusive electron spectrum from $\FourS$ decays, 
shown in Fig.~\ref{fig:pe_f}(b). Above 2.3 \gevc, an excess of events corresponding to the expected
signal $B \to X_u e \nu$ decays is observed above the \BB background.

\begin{figure}[htbp]
\begin{center}
\includegraphics[width=9.5cm]{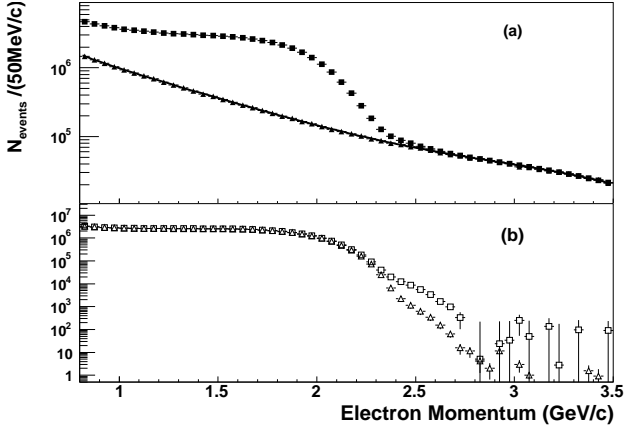}
\caption{
Electron momentum spectra in the $\FourS$ rest frame:
(a) on-resonance data (solid squares), scaled off-resonance data (solid triangles), 
the solid line shows the results of the fit to the continuum component using both on-resonance and off-resonance data.
(b) On-resonance data with non-\BB background subtracted (open squares), 
\BB MC without $B \to X_u e \nu$ decays (open triangles).
}
\label{fig:pe_f}
\end{center}
\end{figure}

\subsubsection{\BB Background and fit to the electron spectrum}
\label{sec:bbfit}

\begin{figure}[htbp]
\begin{center}
\includegraphics[height=5.7cm]{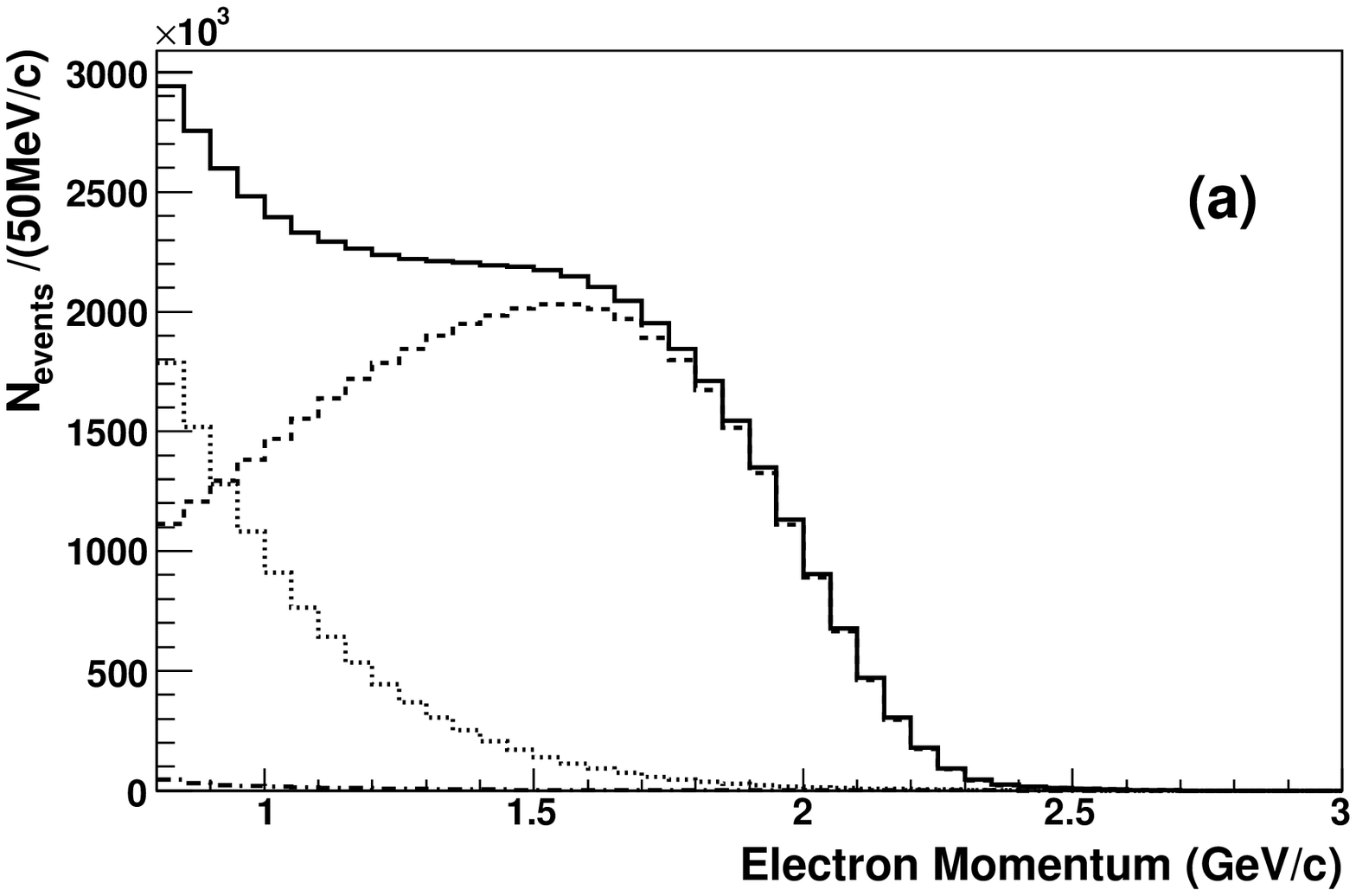}
\includegraphics[height=5.7cm]{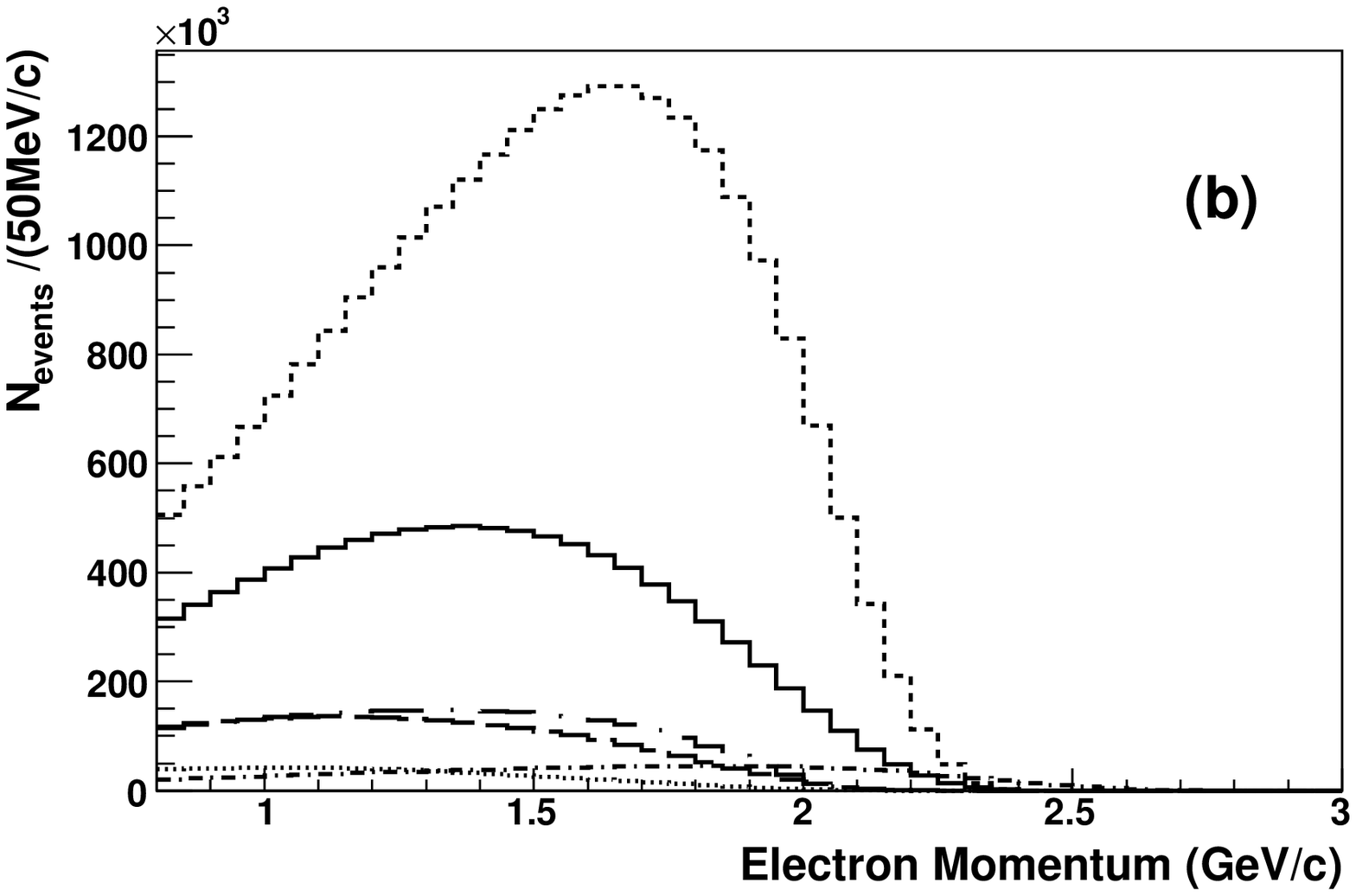}
\includegraphics[height=5.7cm]{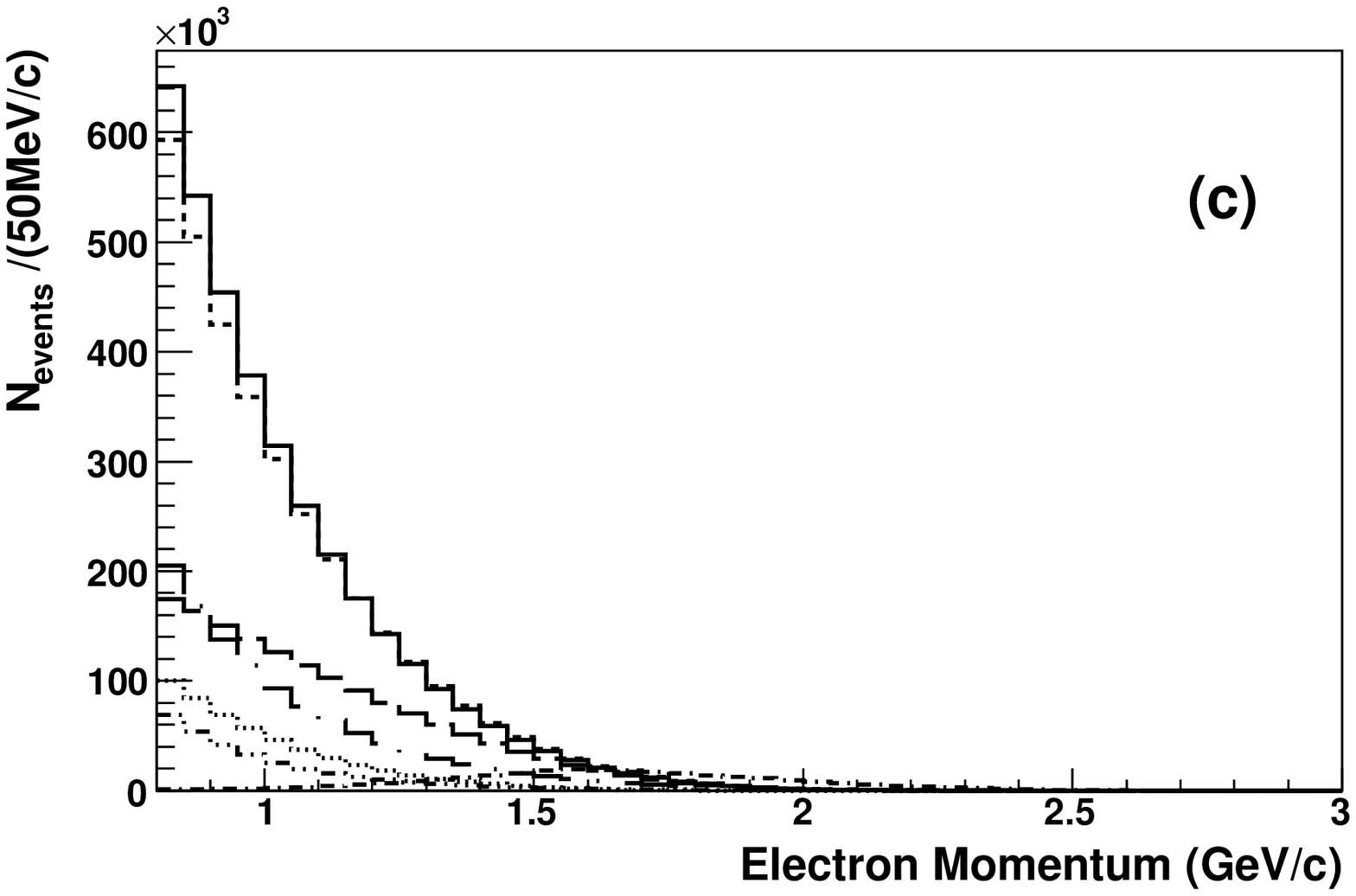}
\caption{The simulated contributions to
\BB events as a function of the momentum for electron candidates
(a) all events (solid histogram), primary electrons (dashed histogram),
secondary electrons (dotted histogram), misidentified hadrons (dash-dotted histogram).
(b) Primary electrons: $B \to D e \nu$(solid histogram),
$B \to D^* e \nu$ (dashed histogram), $B \to D^{(*)}\pi e \nu$ (dotted histogram),
$B \to (D_0^*+D_1^*)e\nu$ (long-dash histogram), $B \to (D_1+D_2^*)e\nu$ (long-dash-dotted histogram),
signal $B \to X_u e \nu$ decays (dash-dotted histogram).
(c) Secondary electrons from: $D^{\pm}$ (solid histogram),
$D^{0}(\bar{D}^{0})$ (dashed histogram), $D_s$ (dotted histogram),
$J/\psi$ (dash-dotted histogram), $\tau$ (long-dash histogram),
$\gamma$ conversion (long-dash-dot histogram), other $e^{\pm}$ (dash-three-dot histogram). 
}
\label{fig:hsel}
\end{center}
\end{figure}

The \BB background spectrum is composed of several contributions, dominated by primary electrons from various 
semileptonic $B$ decays, and secondary electrons from decays of $D$, $D_s$ and $J/\psi$ mesons or photon conversions.
Hadronic $B$ decays contribute mostly via charged-particle misidentification, primarily at low momenta. 
The MC simulated contributions from different background sources are shown in Fig.~\ref{fig:hsel}. 

We estimate the total background by a simultaneous fit to the observed inclusive electron spectra in off- and 
on-resonance data to the sum of the signal and individual background contributions. For the individual signal 
and \BB background contributions, we rely on the MC simulated shapes of the spectra (including some corrections), 
and treat their relative normalization as free parameters in the fit.
For this extended fit, we expand the $\chi^2$ definition as follows, 

\begin{eqnarray}
\label{fitf}
\chi^2 & = & \sum_{i,j}
(f(\vec{a},p_i)+S(\vec{b},\vec{t},p_i)-N_i) V_{ij}^{-1} \nonumber \\
& & (f(\vec{a},p_j)+S(\vec{b},\vec{t},p_j)-N_j)  \nonumber \\
&+& \sum_i \frac{\textstyle (f(\vec{a},p_i)-r_L n_i)^2}{\textstyle r_L^2 n_i}  \nonumber \\
&+&
\sum_{k} \frac{(b_{k}-b_{k}^{(0)})^2}{\sigma_{b_{k}}^2}+
\frac{(r_L-r_L^{(0)})^2}{\sigma_{r_L}^2},
\end{eqnarray}
\noindent with 

\begin{equation}
V_{ij} = (N_i +\sigma_{\textrm{(MC)}\,i}^2) \delta_{ij} + V_{ij}^{\textrm{PID}}. \nonumber
\end{equation}

\noindent Here $n_i$ and $N_j$ refer to the number of selected events in the off- and on-resonance 
samples for momentum bins $i$ and $j$, respectively, and $\vec{a}$ is the set of free parameters of the fit.

The first sum refers to the on-resonance data. The \BB electron 
spectrum is approximated as $S(\vec{b},\vec{t},p)=\sum_k b_k g_k(\vec{t},p_i)$, where the free parameters 
$b_k$ are the BFs for the individual contributions $g_k(\vec{t},p_j)$ representing the signal 
$B\to X_ue\nu$ decays, the background of primary electrons from semileptonic $B$ decays 
($De\nu$, $D^*e\nu$, $D^{(*)} \pi e\nu$, $D^{**} e\nu$ and $D'^{(*)}$), 
and secondary electrons from decays of $D$ and $D_s$ mesons ($D,D_s \to e$ fitted as a scale factor relative to the MC input). 
Smaller contributions to the \BB background are fixed in the fit, for example, 
electrons from $J/\psi$ and $\tau^{\pm}$ decays, photon conversions, and hadrons misidentified as electrons. 
Their simulations and rates rely on independent control samples and are well understood. 

The momentum spectra $g_k(\vec{t},p_j)$ are histograms taken from MC simulations.
The array $\vec{t}$ refers to the form factor parameters $\rho_{D*}^2$, 
$R_1$, $R_2$, and $\rho_D^2$ and other fixed parameters 
such as form factor parameters for $B \to D^{**} e\nu$ and $B \to D'^{(*)} e\nu$ decays. 
$\sigma_{\textrm{(MC)}\,j}$ is the statistical uncertainty of the number of simulated events in the $j$-th bin. 
$V_{ij}^{\textrm{PID}}$ is the covariance matrix for the detection of electrons among charged tracks. It includes 
electron identification and misidentification of pions, kaons, protons and antiprotons,
studied with large data control samples. $V_{ij}^{\textrm{PID}}$ only includes the uncertainty 
for the shape of the momentum distribution due to particle identification (PID) uncertainties. 
The uncertainty of the relative normalization due to PID uncertainties is about $0.5\%$ and is taken as a systematic uncertainty. 
The last two terms of Eq.~(\ref{fitf}) refer to quantities that are well known. 

In the fit, $De\nu$ and $D^{**} e\nu$ contributions are highly correlated. The BF 
for $B \to D e \nu$ is constrained to $0.022 \pm 0.001$~\cite{pdg2014}. 
The luminosity ratio is constrained to the value $r_{L}^{(0)}=9.560 \pm 0.007$~\cite{lumi, detector}. 

The fit is performed in the momentum range from 0.8 to 3.5\gevc, in bins of 50\mevc. At lower momenta, the data determine 
the relative normalization of the various background contributions, allowing for an extrapolation of these backgrounds 
into the endpoint region. This fitting procedure was chosen in recognition of the fact that the current BFs for the individual 
$B \to  X_c \ell \nu$ decays are not sufficiently well measured to perform an adequate background subtraction.
The shape of the signal spectrum is fixed in the fit to one of the theoretical predictions, its normalization is a free parameter.
In a given momentum interval, the excess of events above the sum of the fitted backgrounds is taken as the number of signal events.

To reduce a potential systematic bias from the theoretically predicted shape of the signal spectrum in a region 
where these calculations are less reliable, events in the interval from 2.1 to 2.7 \gevc are combined into a single bin. 
The lower limit of this bin is chosen so as to retain sensitivity to the steeply falling \BB background
distributions, while containing a large fraction of the signal events in a region where the background is low.
The upper limit at 2.7~\gevc\ is chosen to limit the non-\BB background at higher momenta where the signal contributions become 
very small compared to the non-\BB background.

The fits are performed separately for the different theoretical predictions of the signal spectrum, 
introduced in Sec.~\ref{sec:models}. 
The results of these fits are shown in Table~\ref{t_fit}, and for the fit with the GGOU signal spectrum, 
the correlation matrix is presented in Table~\ref{t:cor_matrix}. The differences of the correlation matrices 
for the fit with DN, BLNP, GGOU and DGE signal spectra are small.  
The difference in the fit results for the $B \to X_u e \nu$ BF is primarily due to the difference
between the various predictions for the fraction of the signal spectrum in the high-momentum range. 
The fitted BFs for the dominant $B \to X_c e \nu$ decays agree reasonably well with 
expectations~\cite{pdg2014}. 

\begin{table*}[tbp]
\caption{
Results of the fit to the electron spectrum,  with the non-\BB background subtracted and with all entries 
in the interval from 2.1 to 2.7 \gevc in a single bin, for four different theoretical predictions of the $X_u e\nu$ spectrum.
Fitted BFs(\%), averaged over charged and neutral $B$ mesons, for the signal $X_u e\nu$, the background $De\nu$ (constrained), 
$D^*e\nu$, $D^{(*)} \pi e\nu$, $D^{**} e\nu$, $D^{\prime}(2.55) e\nu$+$D^{\prime *}(2.60) e\nu$, 
and scale factors relative to reweighted MC inputs for secondary $D \to e$, and the luminosity ratio $r_L$ (constrained) are presented.
The contributions to the $\chi^2$ from the on-resonance and the off-resonance data and constraints, and the ratio of the total $\chi^2$
per degree of freedom are listed at the bottom. 
}
\label{t_fit}
{\small
\begin{ruledtabular}
\begin{tabular}{lllll} 

  & DN & BLNP$_{\mu_i=2.0\gev}$ & GGOU & DGE \\
  &    & $m_c$ constraint & $m_c$ constraint &  \\

\hline
$X_u e\nu$          & $0.149 \pm 0.005$  &  $0.240 \pm 0.008$ & $0.166 \pm 0.006$ & $0.153 \pm 0.005$ \\

$De\nu$             & $2.233 \pm 0.090$  &  $2.197 \pm 0.088$ & $2.226 \pm 0.089$ & $2.230 \pm 0.089$ \\

$D^*e\nu$           & $5.612 \pm 0.049$  & $5.424 \pm 0.049$ & $5.579 \pm 0.048$ & $5.611 \pm 0.048$ \\

$D^{(*)}\pi e\nu$   & $< 0.052        $  & $< 0.025$ & $< 0.050$ & $< 0.075$ \\

$D^{**} e\nu$       & $2.285 \pm 0.071$  & $2.540 \pm 0.075$ & $2.331 \pm 0.070$ & $2.287 \pm 0.070$ \\

$D^{\prime(*)} e\nu$& $0.046 \pm 0.011$  & $0.023 \pm 0.011$ & $0.041 \pm 0.011$ & $0.045 \pm 0.011$ \\

$D \to e$           & $0.982 \pm 0.005$  & $0.968 \pm 0.005$ & $0.980 \pm 0.005$ & $0.982 \pm 0.005$ \\

$r_L/r_L^{(0)}$     & $1.0002 \pm 0.0007$ & $1.0002 \pm 0.0007$ & $1.0002 \pm 0.0007$ & $1.0002 \pm 0.0007$ \\ 

\hline
$\chi^2_{\textrm{ON}} + \chi^2_{\textrm{OFF}} + \chi^2_{\textrm{constraints}}$       & 
$27.4+69.7+0.1$ &  $31.9+70.9+0.2$ & $27.8+69.9+0.1$ & $26.8+69.7+0.1$ \\
$\chi^2/N_{d.o.f.}$                     & $97.2/85$ &  $102.9/85$ & $97.8/85$ & $96.6/85$
\\

\end{tabular}
\end{ruledtabular}
}
\end{table*}

\begin{table*}[tbp]
\caption{Correlation matrix for the fit to the total electron spectrum with the GGOU prediction of the signal spectrum, 
with the contributions to the spectrum and the parameters of the background function, $a_0$ through $a_5$.
}
\label{t:cor_matrix}
\begin{center}
\begin{ruledtabular}
\begin{tabular}{>{\scriptsize}r >{\scriptsize}r >{\scriptsize}r >{\scriptsize}r >{\scriptsize}r >{\scriptsize}r >{\scriptsize}r >{\scriptsize}r >{\scriptsize}r >{\scriptsize}r >{\scriptsize}r >{\scriptsize}r >{\scriptsize}r >{\scriptsize}r >{\scriptsize}r}   

& $De\nu$ & $D^*e\nu$ & $D^{(*)}\pi e\nu$ & $D^{**} e\nu$ & $D^{\prime(*)} e\nu$ & $X_u e\nu$ & $D \to e$ &
$a_0$ & $a_1$ & $a_2$ & $a_3$ & $a_4$ & $a_5$ & $r_L/r_L^{(0)}$ \\ \hline
$De\nu$ & 1 & -0.827 & 0.032 & -0.398 & -0.449 & -0.305 & -0.060 & 0.018 & -0.048 & 0.058 & -0.036 & 0.023 & -0.032 & 0.001 \\
$D^*e\nu$ &  & 1 & -0.024 & -0.158 & 0.784 & -0.128 & 0.309 & 0.050 & 0.029 & -0.146 & 0.126 & 0.038 & 0.125 & 0.008 \\
$D^{(*)}\pi e\nu$ &  &  & 1 & -0.031 & 0.004 & 0.027 & 0.012 & -0.066 & 0.033 & 0.033 & -0.048 & -0.044 & -0.052 & -0.028 \\ 
$D^{**} e\nu$ &  &  &  & 1 & -0.601 & 0.598 & -0.361 & -0.062 & 0.030 & 0.055 & -0.063 & -0.055 & -0.066 & -0.012 \\
$D^{\prime(*)} e\nu$ &  &  &  &  & 1 & -0.236 & 0.206 & 0.069 & -0.051 & -0.034 & 0.053 & 0.070 & 0.063 & 0.001 \\
$X_u e\nu$ &  &  &  &  &  & 1 & -0.252 & -0.461 & 0.310 & 0.252 & -0.369 & -0.425 & -0.363 & -0.107 \\
$D \to e$ &  &  &  &  &  &  & 1 & -0.108 & 0.204 & -0.189 & 0.104 & -0.102 & 0.037 & -0.116 \\
$a_0$ &  &  &  &  &  &  &  & 1 & -0.827 & -0.196 & 0.670 & 0.980 & 0.671 & 0.139 \\
$a_1$ &  &  &  &  &  &  &  &  & 1 & -0.315 & -0.190 & -0.870 & -0.209 & -0.103 \\ 
$a_2$ &  &  &  &  &  &  &  &  &  & 1 & -0.801 & -0.122 & -0.818 & 0.012 \\
$a_3$ &  &  &  &  &  &  &  &  &  &  & 1 & 0.610 & 0.947 & 0.035 \\
$a_4$ &  &  &  &  &  &  &  &  &  &  &  & 1 & 0.627 & 0.027 \\
$a_5$ &  &  &  &  &  &  &  &  &  &  &  &  & 1 & -0.006 \\
$r_L/r_L^{(0)}$ &  &  &  &  &  &  &  &  &  &  &  &  &  & 1 \\ 

\end{tabular}
\end{ruledtabular}
\end{center}
\end{table*}

Since the ability of the fit to distinguish between individual $D^{**} e \nu$ decay modes is limited, the sum of the four 
measured decay modes ($D_0^* e\nu$, $D_1^{\prime} e\nu$, $D_1 e\nu$, $D_2^* e\nu$) is treated as single contribution in the fit, 
with the relative BFs for the individual components fixed. Similarly, the contributions from 
$B \to D^{\prime} e \nu$ and $B \to D^{\prime *} e \nu$ are added, and the sum is treated as single contribution, 
with the relative BFs for the two components fixed.

\section{Systematic Uncertainties}
\label{systematics}

The principal sources of systematic uncertainties in the measurement of the lepton
spectrum from 
$B \to X_u e \nu$ decays are the following:
\begin{itemize}
\item the signal selection,
\item the simulation of the signal electron spectrum,
\item the subtraction of the non-\BB background,
\item the subtraction of the \BB background. 
\end{itemize}

To estimate the systematic uncertainties on the signal electron spectrum 
and BFs, we compare the result obtained from the nominal fit 
with results obtained after changes to the MC simulation that reflect 
the uncertainty in the parameters that impact the detector efficiency
and resolution, or the simulation of the signal and background processes.
The sensitivity to the detailed description of the inclusive signal spectrum, 
in particular the theoretical uncertainties in the QCD corrections will 
be discussed in Sec. \ref{results}.

\begin{table}[tbp]
\caption{
Summary of the relative systematic uncertainties (\%) on the partial branching fraction
measurements for  $B \to X_u e \nu$ decays (GGOU), as a function of $p^{\textrm{min}}$, the lower
limit of the signal momentum range, the upper limit is fixed at 2.7 \gevc. The uncertainties 
in the theoretical predictions of the signal spectrum are not included here. 
}
\label{t:t_syst}
{\small
\begin{ruledtabular}
\begin{tabular}{lcccc} %\cline{2-4}
$p^{\textrm{min}} (\gevc)$ &
   $0.8 $  &
   $1.5 $  &
   $2.1 $  &
   $2.3 $  \\
\hline
Single Track efficiency                         & $0.1$ & $0.1$ & $0.1$ & $0.0$ \\
Charged track multiplicity 	                & $1.2$ & $1.9$ & $1.3$ & $1.0$ \\
Particle identification                         & $0.5$ & $0.5$ & $0.5$ & $0.5$ \\
Hadron mis-ID background             		& $0.7$ & $0.7$ & $0.8$ & $0.5$ \\
Photon selection                                & $0.4$ & $0.3$ & $0.4$ & $0.2$ \\
Neural net event selection                      & $^{+3.0}_{-0.8}$ & $^{+3.3}_{-1.2}$ & $^{+3.6}_{-1.2}$ & $^{+3.1}_{-2.1}$ \\
Non-\BB background      			& $0.5$ & $0.5$ & $0.5$ & $0.8$ \\
$B \to X_u e \nu$ exclusive decays              & $0.3$ & $0.3$ & $0.3$ & $0.3$ \\
$B \to D^{(*)} l\nu$ form factors               & $1.1$ & $0.5$ & $1.2$ & $0.2$ \\
$B \to D^{**} e \nu$ form factors               & $0.6$ & $0.4$ & $0.6$ & $0.0$ \\
$B \to D^{**} e \nu$ BF                         & $0.4$ & $1.1$ & $0.5$ & $0.1$ \\
$B \to D^{(\prime)} e \nu$ BF                   & $0.2$ & $0.9$ & $0.2$ & $0.0$ \\
Widths of $D^{(\prime)}$ states                 & $0.2$ & $0.5$ & $0.2$ & $0.0$ \\
$J/\psi$ and $\psi(2S)$ background              & $0.1$ & $0.2$ & $0.1$ & $0.1$ \\
$\tau$ background                               & $0.2$ & $0.7$ & $0.3$ & $0.1$ \\
$B$ momentum                                    & $1.5$ & $1.5$ & $1.6$ & $0.5$ \\
Bremsstrahlung                                  & $0.3$ & $0.1$ & $0.3$ & $0.0$ \\
Final state radiation                           & $0.6$ & $0.6$ & $0.5$ & $0.6$ \\
Width of wide bin                               & $0.4$ & $0.4$ & $0.3$ & $0.0$ \\
$N_{\BB}$ normalization                    & $1.1$ & $1.1$ & $1.1$ & $1.1$ \\
\hline
\noalign{\vskip2pt}
Total exp. systematic uncertainty             & $^{+4.2}_{-3.1}$ & $^{+4.8}_{-3.7}$ & $^{+4.7}_{-3.3}$ & $^{+3.8}_{-3.0}$ \\
Total exp. statistical uncertainty		& $3.8$            & $5.0$            & $3.5$            & $2.8$            \\
\hline
\noalign{\vskip2pt}
Total exp. uncertainty 			& $^{+5.7}_{-4.9}$ & $^{+7.0}_{-6.2}$ & $^{+5.9}_{-4.8}$ & $^{+4.7}_{-4.1}$ \\
\noalign{\vskip2pt}
\end{tabular}
\end{ruledtabular}
}
\end{table}

A summary of the experimental systematic uncertainties is given in Table~\ref{t:t_syst} 
for four intervals of the electron momentum with different lower limits, 
and a common upper limit of $2.7 \gevc$. The uncertainty in the simulation of the detector 
performance and its impact on the momentum dependence of the efficiencies for signal and
background are the experimental limitations of the current analysis.
The largest source of uncertainties in the fit to the observed electron spectrum,
is the event selection, primarily the suppression of the non-\BB background
using a neural network. The impact of the uncertainties in the theoretical predictions of the spectrum 
are not included in this Table.

\subsection{Event selection}

The principal sources of uncertainties in the event selection arise from the efficiency of reconstruction of 
charged-particle tracks, the restriction on the number of charged particles, the electron identification, 
and the application of a neural network.

The single charged-particle tracking efficiency inside the detector acceptance exceeds $96\%$ 
and it is largely independent of momentum. It is well reproduced by the MC simulation.  
To estimate the impact of the uncertainty in the detection of charged-particle tracks, the 
track finding efficiency is decreased by 1 standard deviation ($\sigma$) and the fit is repeated.
The observed impact is $0.1\%$ for electron momenta below $2.3~\gevc$ and increases 
slightly at higher momenta.

The electron identification efficiency has been studied on a high statistics sample of   
radiative Bhabha events, with electron momenta in the laboratory frame extending up to 
10~\gevc. The ratio of efficiencies from Bhabha data and simulated events is measured 
in bins of the polar angle $\theta$ and laboratory momentum and used to correct 
the identification efficiency of electrons in \BB events. 
The uncertainty of this correction is about $0.5\%$. 
The fit to the electron spectrum, in bins of 50 \mevc, incorporates the uncertainty in the 
shape of the momentum distribution, and the momentum dependence of the electron efficiency and 
hadron misidentification using a covariance matrix to account for bin-to-bin correlations.

The requirement of at least four charged tracks in the event suppresses primarily QED processes
in non-\BB background, but also impacts both signal and other background events with low charged-particle 
multiplicity. To estimate the systematic uncertainty we increase the requirement on the minimum 
number of charged tracks in an event from four to five and observe changes in the partial BF of up to $1.9\%$. 
This estimation is rather conservative because the default requirement rejects less than 2\% of reconstructed \BB events. 

To estimate the systematic uncertainty on the modeling of photons, the range of rejected photon
energies was doubled. We increase the requirement on the minimum photon energy from the nominal 
value of $50$ to $100 \mev$ and observe changes in the partial BF of up to $0.4\%$ below 2.3~\gevc, 
increasing at higher electron momenta.

The neural network is primarily designed to suppress non-\BB backgrounds. Its input 
variables depend on the momenta of all tracks and the energies of all photons in the event.  
We have examined the sensitivity of the fit results to the requirement on the NN output value 
(default requirement at 0.0), and observe that for an increase to +0.15, the non-\BB background decreases 
and the signal BF changes by up to $-2.1\%$, whereas for a decrease to $-$0.15, the non-\BB background 
increases and the signal BF changes by up to $+3.6\%$. Decreasing the requirement on the NN output value 
increases the fraction of non-\BB events, worsening the fit quality. 
Specifically, the fit probability changes from $20\%$ for the default requirement to about 
$1\%$ for a requirement at $-$0.15 and to about $10^{-7}$ for a requirement at $-$0.20. Increasing the requirement 
beyond +0.15 increases the statistical uncertainty by an amount consistent
with the observed shifts in the branching ratio relative to the default value.

\subsection{Signal electron spectrum}

The momentum spectrum of the electrons from charmless semileptonic decays is not precisely known. 
It is composed of contributions from exclusive and inclusive decays.

Many of the exclusive decay modes are still unobserved or poorly measured due to small event samples, 
and the form factors for many of the observed exclusive decay modes are not measured, thus we have to 
rely on theoretical predictions. To estimate the sensitivity of the signal BF to the relative contributions 
of the different exclusive decay modes, the BFs for $B \to \pi l\nu$, $B \to \rho l\nu$, 
$B \to \omega l\nu$, $B \to \eta l\nu$, $B \to \eta^{\prime} l\nu$ and $B \to X^{\rm nr}_ul\nu$ 
are varied by $5\%$, $10\%$, $15\%$, $20\%$, $35\%$ and $15\%$ \cite{pdg2014}, respectively. 
The $B \to \pi l\nu$ and $B \to \rho l\nu$ form factors are varied between the ISGW2 model and current
fits to measurements plus lattice QCD predictions. The observed impact 
of these variations on the fit result does not exceed $0.3\%$. 

The systematic uncertainties inherent to the modeling of the inclusive lepton spectrum in charmless decays 
and their impact on the signal yield are studied by varying the SF parameters. 
For each set of SF parameters, we recalculate the signal momentum spectrum and repeat the fit to the data. 
Taking into account the uncertainties and correlations of the measured SF parameters, 
we derive the uncertainties. This  is the largest source of systematic uncertainty for
the measurement of the partial BFs and is discussed separately for each theoretical calculation
in Sec. \ref{results}.

\subsection{Non-\boldmath{\BB} background}

Systematic uncertainties associated with the subtraction of the non-\BB background originate from the choice of
the function describing the lepton momentum spectrum and the relative normalization of the on- and off-resonance data samples.
We assess the impact of the fit function by replacing the default ansatz in Eq.~(\ref{function:cf}) with the following: 

\begin{eqnarray}
\label{function:cf1}
f(\vec{a},p) & = & (a_0 + a_1 p + a_2 p^2) \\
&\times& \exp(a_3 p + a_4 p^2 + a_5 p^3 + a_6 p^4 + a_7 p^5). \nonumber 
\end{eqnarray}

\noindent The observed difference is taken as the uncertainty, 
which is about $0.5\%$ below $ 2.3$~\gevc and increases for higher momenta where this contribution dominates.
The uncertainty in the relative normalization is taken as a constraint in the fit.

\subsection{\boldmath{\BB} background}

The momentum spectra of the dominant \BB backgrounds are derived from MC simulations. 
Their relative contributions are adjusted in the fit. 

The uncertainty in the lepton spectrum from the dominant decay modes was estimated 
by varying the form factors in $B\to D^* e \nu$ and $B\to D e \nu$ decays. 
The change in the partial BFs from these variations is less than $1.2\%$, and decreases for momenta above 2.3~\gevc.  

The sum of the BFs for the four decays to $D^{**}$ resonances are included as a free parameter in the fit 
and the uncertainty is estimated by changing the relative BFs for decays to individual  
$D^{**}$ resonances. The uncertainty due to the theoretical predictions of the form factors is estimated 
by varying the form factors for $B\to D^*_0 e \nu$, $D^{\prime}_1 e \nu$, $D_1 e \nu$ and $D^*_2 e \nu$. 
The impact on the fitted signal yield of each of these two sets of uncertainties in the $D^{**} e \nu$ spectrum 
is estimated to be less than $0.6\%$ over most of the spectrum, and decreases for momenta above 2.3~\gevc. 

Similarly, the impact of the uncertainties in the relative BFs of $D^{\prime}$ and $D^{\prime *}$ resonances 
and their widths are evaluated. They change the fit results by less than $1\%$ for momenta below 2.3~\gevc, 
and are negligible above.

Electrons from $J/\psi \to e^+e^-$ decays are one of the main sources of background near the endpoint, 
because unpaired $e^{\pm}$ from these decays are unaffected by the veto on the invariant mass of $e^+e^-$ pairs.  
We observe a difference between the veto efficiency for electron pairs in data and simulation of $(4.9 \pm 0.9)\%$ 
and also a difference in the observed momentum spectrum of the $J/\psi$. The \BB MC simulation is
corrected to match the measured spectrum and yield. To assess the impact of the veto efficiency, 
the number of $J/\psi$ events is varied by $0.9\%$; this variation changes the signal BF by less than $0.2\%$.
Background from $\psi(2S) \to e^+e^-$ decays is significantly smaller, and its uncertainty is negligible.

Varying the relative BFs for semileptonic $B$ decays involving $\tau^{\pm}$ leptons by $20\%$  
changes the fitted signal yield by up to $0.7\%$. 

For background from hadronic $B$ decays, the uncertainty in the spectrum is primarily due to the uncertainty 
in the momentum-dependent hadron misidentification. They have been studied and 
uncertainties in the normalization of $\pi$, $K$ and $p$ 
spectra are estimated to be $15\%$, $30\%$ and $100\%$, respectively. Variations of the charged pion spectrum change the
signal BFs by less than $0.8\%$, they are negligible for kaons and protons. The uncertainties in 
the momentum-dependent misidentification of pions, kaons and protons and antiprotons are 
included in the fit of the electron spectrum (uncertainty for shape of spectrum), taking into account 
bin-to-bin correlations.

\subsection{\label{b_meson_momentum} 
{\boldmath{$B$} meson momentum spectrum}}

The $B$ momentum is sensitive to the energies of the colliding beams which may vary slightly with time. 
Any variation in the momentum of the $B$ meson in the \FourS\ rest frame affects the shape of the electron 
spectrum near the endpoint. We have compared the simulated and measured momentum spectra for fully
reconstructed hadronic decays of charged $B$ mesons for different data taking periods. The widths of the 
total energy distributions agree well for all data, but for some of the data sets we observed a shift in the
central value of up to 3.4 \mev relative to the simulation, which assumes a fixed c.m. energy. 
We correct the simulation for the observed shifts. The uncertainty of 0.1 \mev\ in this correction results 
in an estimated uncertainty on the signal BF of up to $1.6\%$.

\subsection{Bremsstrahlung and radiative corrections}
\label{Radiative}

The MC simulations include the effects of bremsstrahlung and  final state radiation. 
Corrections for QED radiation in the decay process are simulated using PHOTOS~\cite{photos}. 
This simulation includes multiple-photon emission from the electron, but does not include 
electroweak corrections for quarks. The accuracy of this simulation has been compared to analytical 
calculations performed to ${\cal O}(\alpha)$~\cite{photos}. Based on this comparison
we assign an uncertainty of $20\%$ to the PHOTOS correction, leading to an uncertainty in the signal 
yield of about $0.6\%$.

The uncertainty in the energy loss of electrons due to bremsstrahlung in the beam pipe and 
tracking system is determined by the uncertainty in the thickness of the detector material, 
estimated to be $(0.0450 \pm 0.0014) X_0$ at normal incidence. The thickness of the material
was verified using electrons from Bhabha scattering as a function of the polar angle relative 
to the beam. The impact of the uncertainty in the energy loss on the signal rate was estimated 
by calculating the impact of an additional $0.0014 X_0$ of material. Relative shifts in the 
$B\to X_u e\nu$ BF due to variations of bremsstrahlung  with respect to the default simulation 
are estimated to be about $0.3\%$ up to $2.4$~\gevc and less at higher momenta.

\subsection{\label{WB}
{Width of wide bin}}

The lower boundary of the large bin noticeably effects the fitted signal yield and uncertainty.  
For values less than $2.05$~\gevc, the uncertainty on the BF increases significantly, 
and for values greater than $2.1$~\gevc, the fit quality diminishes because the data and the 
predicted spectra differ at higher momenta. The impact of this sensitivity is estimated as 
the difference between the nominal fit with $2.1-2.7$~\gevc and a fit with a slightly lower 
boundary of the bin, $2.05-2.7$~\gevc. The relative change in the $B\to X_u e\nu$ BFs due to 
this variation is estimated to be $0.4\%$ below $2$~\gevc and much lower at higher momenta.

\section{Results}
\label{results}

The primary results of this study of the electron spectrum for inclusive semileptonic decays
of $B$ mesons are the total inclusive $B\to X e \nu$ spectrum and BF, the extraction of the spectrum 
and partial and total BF for the charmless $B\to X_u e \nu$ decays and the determination of 
the CKM element $|V_{ub}|$, using four theoretical predictions based on different approaches 
to estimate the QCD corrections to the decay rate. All results are based on the full \babar\ 
data sample and averaged over charged and neutral $B$ mesons produced in $\FourS$ decays.

\subsection{Total semileptonic spectrum and branching fraction}

\begin{figure}[htbp]
\begin{center}
\includegraphics[height=6.cm]{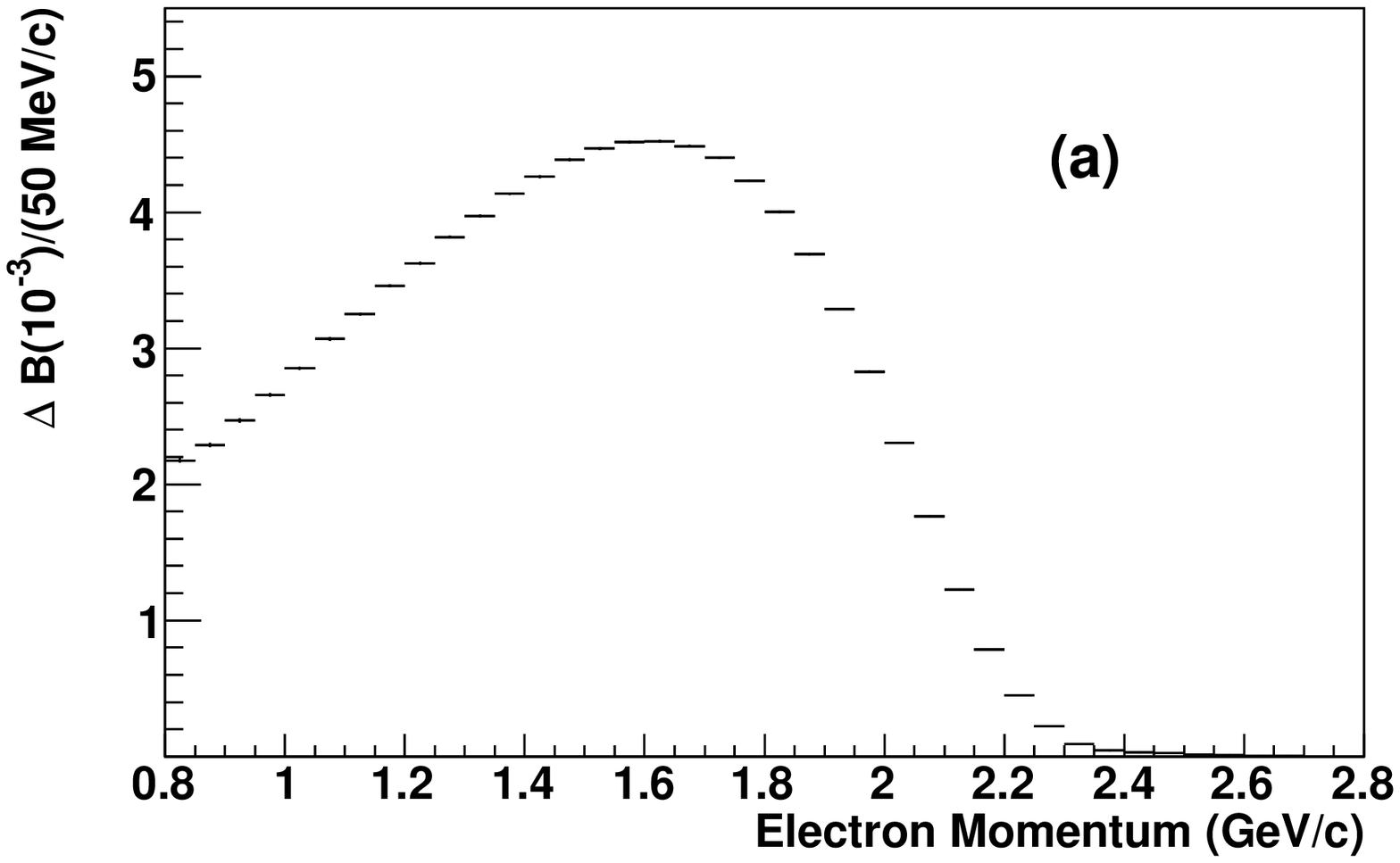}
\includegraphics[height=6.cm]{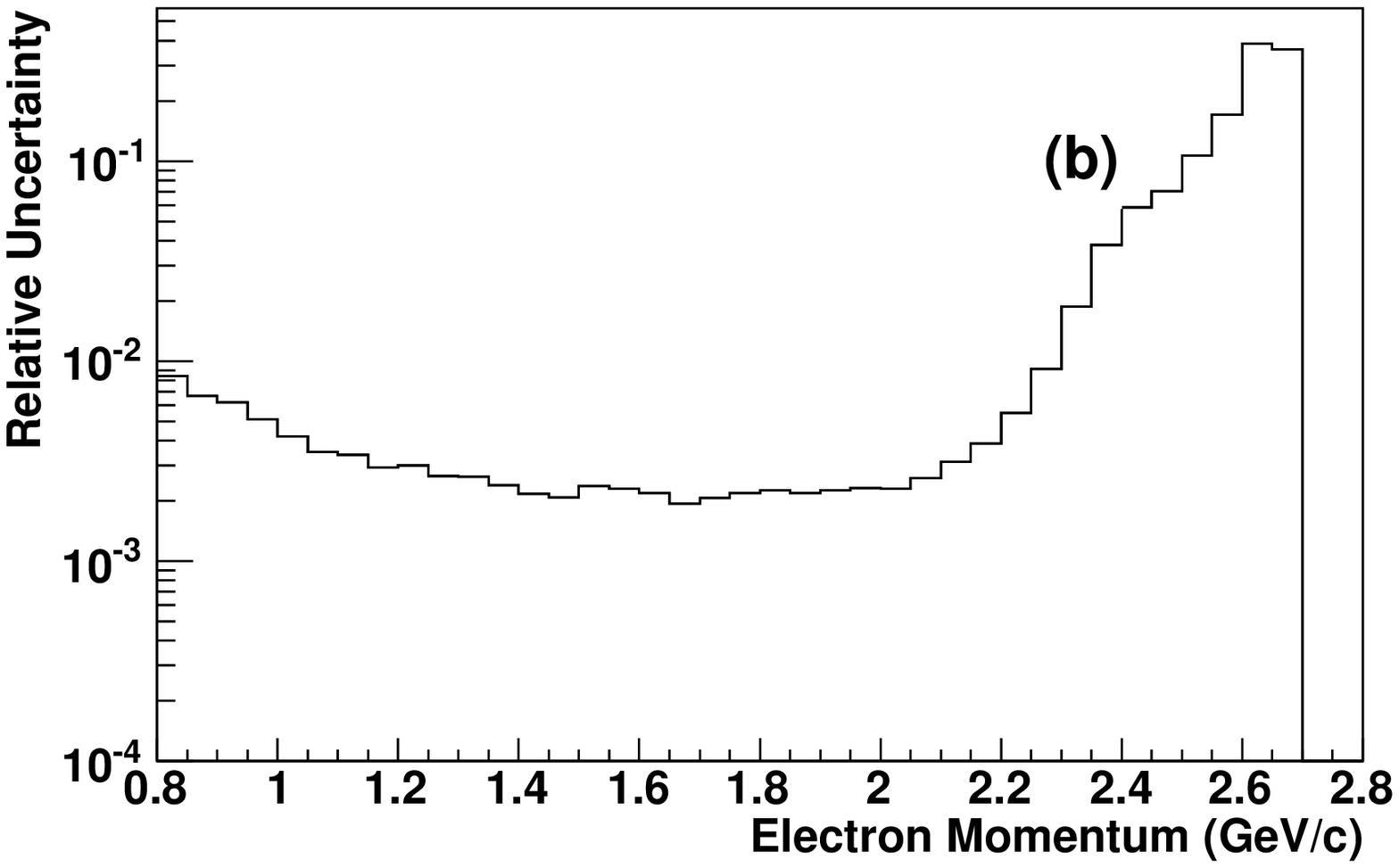}
\caption{
Inclusive $B \rightarrow X e\nu$ decays:
(a) The differential branching fraction as a function of the electron momentum 
(in the $\FourS$ rest frame) after background subtraction and corrections for 
bremsstrahlung and final state radiation. 
(b) The relative statistical uncertainties on the background subtraction combined 
with the uncertainties of the fit parameters.  
}
\label{fig:pe_fc_br_X}
\end{center}
\end{figure}

The differential BF for primary electrons in $B \to X e\nu$ decays
as a function of the electron momentum in the \FourS\ rest frame is shown in
Fig.~\ref{fig:pe_fc_br_X}. It is derived from the fit to the total observed 
electron spectrum for GGOU calculations of the $B \to X_u e \nu$ spectrum.
It is fully corrected for efficiencies and radiative effects. The data are normalized 
to the total number of produced $B^+B^-$ and $B^0 {\overline B}^0$ pairs,
$N_{\BB} = (466.48\pm 0.11_{\textrm{stat}} \pm 2.39_{\textrm{syst}}) \times 10^6$. 
The uncertainties shown represent the statistical uncertainties of the measurement,
including the background subtraction and the uncertainties in the fit parameters. 
These uncertainties do not include systematic uncertainties, in particular 
those related to the prediction of the $B \to X_u e \nu$ spectrum.
This spectrum and its uncertainties and the correlation matrix of the fit are available 
in the Supplemental Material.
% at [URL will be inserted by publisher]. 

The total inclusive semileptonic BF, averaged over charged and neutral $B$ mesons,
is obtained as the sum of the individual semileptonic BFs determined by the fit to the observed 
electron spectrum: 
\begin{equation}
{\cal B}(B\rightarrow X e\nu) = (10.34 \pm 0.04_{\textrm{stat}} \pm 0.26_{\textrm{syst}}) \%. 
\end{equation}
Using GGOU for the predicted contribution from $B \to X_u e \nu$ decays, we obtain 
\begin{equation}
{\cal B}(B\rightarrow X_c e\nu) = (10.18 \pm 0.03_{\textrm{stat}} \pm 0.24_{\textrm{syst}}) \%, 
\end{equation} 
\noindent 
where the stated systematic uncertainty takes into account the differences of about $1\%$ between 
this result and those obtained with predictions of the $B \to X_u e \nu$ spectrum by DN, BLNP, and DGE. 
The results, which are dominated by systematic uncertainties, are consistent 
with the most recent HFAG average of 
${\cal B}(B\rightarrow X e\nu) = (10.86 \pm\ 0.16)\%$ and 
${\cal B}(B\rightarrow X_c e\nu) = (10.65 \pm\ 0.16)\%$~\cite{hfag14}.

\subsection{Differential \boldmath{$B\to X_u e \nu$} branching fractions}

\begin{figure}[htbp]
\begin{center}
\includegraphics[height=4.6cm]{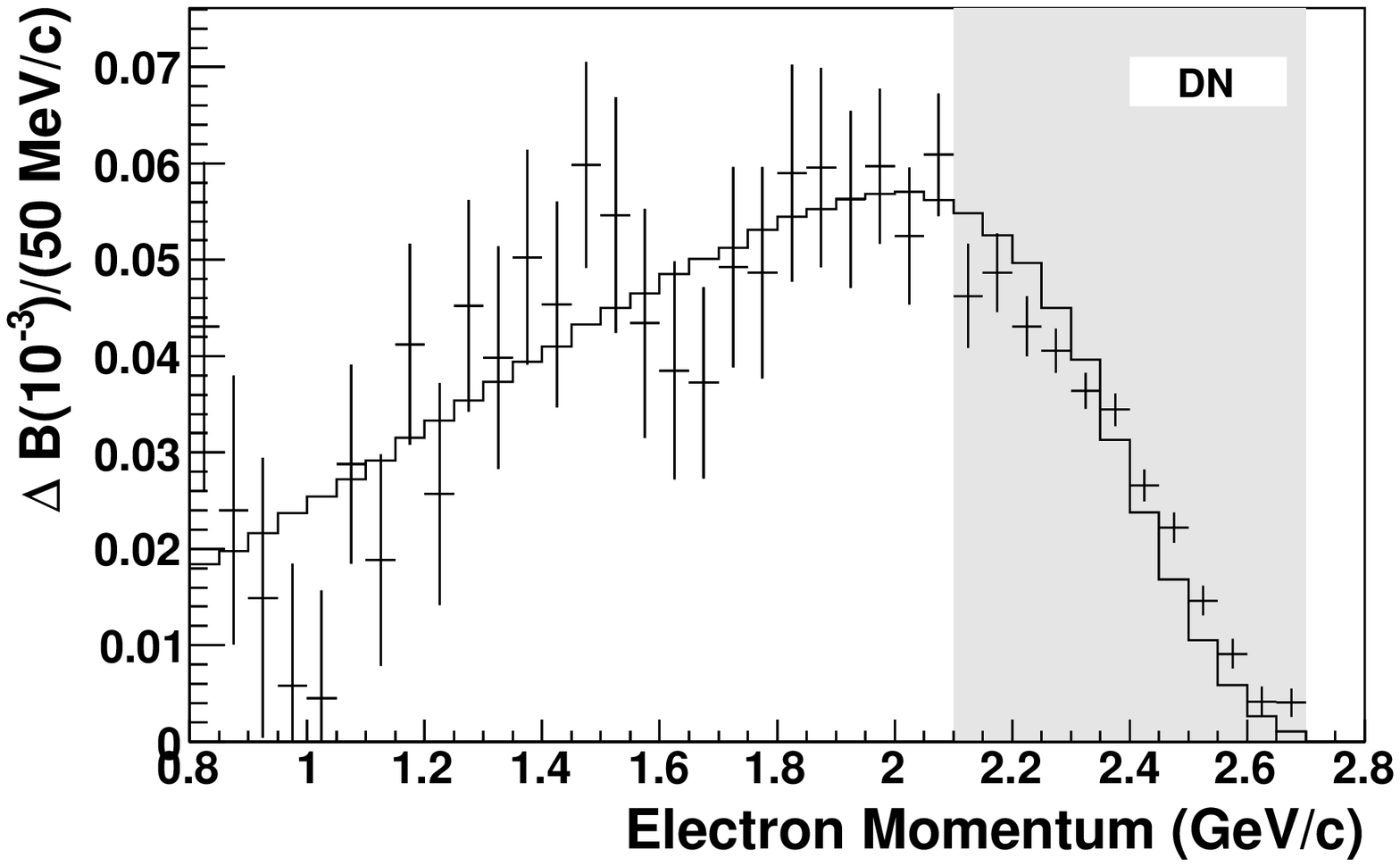}
\includegraphics[height=4.6cm]{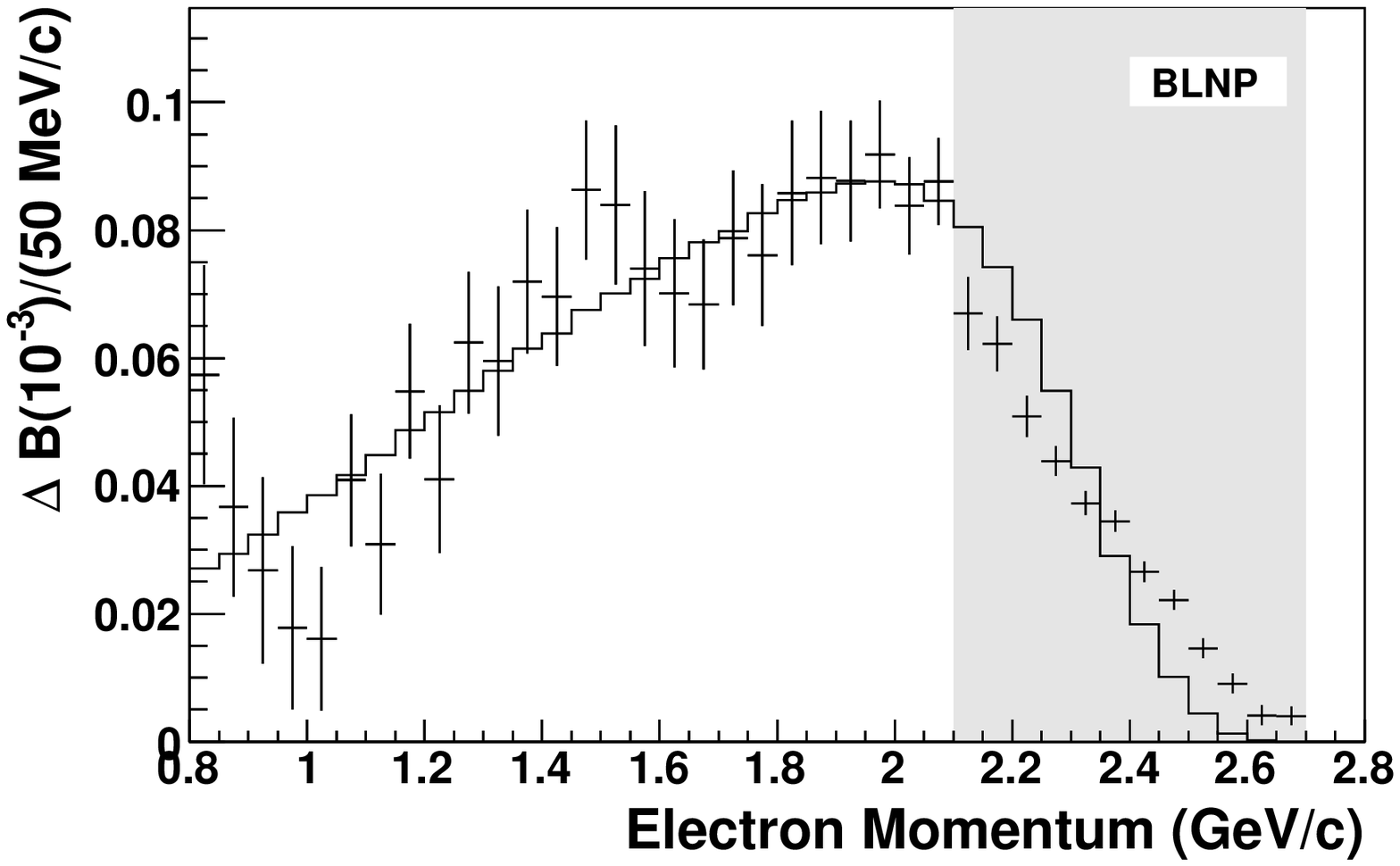}
\includegraphics[height=4.6cm]{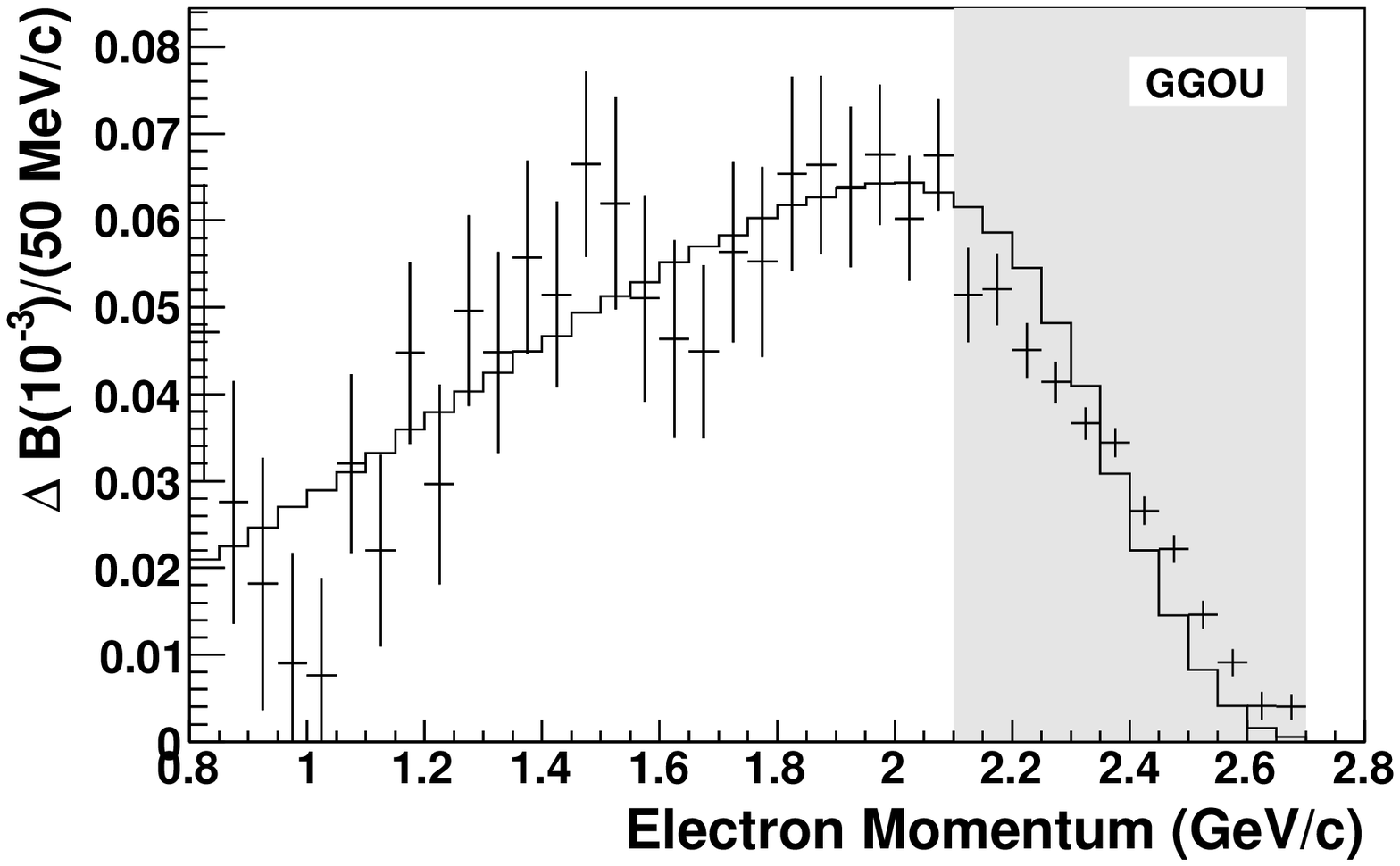}
\includegraphics[height=4.6cm]{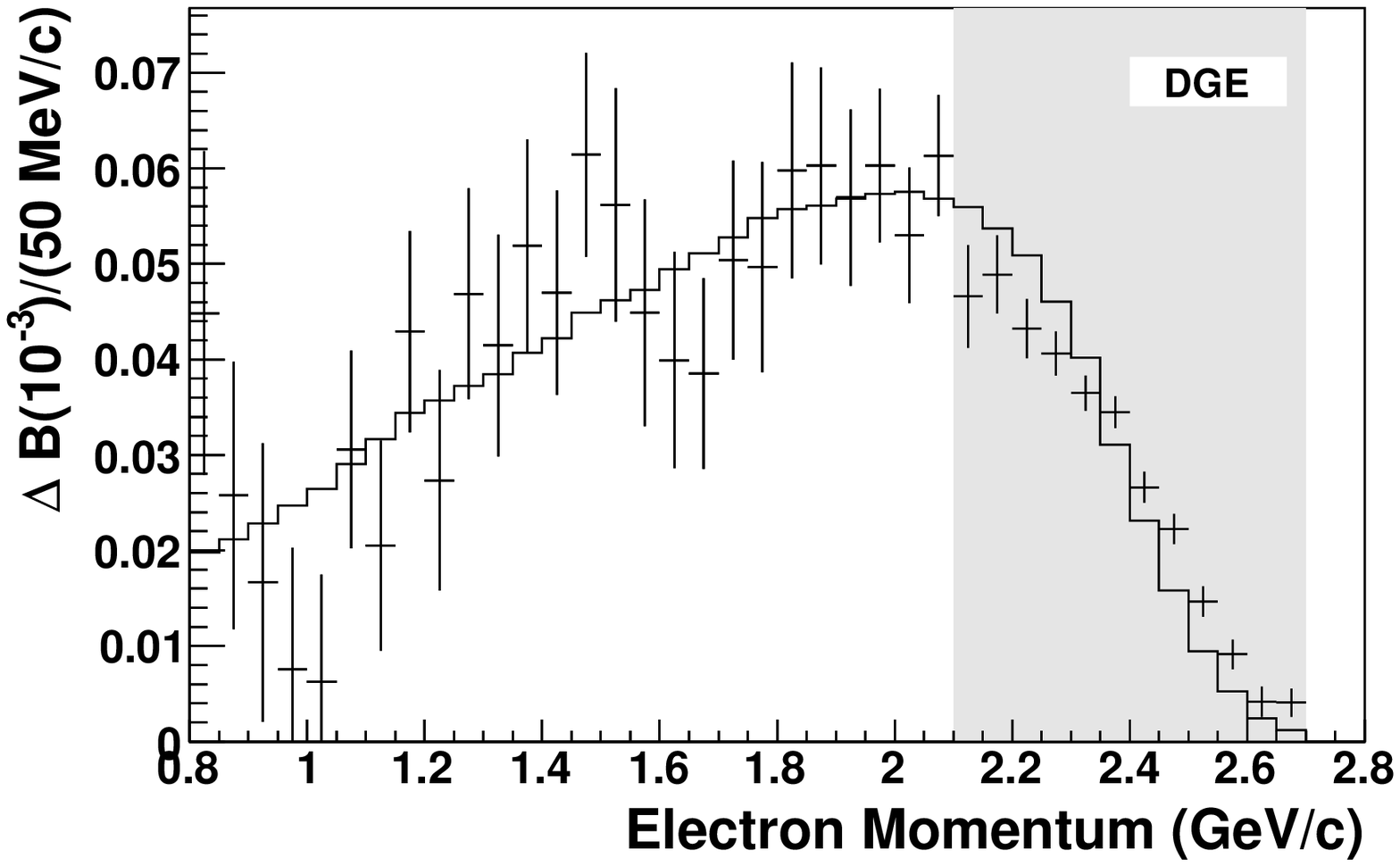}
\caption{
The differential branching fraction for charmless semileptonic $B$ decays (data points)
as a function of the electron momentum [in the $\FourS$ rest frame] after background 
subtraction and corrections for bremsstrahlung and final state radiation, compared 
to the Monte Carlo simulation (histogram). The uncertainties indicate the statistical uncertainties 
on the background subtraction, including the uncertainties of the fit parameters. 
The shaded area indicates the momentum interval for which the on-resonance data 
are combined into a single bin. 
}
\label{fig:pe_fc_br}
\end{center}
\end{figure}

\begin{figure}[htbp]
\begin{center}
\includegraphics[height=4.6cm]{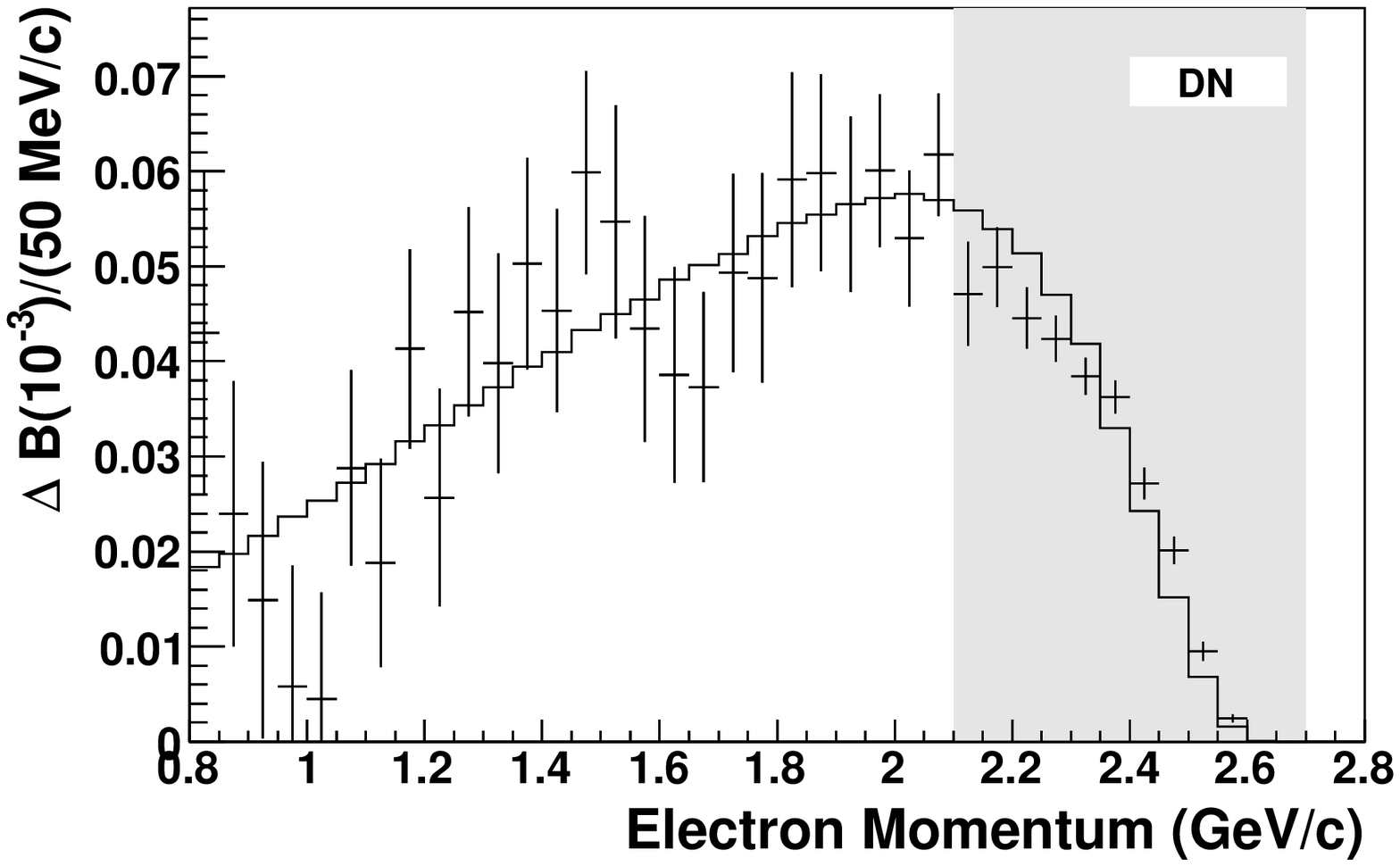}
\includegraphics[height=4.6cm]{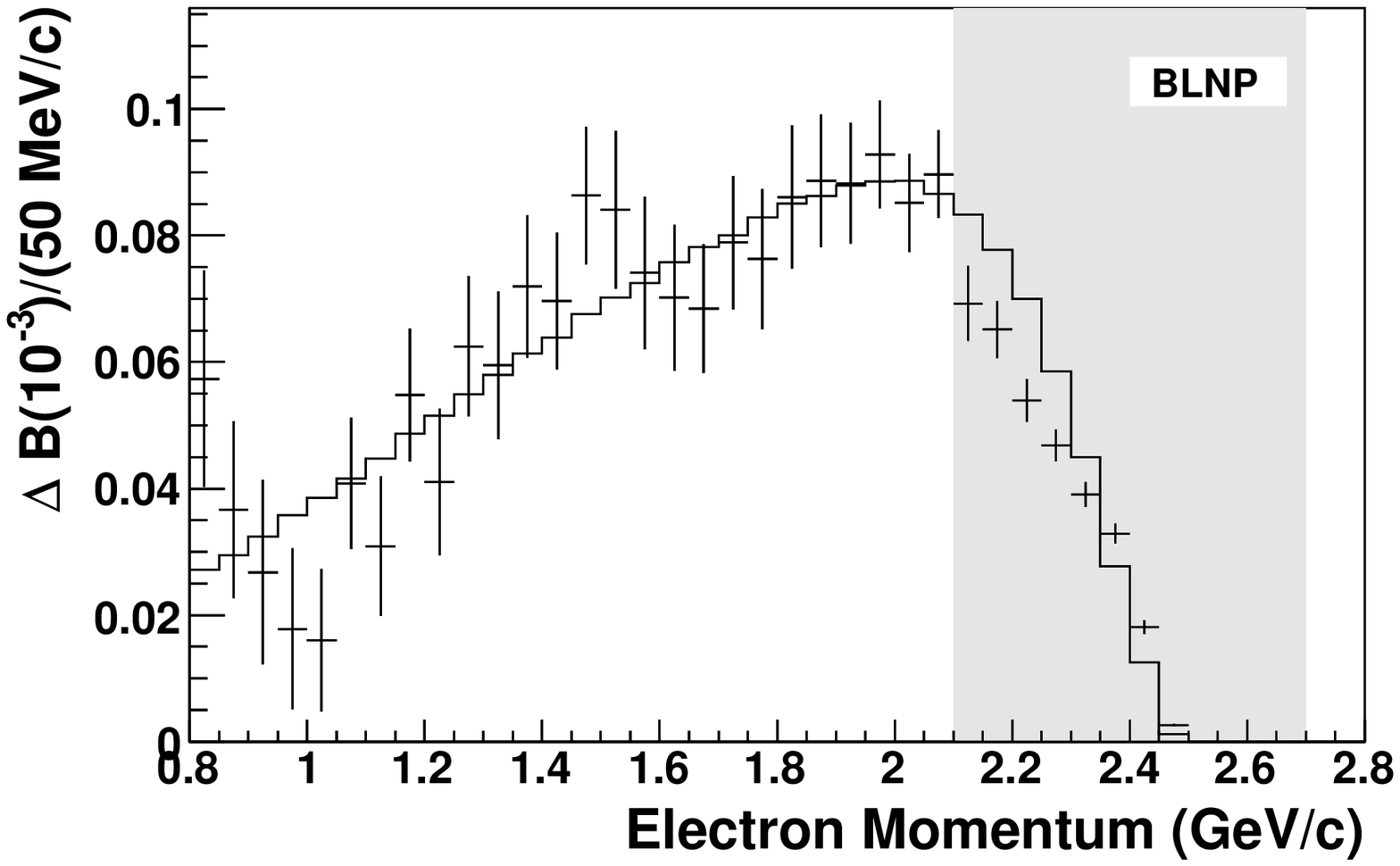}
\includegraphics[height=4.6cm]{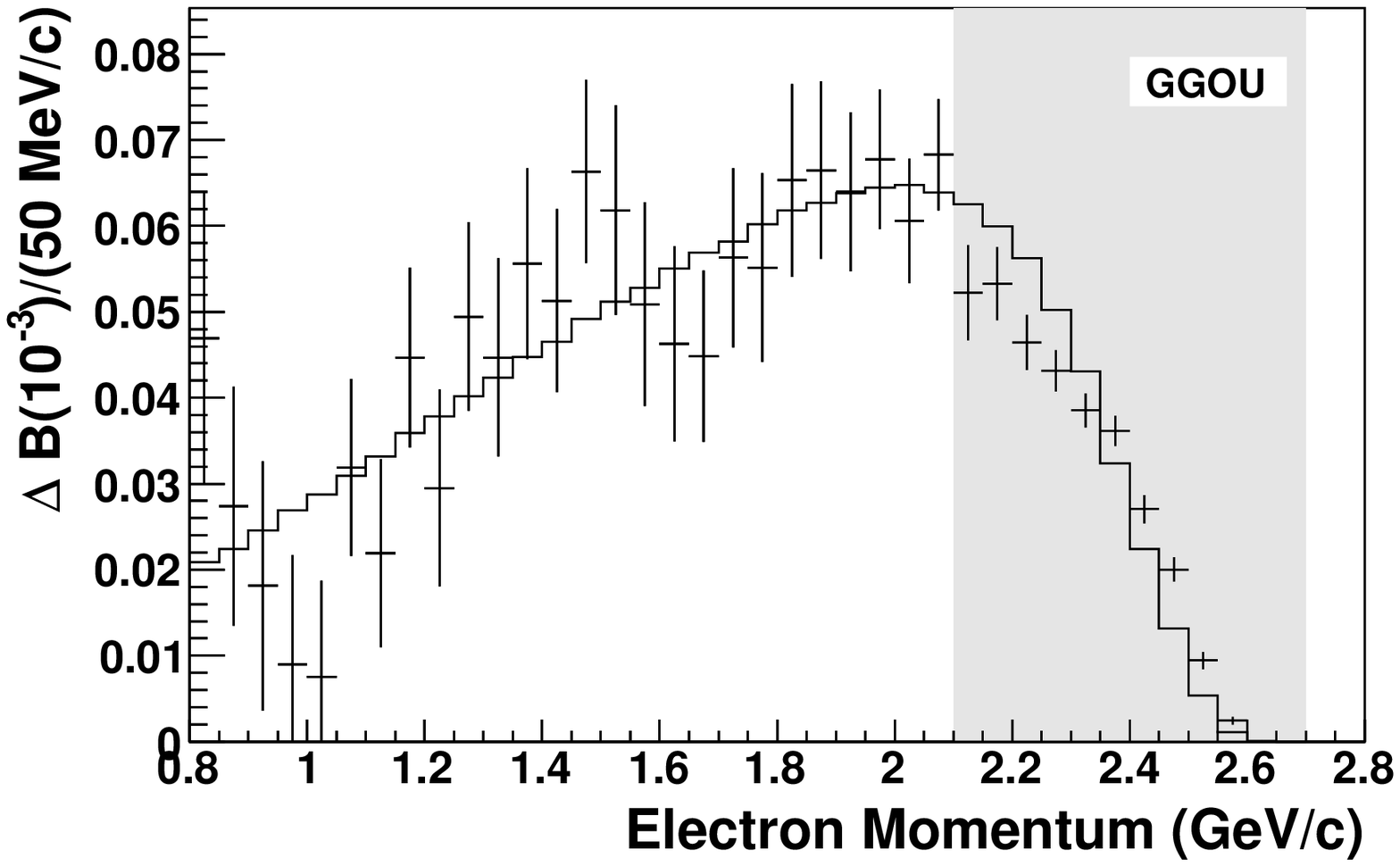}
\includegraphics[height=4.6cm]{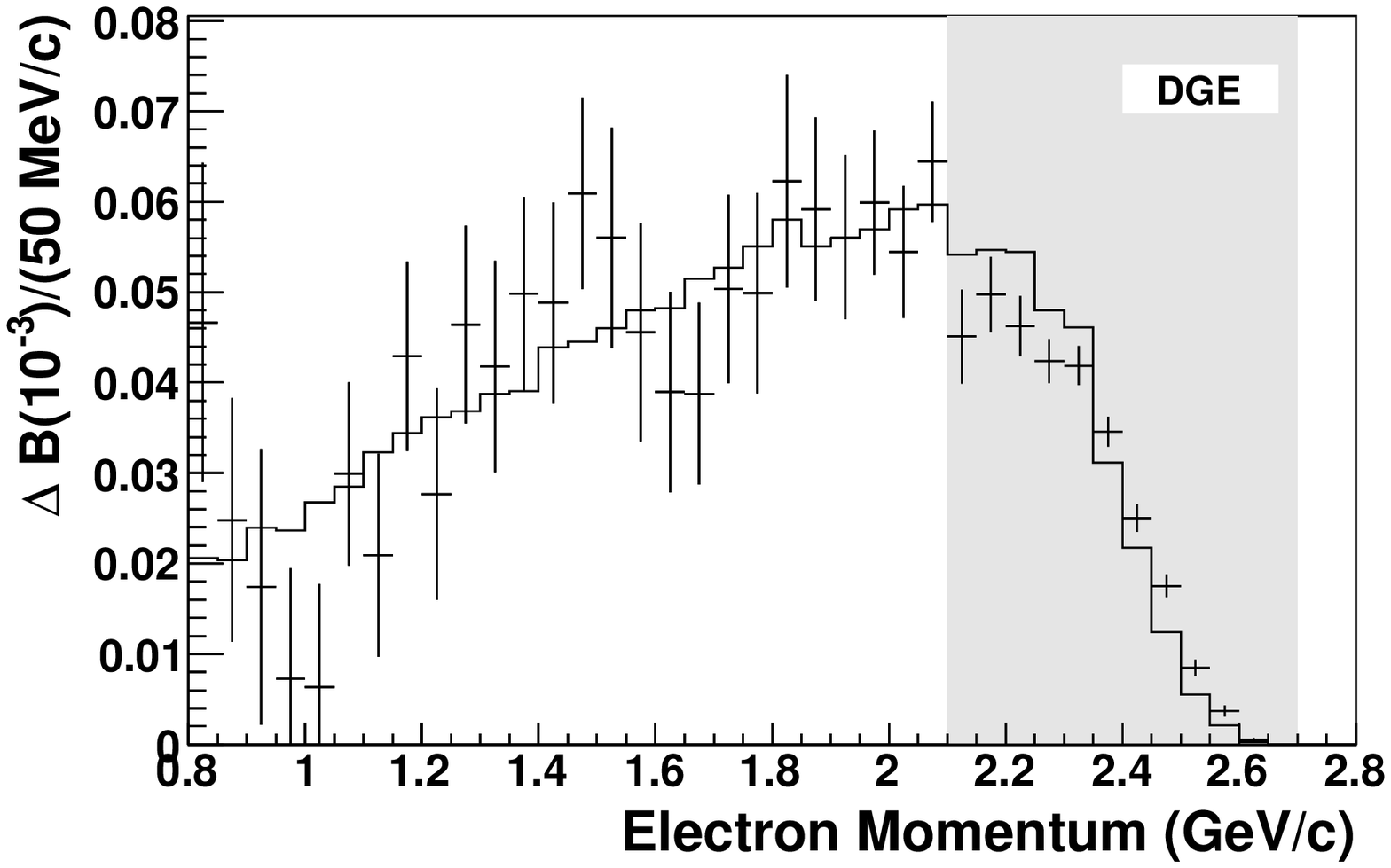}
\caption{
The differential branching fraction for charmless semileptonic $B$ decays (data points)
as a function of the electron momentum [in the $B$ rest frame] after background subtraction 
and corrections for bremsstrahlung and final state radiation, compared to the Monte Carlo 
simulation (histogram). The uncertainties indicate the statistical uncertainties on the background subtraction,
including the uncertainties of the fit parameters. The shaded area indicates the momentum 
interval for which the on-resonance data are combined into a single bin. 
}
\label{fig:pe_fc_br_B}
\end{center}
\end{figure}

The partial $B\to X_u e \nu$  BF for a given electron momentum interval $\Delta p$ 
is determined as 

\begin{equation}
\Delta {\cal B} (\Delta p) = \frac{N_{\textrm{tot}}(\Delta p)-N_{\textrm{bg}} (\Delta p) } 
     {2\epsilon(\Delta p) N_{\BB}} (1+\delta_{\textrm{rad}}(\Delta p)).
\end{equation}
 
\noindent Here $N_{\textrm{tot}}$ refers to the total number of selected
electron candidates from the on-resonance data and $N_{\textrm{bg}}$ refers to 
the total non-\BB and \BB background, as determined from the fit to the spectrum.
$\epsilon(\Delta p)$ is the total efficiency for selecting a signal electron from 
$B \to X_u e\nu$ decays (including bremsstrahlung in the detector 
material), and $\delta_{\textrm{rad}}$ accounts for the impact of
final state radiation on the electron spectrum. This momentum-dependent correction is derived 
from the MC simulation based on PHOTOS~\cite{photos}.

The differential BF for $B \rightarrow X_u e\nu$ decays, fully corrected for efficiencies and radiative effects,
as a function of the electron momentum in the \FourS\ rest frame is shown in 
Fig.~\ref{fig:pe_fc_br}, and in the $B$ meson rest frame in Fig.~\ref{fig:pe_fc_br_B}. 
The error bars represent the statistical uncertainties of the measurement. 
They do not include the systematic uncertainties, nor the uncertainty due to the $B \to X_u e \nu$ predictions.
For fits using the GGOU prediction for $B \to X_u e \nu$ the results for the differential BFs
and the correlation matrix are available in the Supplemental Material.
% at [URL will be inserted by publisher]. 
Differences of the fitted spectra and the data are clearly visible inside the wide bin, and 
are most pronounced for BLNP, for which the predicted rate is negative above 2.4\gevc. 
In all cases the data exceed the predictions above 2.3 \gevc, and are lower below, such that 
the data summed over the wide bin agree with the predictions in this momentum range.

\begin{figure}[htbp]
\begin{center}
\includegraphics[width=9.5cm]{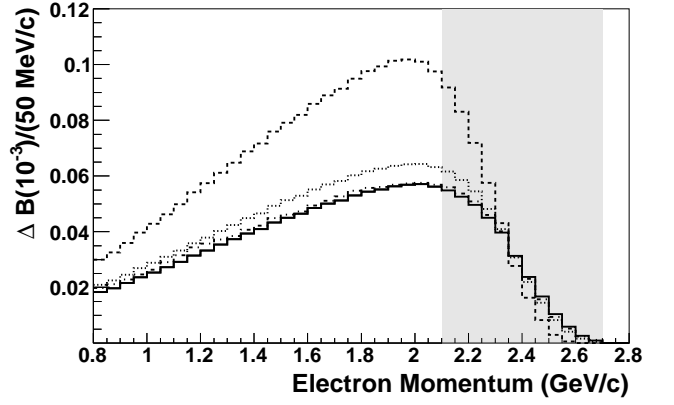}  
\caption{
The comparison of the theoretical differential branching fraction for charmless 
semileptonic $B$ decays with normalization based on the fit as a function of 
the electron momentum [in the $\FourS$ rest frame] for DN (solid), BLNP (dashed), 
GGOU (dotted) and DGE (dash-dotted). The shaded area indicates the momentum interval 
for which the on-resonance data are combined into a single bin.
}
\label{f:pe_fc_model_MC}
\end{center}   
\end{figure}

A comparison of the predicted $B\to X_u e \nu$ electron spectra, each normalized to the
fitted rate is presented in Fig.~\ref{f:pe_fc_model_MC}. While these spectra agree 
reasonably well for DN, GGOU and DGE, the BLNP prediction deviates substantially.
This is explained by a lower predicted rate for momenta above 2.1 \gevc, which leads to a
significantly larger fitted normalization of this spectrum.

\subsection{Total charmless branching fraction} 

The total BF for charmless $B\to X_u e \nu$ decays is determined from the partial BF 
$\Delta{\cal B}(\Delta p)$ in a given momentum range $\Delta p$, as follows:

\begin{equation}
{\cal B}(B\rightarrow X_u e\nu) = \Delta{\cal B}(\Delta p)/f_u(\Delta p),
\end{equation}

\noindent where $f_u(\Delta p)$ is the theoretically predicted fraction of the electron spectrum.
These total BF which have been determined as a function of $p_{\textrm{min}}$, the lower limit of 
the momentum range $\Delta p = [p_{\textrm{min}}, 2.7 \gevc]$,
(with fixed upper limit of $2.7$ \gevc) and their relative uncertainties are shown in Figs.~\ref{fig:branching_1} 
and \ref{fig:branching_2}, for the four different theoretical predictions.
Up to 2.1 \gevc, the resulting BFs are independent of $p_{\textrm{min}}$, above 2.1 \gevc,
the BFs and their uncertainties increase significantly.

\begin{figure}[htbp]
\begin{center}
\includegraphics[height=4.9cm]{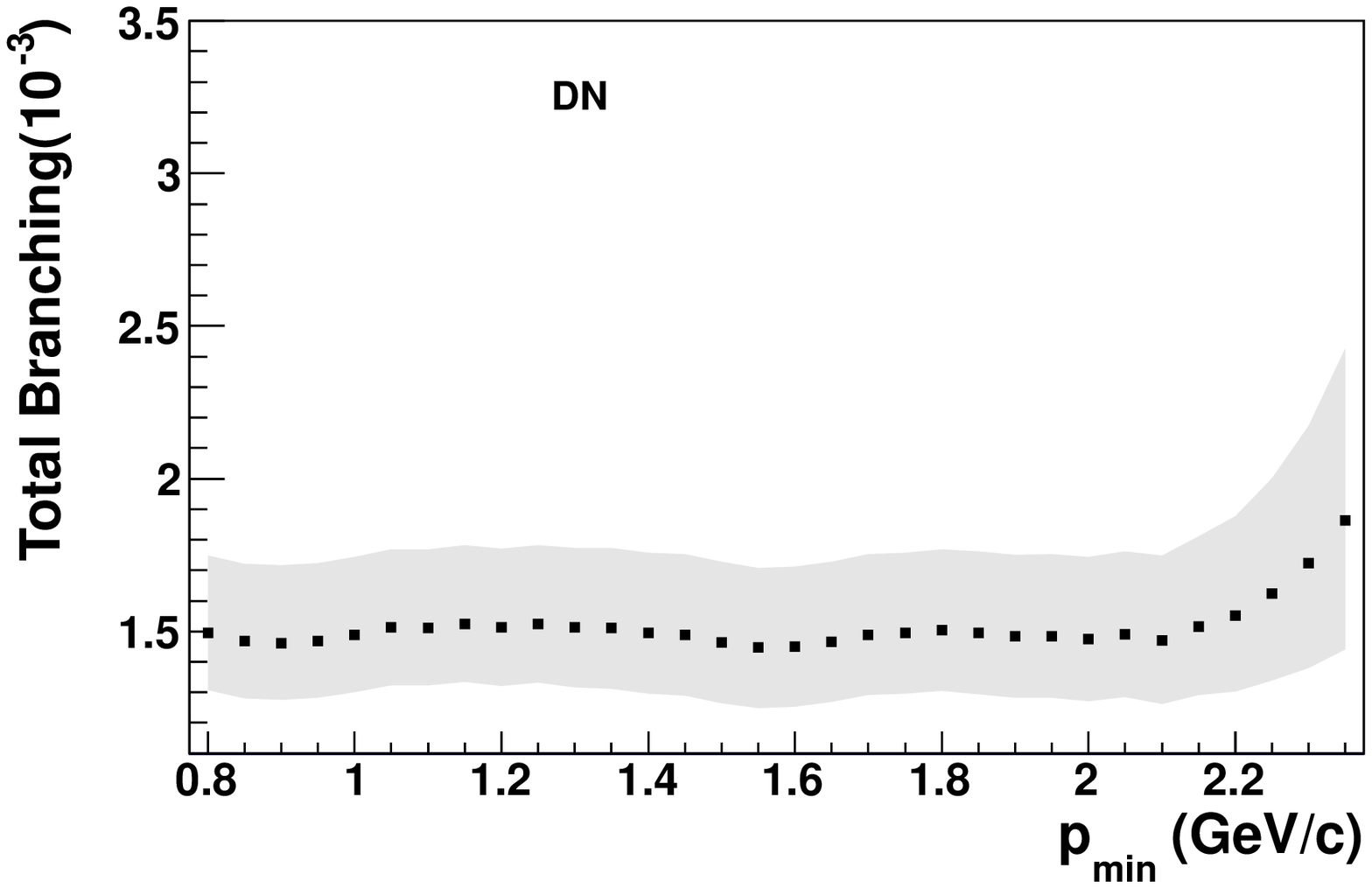}
\includegraphics[height=4.9cm]{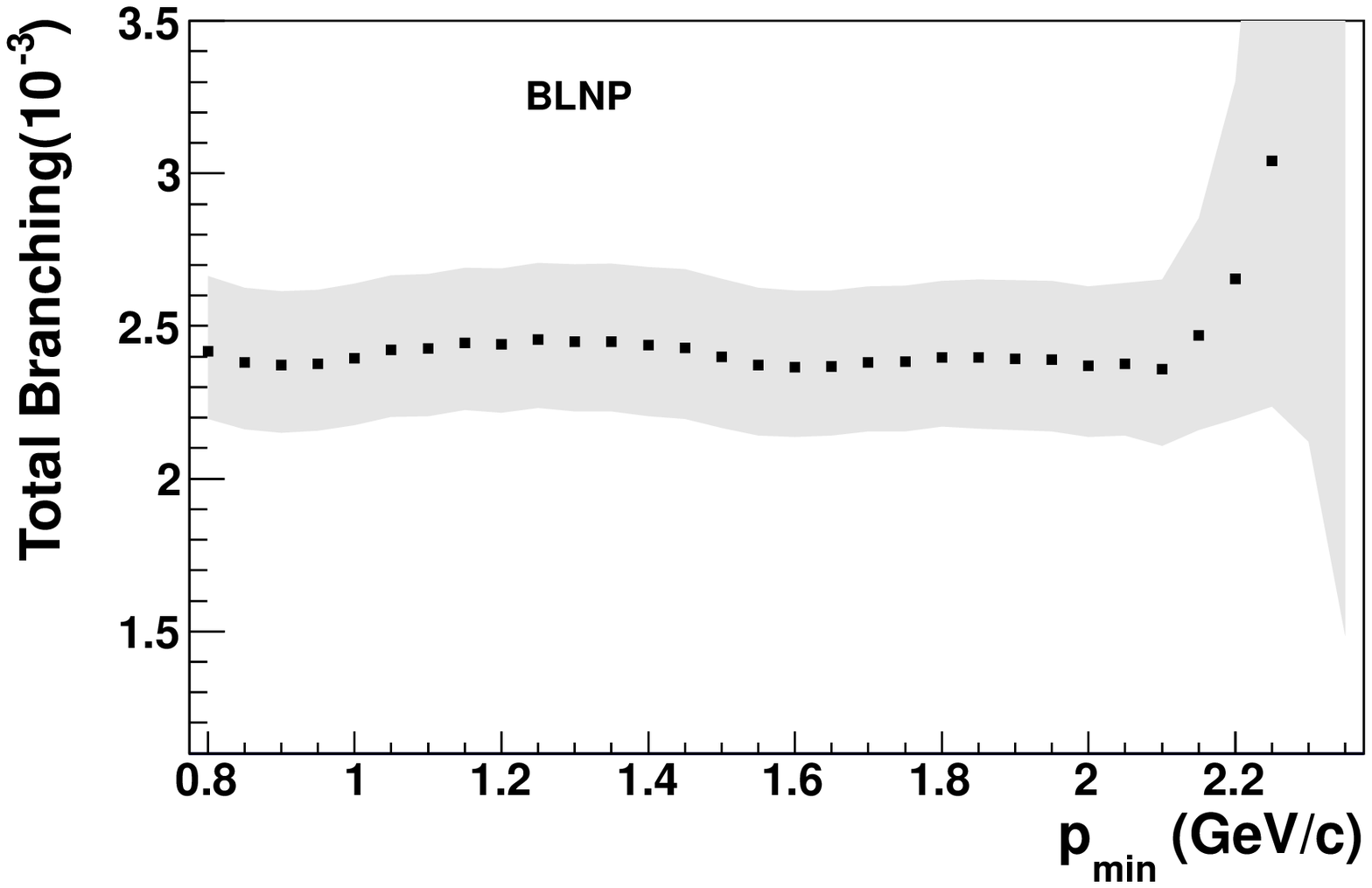}
\includegraphics[height=4.9cm]{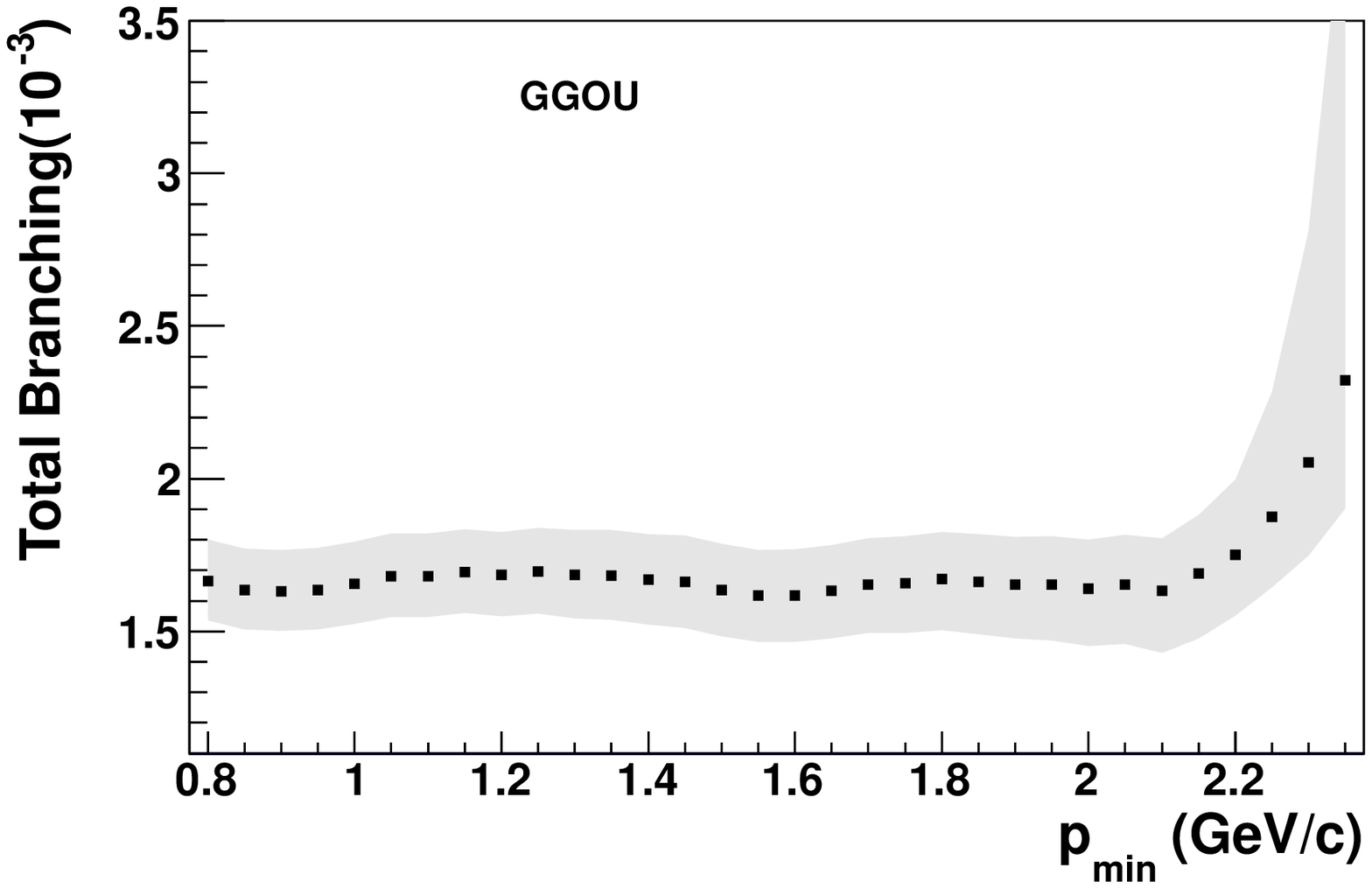}
\includegraphics[height=4.9cm]{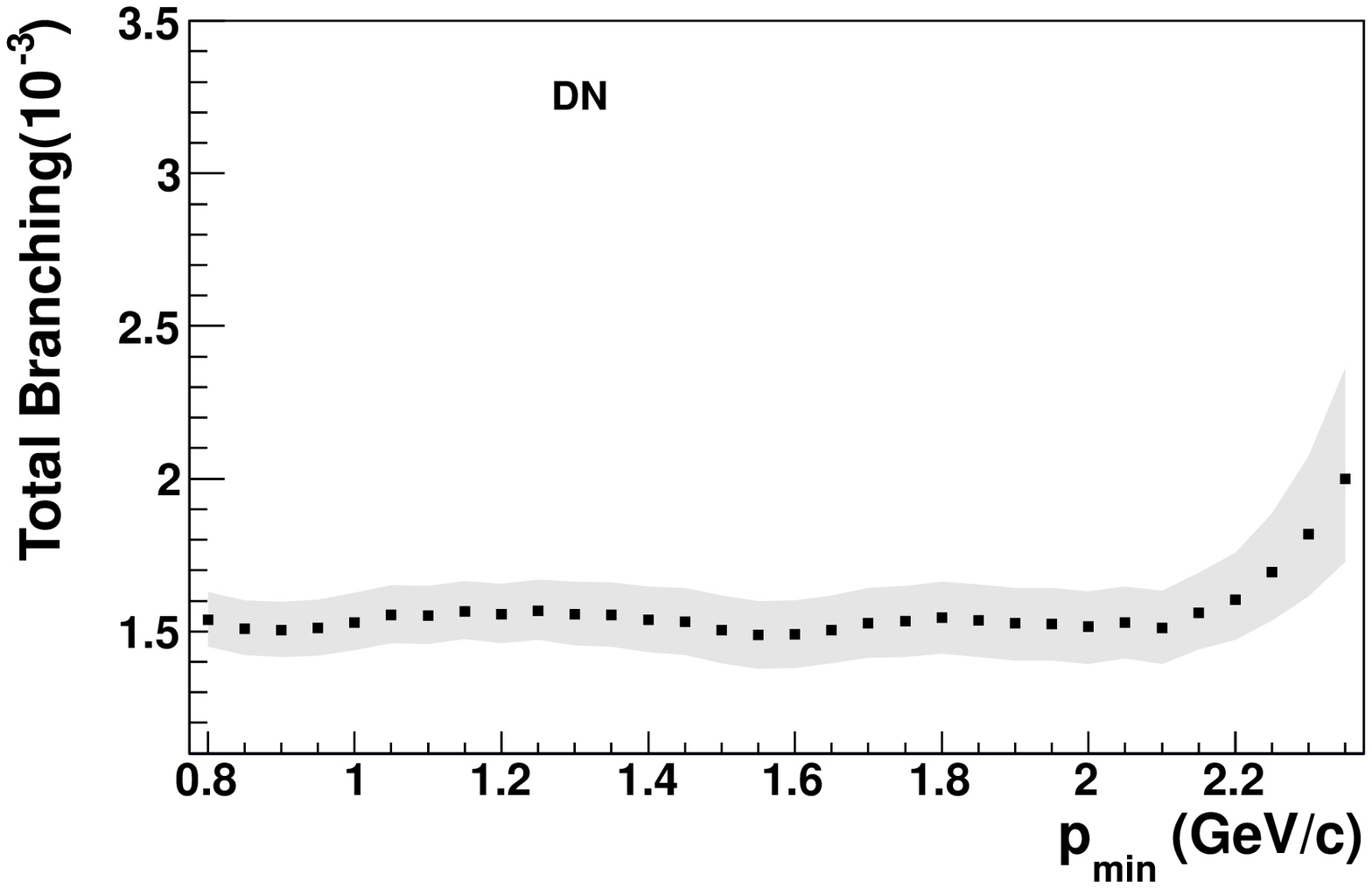}
\caption{
Total branching fraction for $B \to X_u e \nu$ decays as a function  
$p_{\textrm{min}}$, the lower limit of the electron momentum range used in the extraction 
of the signal and the total uncertainty which include experimental, SF parametrization and 
theoretical uncertainties, separately for DN, BLNP$_1$, GGOU$_1$, and DGE predictions 
of the decay rate used in the fit. 
}
\label{fig:branching_1}
\end{center}
\end{figure}

\begin{figure}[htbp]
\begin{center}
\includegraphics[height=4.9cm]{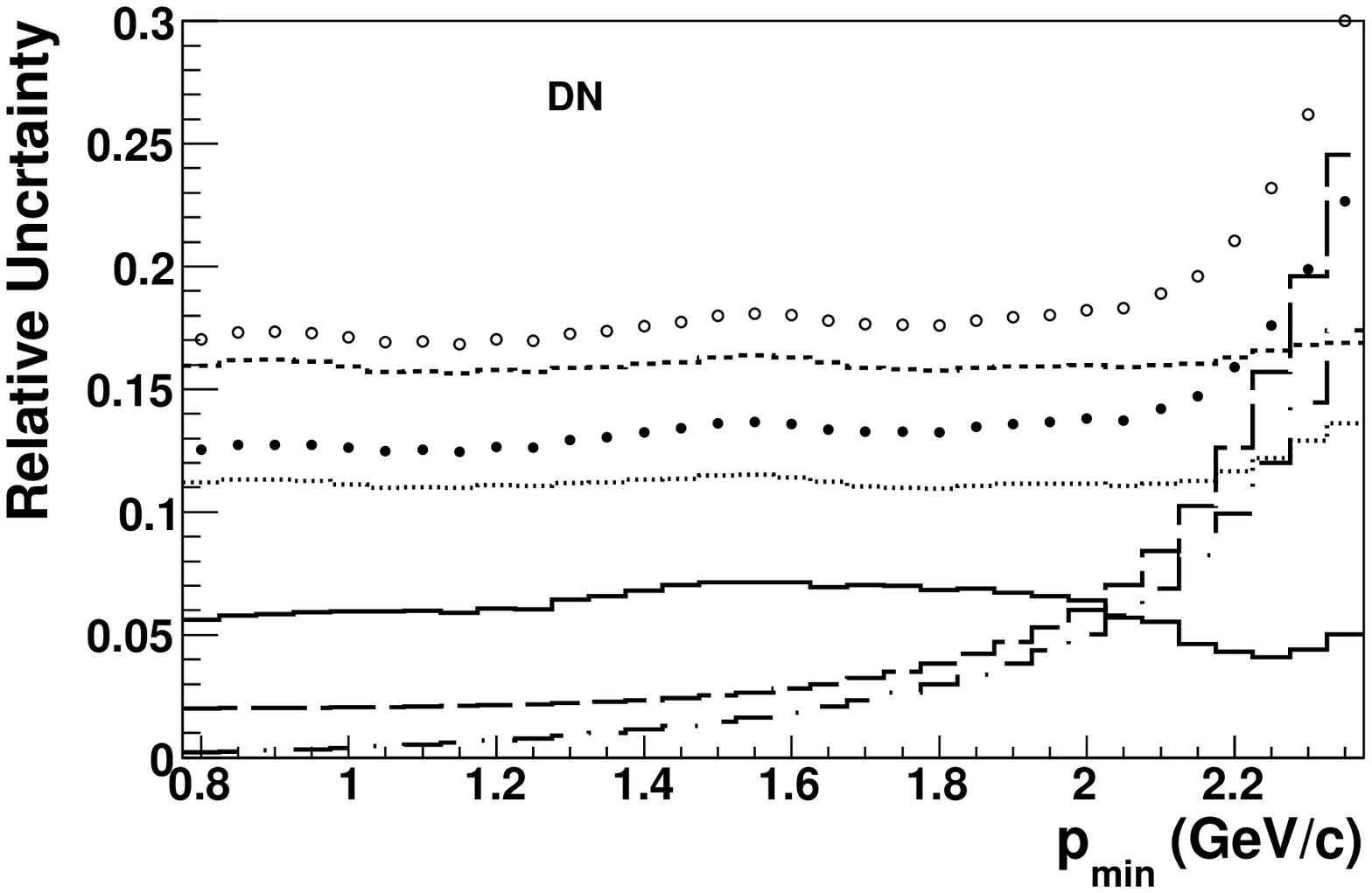}
\includegraphics[height=4.9cm]{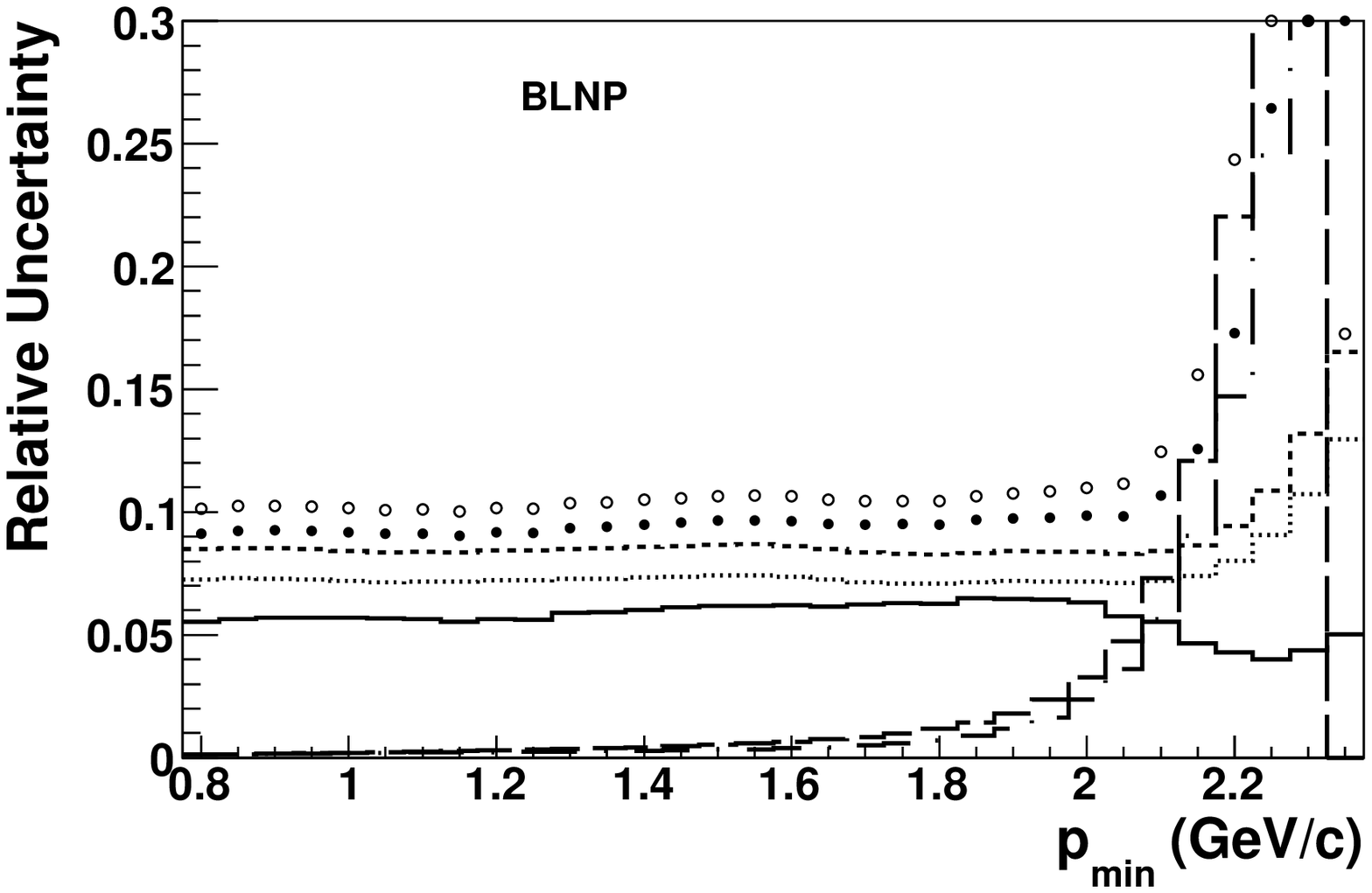}
\includegraphics[height=4.9cm]{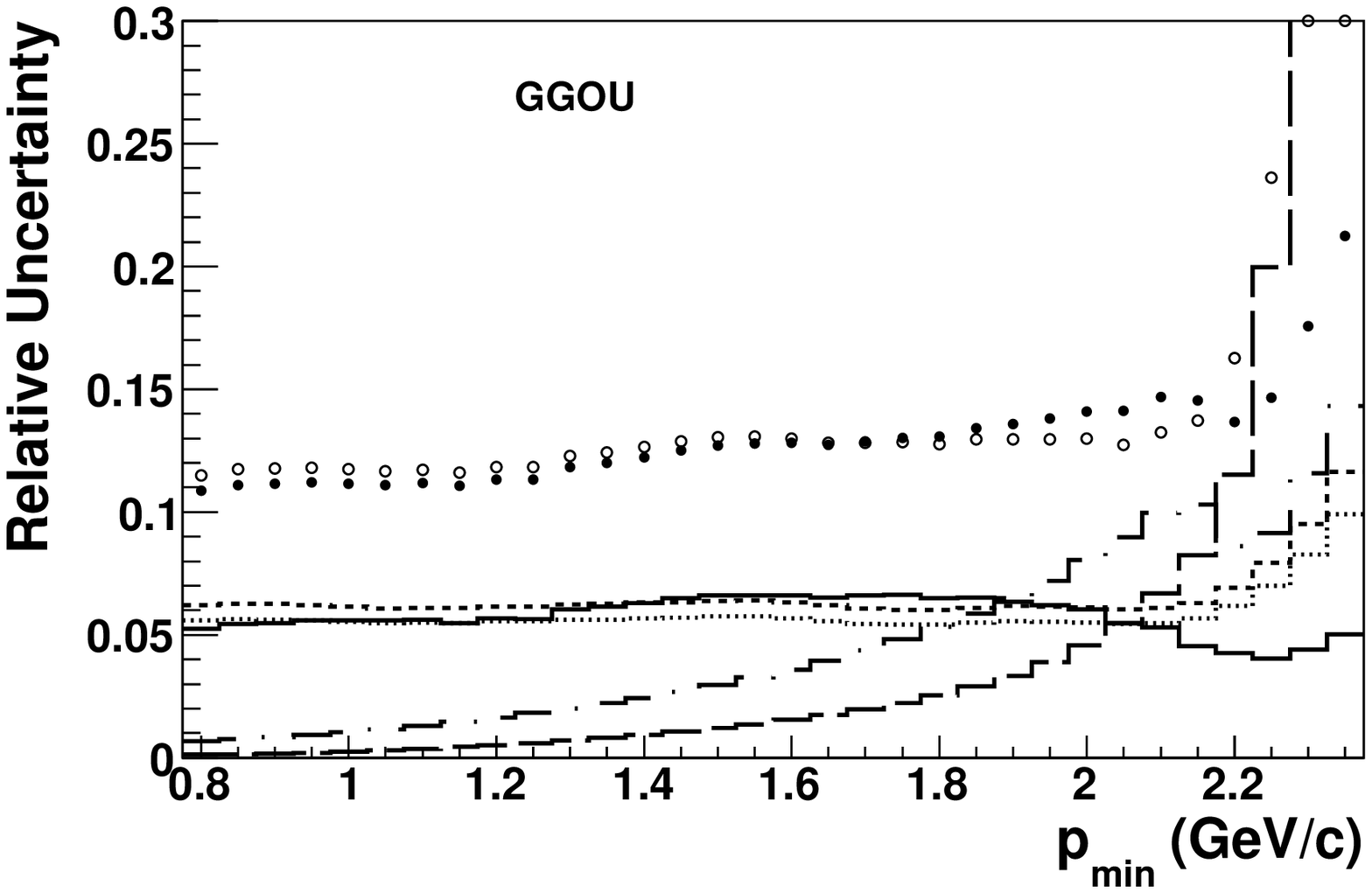}
\includegraphics[height=4.9cm]{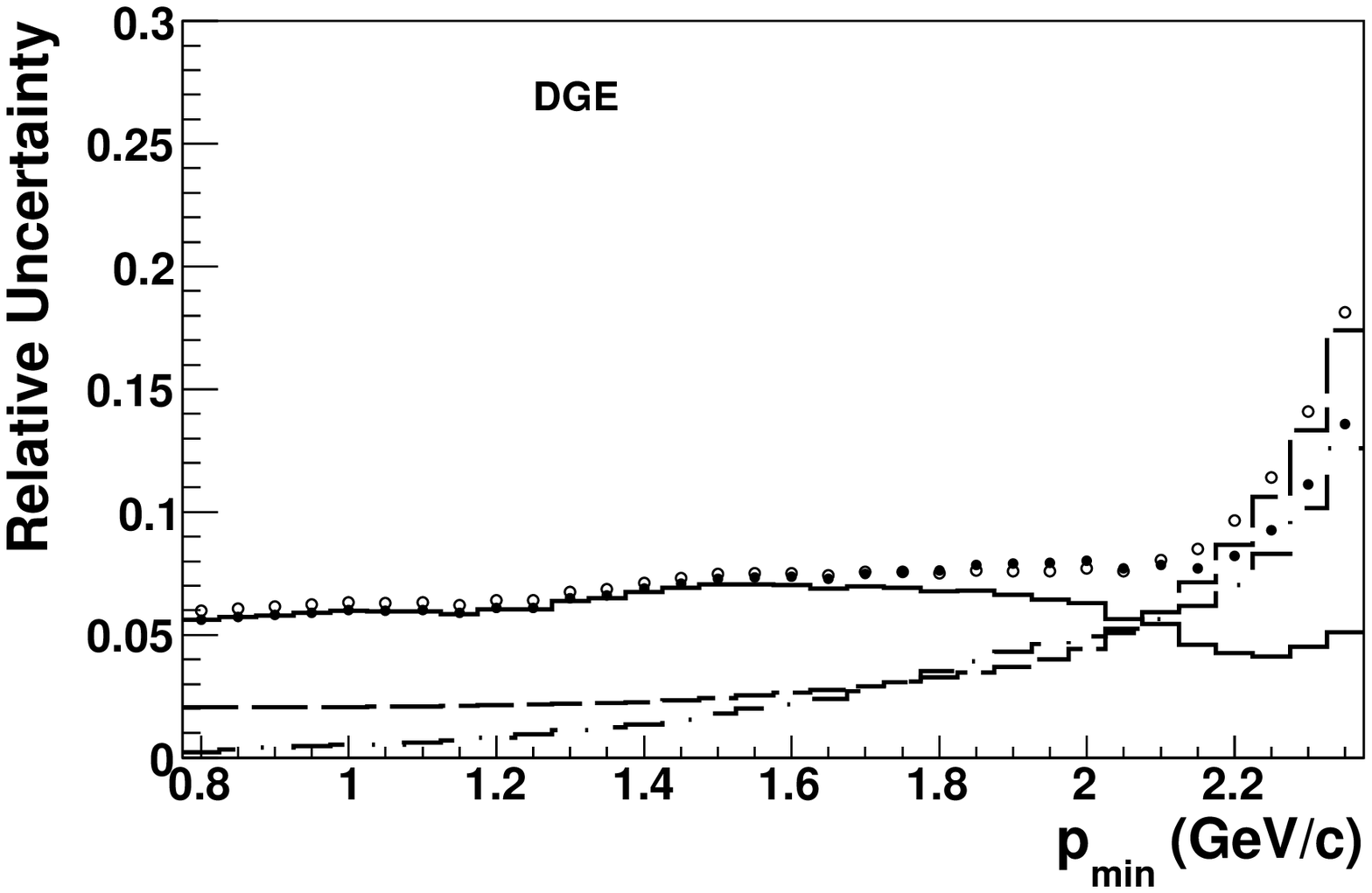}
\caption{
Relative uncertainties on the signal BF for DN, BLNP$_1$, GGOU$_1$, and DGE 
calculations as a function of $p_{\textrm{min}}$,
the lower limit of the electron momentum range used in the signal extraction, 
open and solid circles: total uncertainties (positive and negative uncertainties), 
solid line: experimental, long dashed and long dash-dotted lines: theoretical, 
dashed and dotted lines: SF uncertainty. 
}
\label{fig:branching_2}
\end{center}
\end{figure}

\subsection{ Extraction of \boldmath{$|V_{ub}|$}}

We rely on four  different theoretical calculations to extract $|V_{ub}|$ from the inclusive electron 
spectrum for $B\to X_u e \nu$ decays. The $|V_{ub}|$ and relative uncertainties are shown in
Figs.~\ref{fig:vub_1} and \ref{fig:vub_2}, respectively.
The experimental uncertainty includes statistical uncertainty and the uncertainty of the background subtraction. 
The SF uncertainty includes stated uncertainties on the SF parameters and their correlation. Specifically,  
we adopt the maximum deviation of the central value of $|V_{ub}|$ from selected SF parameter values on the error ellipse. 
The resulting values of $|V_{ub}|$ and their uncertainties are largely constant for lower values of $p_{\textrm{min}}$, 
and rise sharply above 2.1 \gevc.

\begin{figure}[htbp]
\begin{center}
\includegraphics[height=4.74cm]{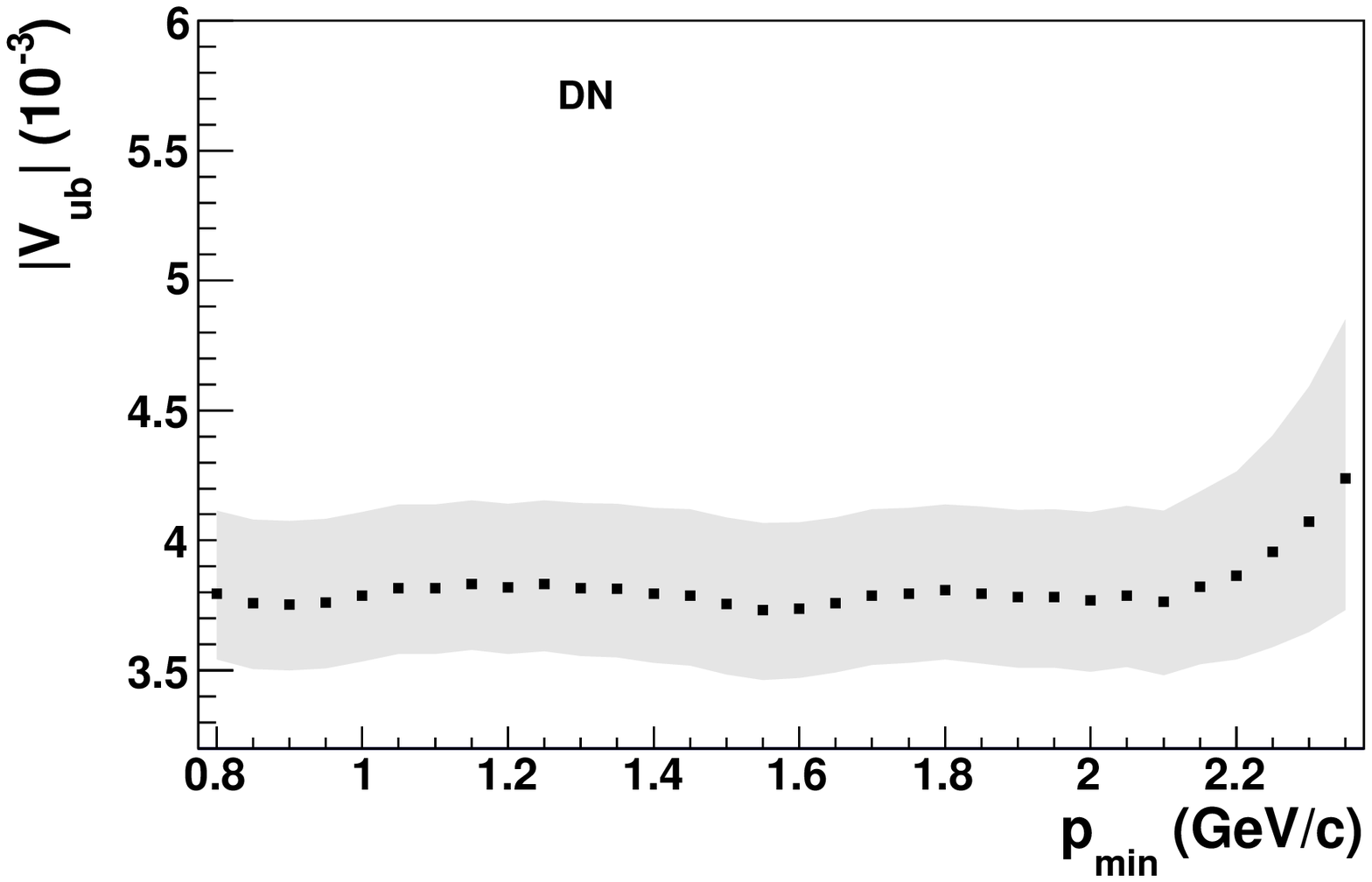}
\includegraphics[height=4.74cm]{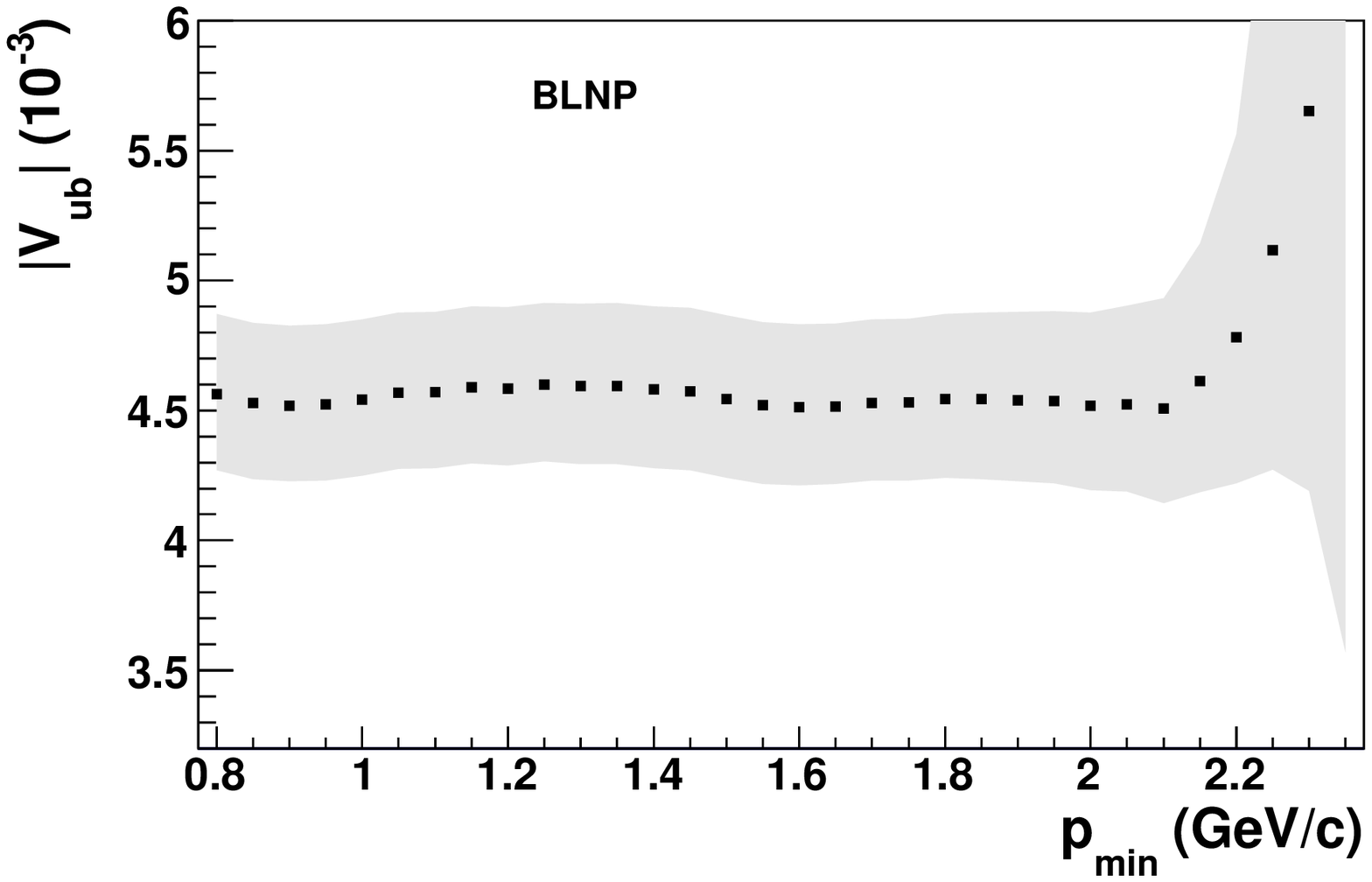}
\includegraphics[height=4.74cm]{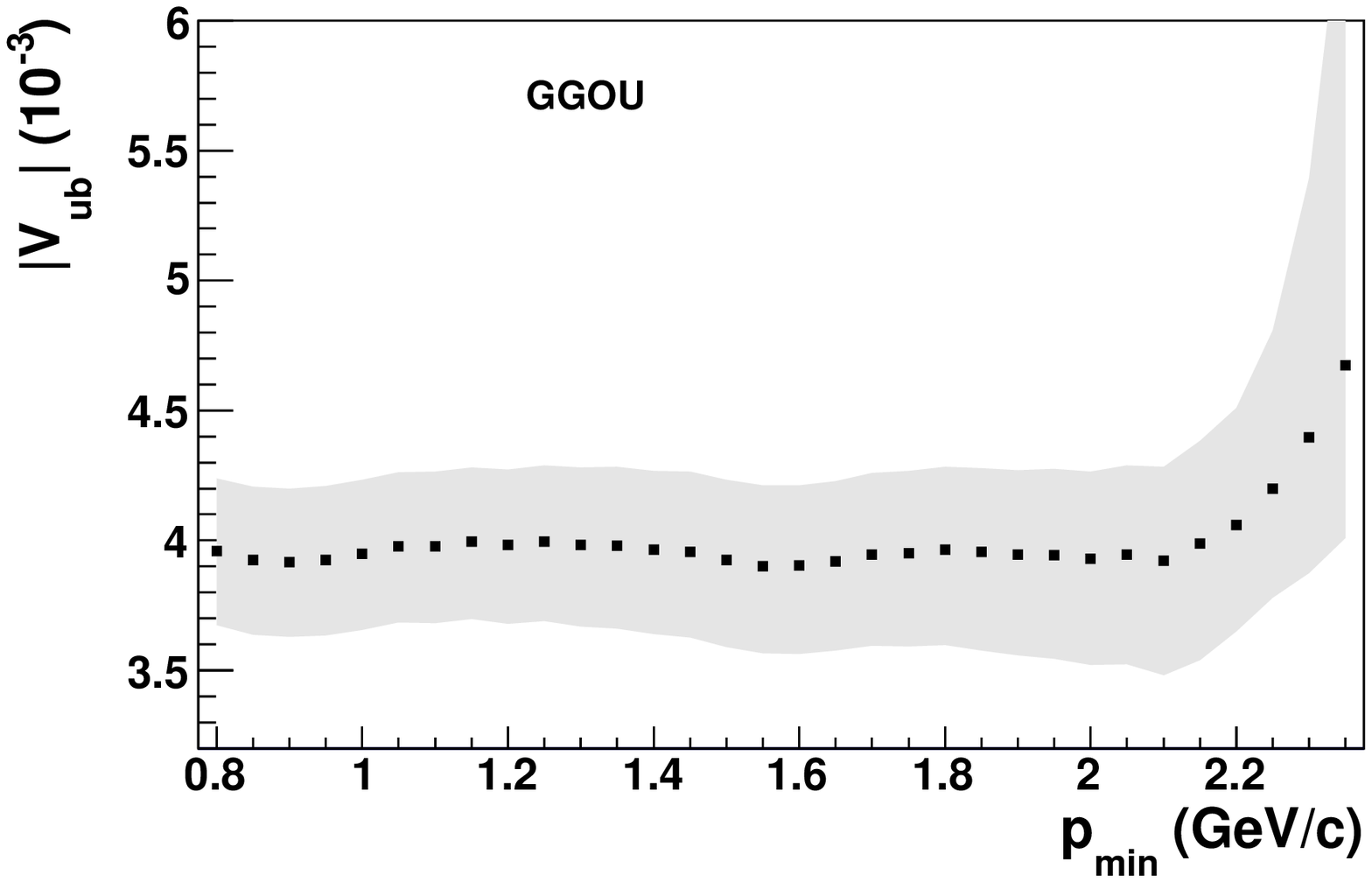}
\includegraphics[height=4.74cm]{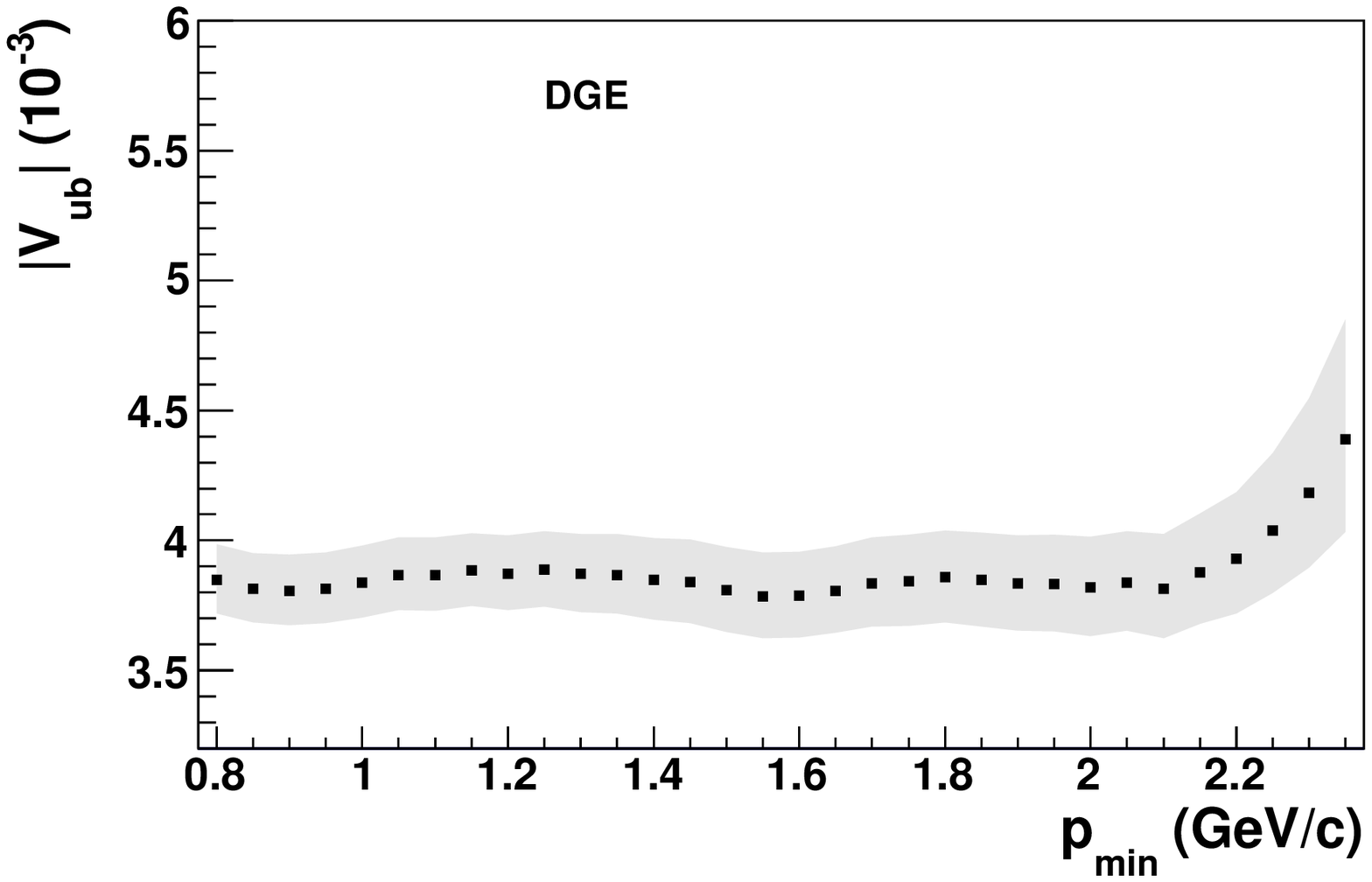}
\caption{
$|V_{ub}|$ as a function of $p_{\textrm{min}}$, the lower limit of the momentum range used 
in the extraction of the signal, for DN, BLNP$_1$, GGOU$_1$, and DGE predictions of the decay rate.
}
\label{fig:vub_1}
\end{center}
\end{figure}

\begin{figure}[htbp]
\begin{center}
\includegraphics[height=4.74cm]{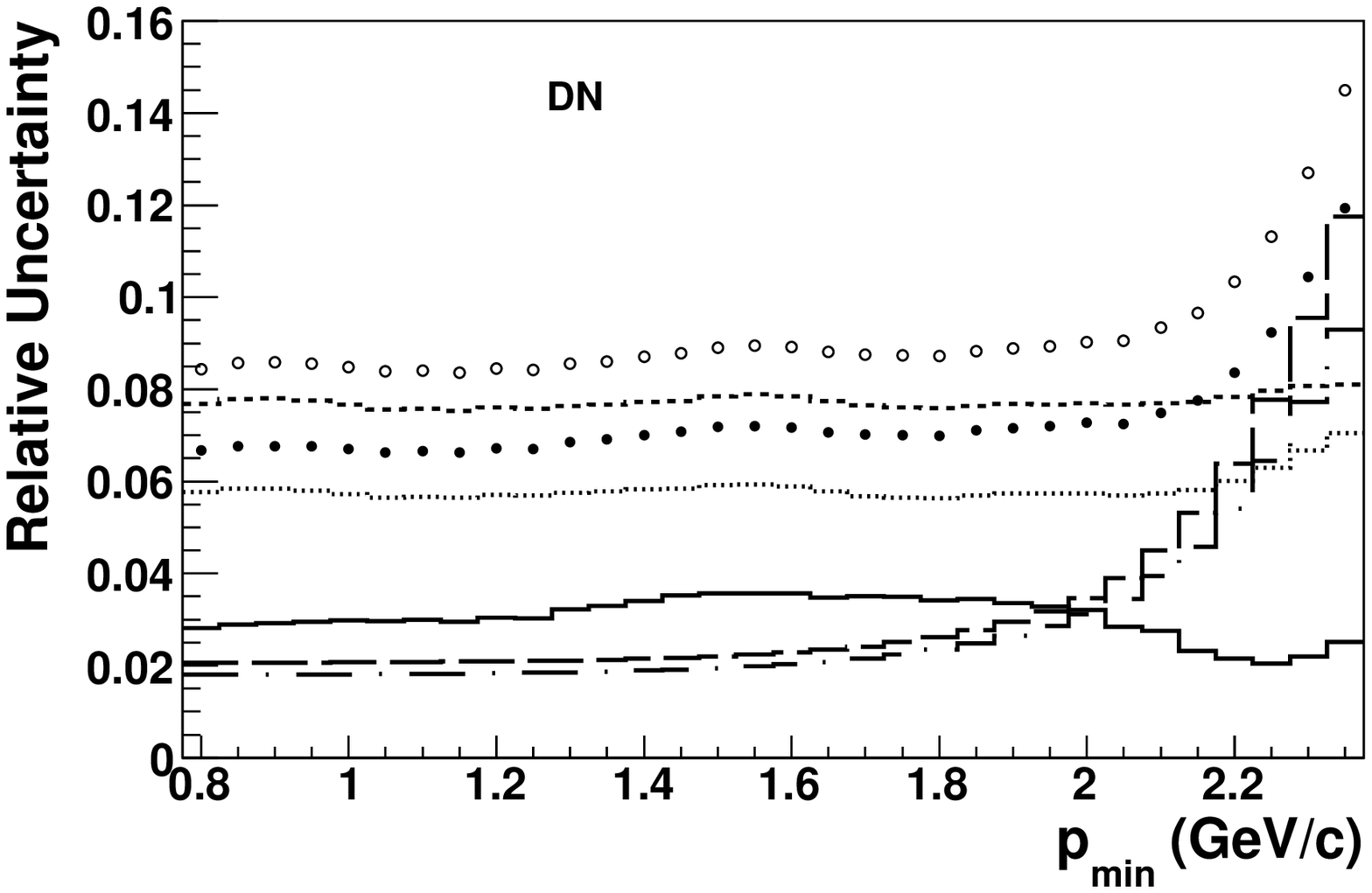}
\includegraphics[height=4.74cm]{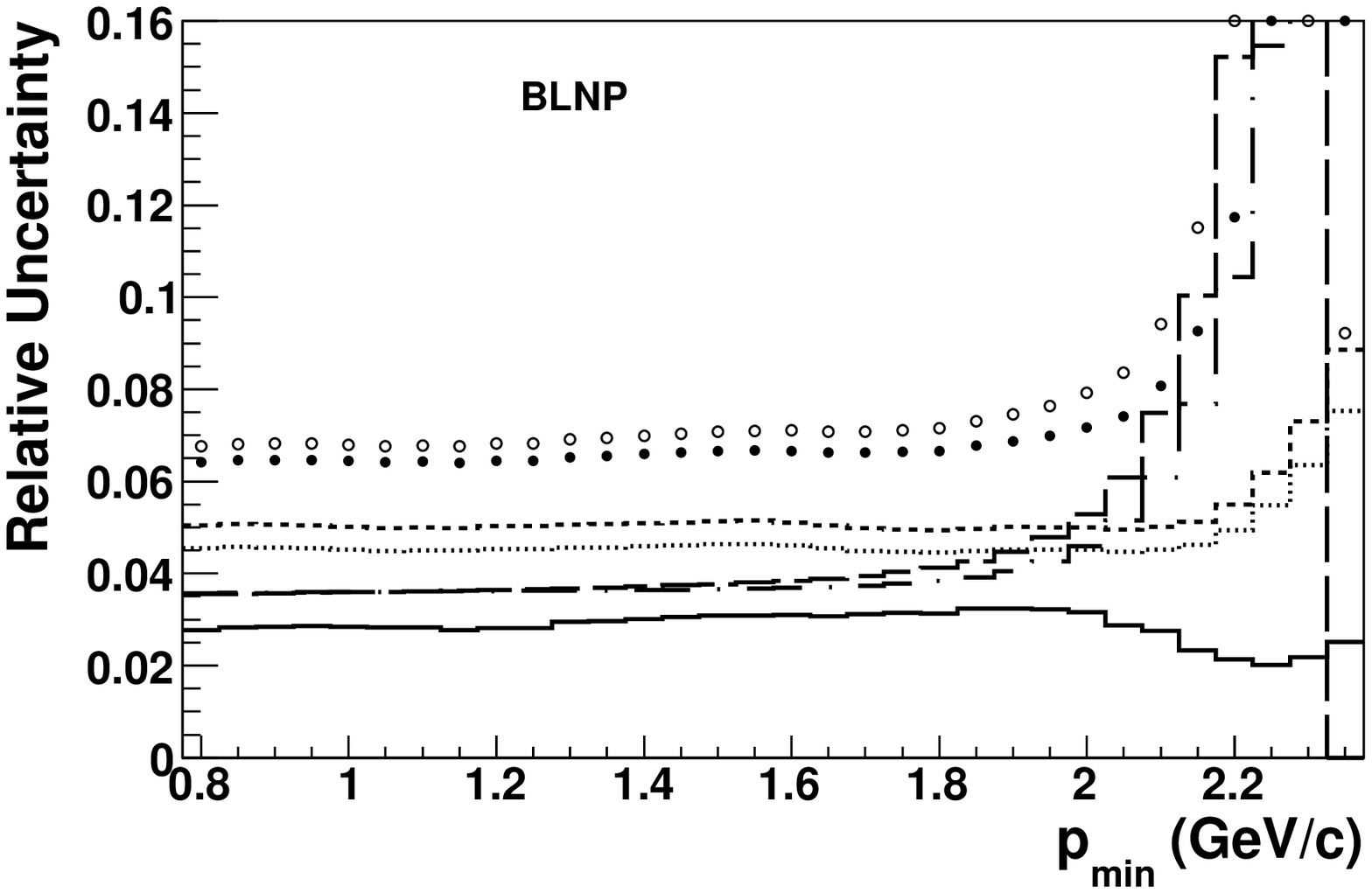}
\includegraphics[height=4.74cm]{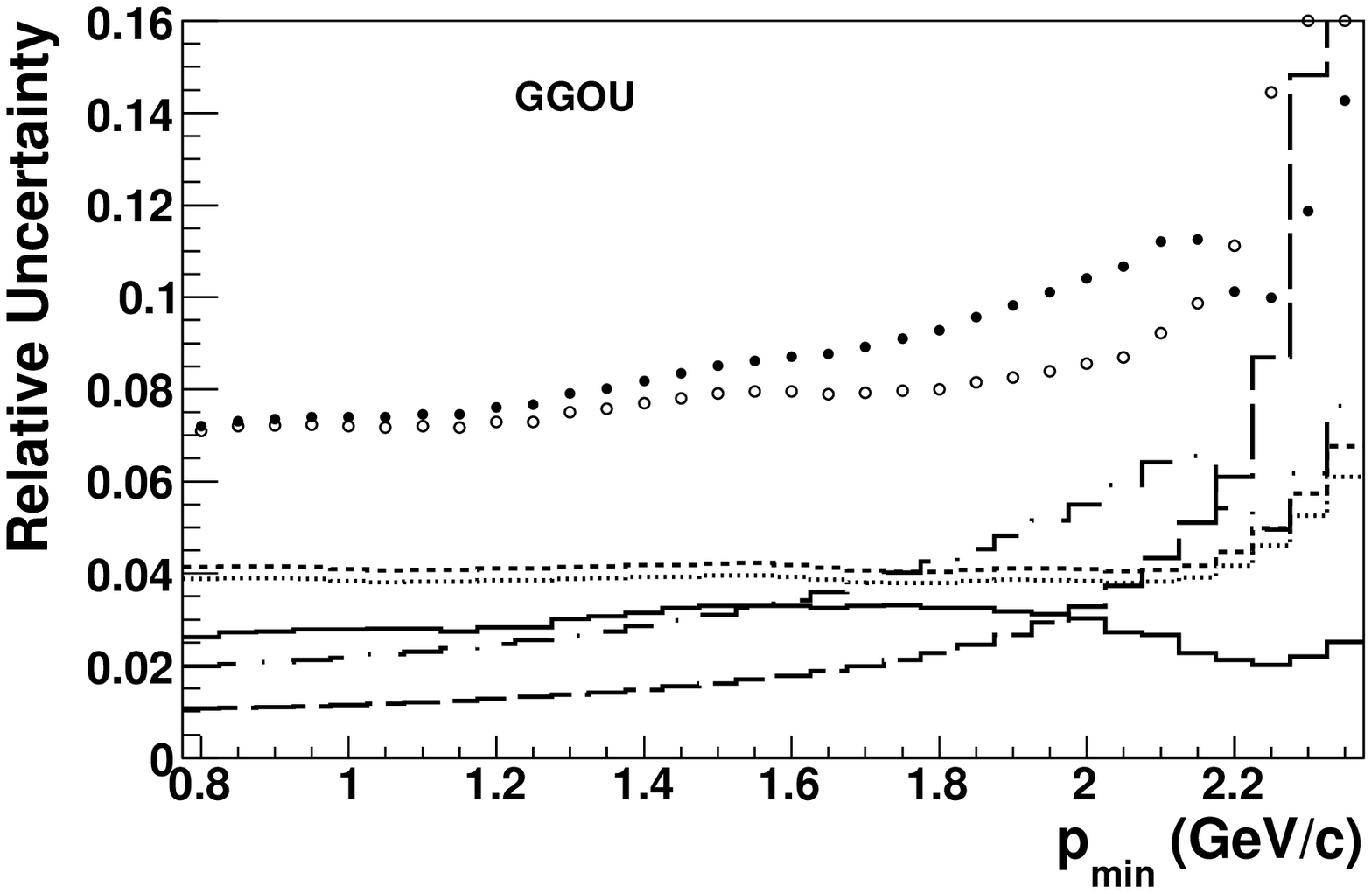}
\includegraphics[height=4.74cm]{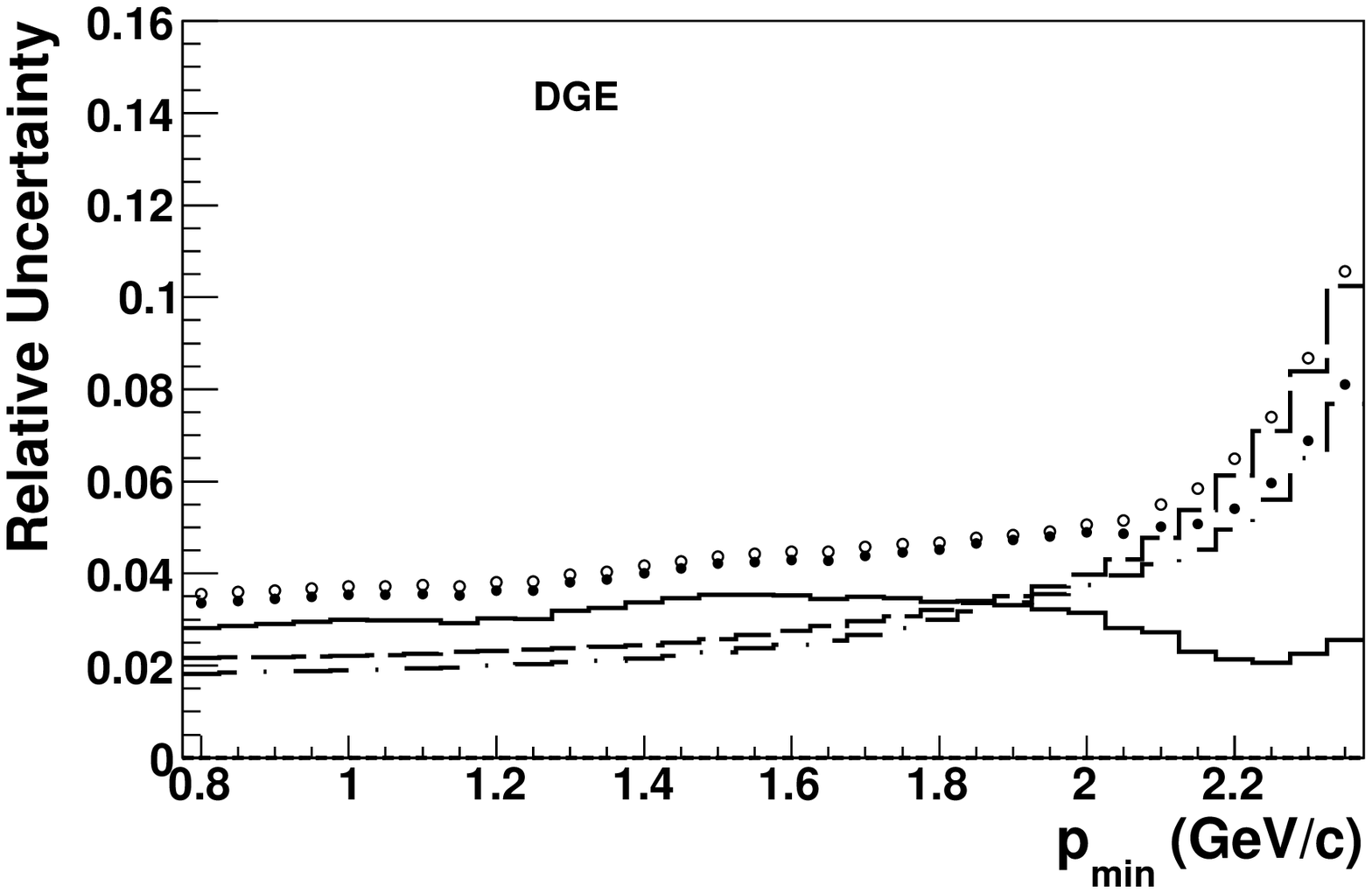}
\caption{
Relative uncertainties on $|V_{ub}|$ for four different predictions as a function of the lower limit $p_{\textrm{min}}$
of the momentum range used to extract the signal, for DN, BLNP$_1$, GGOU$_1$, and DGE predictions of the signal rate: 
total uncertainties (open and solid dots for positive and negative uncertainties), 
experimental (solid histogram), SF parametrization (dashed and dotted histograms for positive and negative uncertainties), and
theoretical (long dashed and long-dotted histograms for positive and negative uncertainties).
}
\label{fig:vub_2}
\end{center}
\end{figure}

\subsection{ Summary of results}

\begin{table*}[tbp]
\caption{
Results for ${\Delta \cal B}(B \to X_u e \nu)$, ${\cal B}(B \to X_u e \nu)$, $|V_{ub}|$ and $\Delta \zeta (\Delta p)$ 
based on the electron momentum range $\Delta p =0.8-2.7$~\gevc 
for different theoretical predictions, with experimental, SF, and theory uncertainties.  
}
\label{t:vub}
{\small
\begin{ruledtabular}
\begin{tabular}{llll}
${\Delta \cal B} [10^{-3}]$ & ${\cal B} [10^{-3}]$ & $|V_{ub}| [10^{-3}]$ & $\Delta \zeta (\Delta p) [\textrm{ps}^{-1}]$\\ \hline

\noalign{\vskip2pt}
\multicolumn{4}{c}{DN} \\
\rule[-0mm]{0mm}{4mm}
$1.397 \pm 0.078_{\textrm{exp}} \,^{+0.214}_{-0.153\,{\textrm{SF}}}$ &
$1.494 \pm 0.084_{\textrm{exp}} \,^{+0.239}_{-0.167\,{\textrm{SF}}} \,^{+0.030}_{-0.003\,{\textrm{theory}}}$ & 
$3.794 \pm 0.107_{\textrm{exp}} \,^{+0.292}_{-0.219\,{\textrm{SF}}} \,^{+0.078}_{-0.068\,{\textrm{theory}}}$ & 
$61.43 \,^{+0.20}_{-0.33\,{\textrm{SF}}} \,^{+2.28}_{-2.45\,{\textrm{theory}}}$ \\ \\ \hline

\noalign{\vskip2pt}
\multicolumn{4}{c}{DGE} \\
\rule[-0mm]{0mm}{4mm}
$1.433 \pm 0.081_{\textrm{exp}}$ &
$1.537 \pm 0.086_{\textrm{exp}} \,^{+0.031}_{-0.003\,{\textrm{theory}}}$ & 
$3.848 \pm 0.108_{\textrm{exp}} \,^{+0.084}_{-0.070\,{\textrm{theory}}}$ & 
$61.26 \,^{+2.30}_{-2.58\,{\textrm{theory}}}$ \\ \\ \hline 

\noalign{\vskip2pt}
\multicolumn{4}{c}{$X_c \ell \nu$, $m_c$ constraint fit of SF parameters, GGOU$_1$} \\
\rule[-0mm]{0mm}{4mm}
$1.554 \pm 0.082_{\textrm{exp}} \,^{+0.095}_{-0.086\,{\textrm{SF}}}$ &
$1.665 \pm 0.087_{\textrm{exp}} \,^{+0.103}_{-0.093\,{\textrm{SF}}} \,^{+0.002}_{-0.011\,{\textrm{theory}}}$ & 
$3.959 \pm 0.104_{\textrm{exp}} \,^{+0.164}_{-0.154\,{\textrm{SF}}} \,^{+0.042}_{-0.079\,{\textrm{theory}}}$ & 
$62.76 \,^{+1.59}_{-1.55\,{\textrm{SF}}} \,^{+2.58}_{-1.32\,{\textrm{theory}}}$ \\ \\

\multicolumn{4}{c}{$X_c \ell \nu$, $X_s\gamma$ constraint fit of SF parameters, GGOU$_2$} \\
\rule[-0mm]{0mm}{4mm}
$1.551 \pm 0.081_{\textrm{exp}} \,^{+0.117}_{-0.100\,{\textrm{SF}}}$ &
$1.661 \pm 0.086_{\textrm{exp}} \,^{+0.127}_{-0.109\,{\textrm{SF}}} \,^{+0.008}_{-0.011\,{\textrm{theory}}}$ & 
$3.936 \pm 0.102_{\textrm{exp}} \,^{+0.202}_{-0.188\,{\textrm{SF}}} \,^{+0.168}_{-0.086\,{\textrm{theory}}}$ & 
$63.38 \,^{+2.15}_{-2.15\,{\textrm{SF}}} \,^{+2.87}_{-5.08\,{\textrm{theory}}}$ \\ \\ \hline

\noalign{\vskip2pt}
\multicolumn{4}{c}{$X_c \ell \nu$, $m_c$ constraint fit of SF parameters with $\mu_i=2.0 \gev$, BLNP$_1$} \\
\rule[-0mm]{0mm}{4mm}
$2.268 \pm 0.125_{\textrm{exp}} \,^{+0.191}_{-0.163\,{\textrm{SF}}}$ &
$2.418 \pm 0.134_{\textrm{exp}} \,^{+0.205}_{-0.176\,{\textrm{SF}}} \,^{+0.003}_{-0.003\,{\textrm{theory}}}$ & 
$4.563 \pm 0.126_{\textrm{exp}} \,^{+0.230}_{-0.208\,{\textrm{SF}}} \,^{+0.162}_{-0.163\,{\textrm{theory}}}$ & 
$68.93 \,^{+1.88}_{-1.85\,{\textrm{SF}}} \,^{+5.20}_{-4.65\,{\textrm{theory}}}$ \\ \\

\multicolumn{4}{c}{$X_c \ell \nu$, $m_c$ constraint fit of SF parameters with $\mu_i=1.5 \gev$, BLNP$_2$} \\ 
\rule[-0mm]{0mm}{4mm}
$2.924 \pm 0.112_{\textrm{exp}} \,^{+0.152}_{-0.137\,{\textrm{SF}}}$ &
$2.160 \pm 0.120_{\textrm{exp}} \,^{+0.164}_{-0.148\,{\textrm{SF}}} \,^{+0.002}_{-0.003\,{\textrm{theory}}}$ & 
$4.406 \pm 0.122_{\textrm{exp}} \,^{+0.203}_{-0.193\,{\textrm{SF}}} \,^{+0.176}_{-0.189\,{\textrm{theory}}}$ & 
$66.00 \,^{+1.88}_{-1.82\,{\textrm{SF}}} \,^{+6.06}_{-4.96\,{\textrm{theory}}}$ \\ \\

\multicolumn{4}{c}{$X_c \ell \nu$, $X_s\gamma$ constraint fit of SF parameters with $\mu_i=2.0 \gev$, BLNP$_3$} \\
\rule[-0mm]{0mm}{4mm}
$2.157 \pm 0.120_{\textrm{exp}} \,^{+0.207}_{-0.176\,{\textrm{SF}}}$ &
$2.298 \pm 0.128_{\textrm{exp}} \,^{+0.222}_{-0.189\,{\textrm{SF}}} \,^{+0.003}_{-0.003\,{\textrm{theory}}}$ & 
$4.420 \pm 0.123_{\textrm{exp}} \,^{+0.273}_{-0.251\,{\textrm{SF}}} \,^{+0.156}_{-0.158\,{\textrm{theory}}}$ & 
$66.94 \,^{+2.69}_{-2.62\,{\textrm{SF}}} \,^{+6.15}_{-5.02\,{\textrm{theory}}}$ \\ \\

\multicolumn{4}{c}{$X_c \ell \nu$, $X_s\gamma$ constraint fit of SF parameters with $\mu_i=1.5 \gev$, BLNP$_4$} \\ 
\rule[-0mm]{0mm}{4mm}
$1.931 \pm 0.108_{\textrm{exp}} \,^{+0.172}_{-0.147\,{\textrm{SF}}}$ &
$2.059 \pm 0.115_{\textrm{exp}} \,^{+0.185}_{-0.158\,{\textrm{SF}}} \,^{+0.002}_{-0.002\,{\textrm{theory}}}$ & 
$4.273 \pm 0.119_{\textrm{exp}} \,^{+0.263}_{-0.243\,{\textrm{SF}}} \,^{+0.170}_{-0.184\,{\textrm{theory}}}$ & 
$69.88 \,^{+2.70}_{-2.64\,{\textrm{SF}}} \,^{+5.26}_{-4.69\,{\textrm{theory}}}$ \\ \\

\end{tabular}
\end{ruledtabular}
}
\end{table*}

\begin{table*}[tbp]
\caption{
Results for ${\Delta \cal B}(B \to X_u e \nu)$, ${\cal B}(B \to X_u e \nu)$, $|V_{ub}|$ and $\Delta \zeta (\Delta p)$
based on the electron momentum range $\Delta p =2.1-2.7$~\gevc 
for different theoretical predictions, with experimental, SF, and theory uncertainties. 
}
\label{t:vub_20}
{\small
\begin{ruledtabular}
\begin{tabular}{llll}
${\Delta \cal B} [10^{-3}]$ & ${\cal B} [10^{-3}]$ & $|V_{ub}| [10^{-3}]$ & $\Delta \zeta (\Delta p)[\textrm{ps}^{-1}]$ \\ \hline

\noalign{\vskip2pt}
\multicolumn{4}{c}{DN} \\
\rule[-0mm]{0mm}{4mm}
$0.330 \pm 0.018_{\textrm{exp}} \,^{+0.009}_{-0.009\,{\textrm{SF}}}$ &
$1.471 \pm 0.081_{\textrm{exp}} \,^{+0.235}_{-0.164\,{\textrm{SF}}} \,^{+0.124}_{-0.101\,{\textrm{theory}}}$ & 
$3.764 \pm 0.104_{\textrm{exp}} \,^{+0.290}_{-0.216\,{\textrm{SF}}} \,^{+0.170}_{-0.148\,{\textrm{theory}}}$ & 
$14.75 \,^{+1.41}_{-1.70\,{\textrm{SF}}} \,^{+1.23}_{-1.24\,{\textrm{theory}}}$ \\ \\ \hline

\noalign{\vskip2pt}
\multicolumn{4}{c}{DGE} \\
\rule[-0mm]{0mm}{4mm}
$0.331 \pm 0.018_{\textrm{exp}}$ & 
$1.511 \pm 0.082_{\textrm{exp}} \,^{+0.090}_{-0.085\,{\textrm{theory}}}$ & 
$3.815 \pm 0.104_{\textrm{exp}} \,^{+0.182}_{-0.160\,{\textrm{theory}}}$ & 
$14.40 \,^{+1.29}_{-1.28\,{\textrm{theory}}}$ \\ \\ \hline

\noalign{\vskip2pt}
\multicolumn{4}{c}{$X_c \ell \nu$, $m_c$ constraint fit of SF parameters, GGOU$_1$} \\
\rule[-0mm]{0mm}{4mm}
$0.342 \pm 0.018_{\textrm{exp}} \,^{+0.007}_{-0.006\,{\textrm{SF}}}$ & 
$1.634 \pm 0.087_{\textrm{exp}} \,^{+0.100}_{-0.090\,{\textrm{SF}}} \,^{+0.109}_{-0.163\,{\textrm{theory}}}$ & 
$3.922 \pm 0.104_{\textrm{exp}} \,^{+0.160}_{-0.150\,{\textrm{SF}}} \,^{+0.170}_{-0.251\,{\textrm{theory}}}$ & 
$14.06 \,^{+0.87}_{-0.82\,{\textrm{SF}}} \,^{+1.99}_{-1.14\,{\textrm{theory}}}$ \\ \\

\multicolumn{4}{c}{$X_c \ell \nu$, $X_s\gamma$ constraint fit of SF parameters, GGOU$_2$} \\
\rule[-0mm]{0mm}{4mm}
$0.342 \pm 0.018_{\textrm{exp}} \,^{+0.008}_{-0.007\,{\textrm{SF}}}$ & 
$1.630 \pm 0.086_{\textrm{exp}} \,^{+0.122}_{-0.105\,{\textrm{SF}}} \,^{+0.188}_{-0.189\,{\textrm{theory}}}$ & 
$3.899 \pm 0.103_{\textrm{exp}} \,^{+0.198}_{-0.185\,{\textrm{SF}}} \,^{+0.381}_{-0.289\,{\textrm{theory}}}$ & 
$14.23 \,^{+1.12}_{-1.08\,{\textrm{SF}}} \,^{+2.37}_{-2.42\,{\textrm{theory}}}$ \\ \\ \hline

\noalign{\vskip2pt}
\multicolumn{4}{c}{$X_c \ell \nu$, $m_c$ constraint fit of SF parameters with $\mu_i=2.0 \gev$, BLNP$_1$} \\
\rule[-0mm]{0mm}{4mm}
$0.397 \pm 0.022_{\textrm{exp}} \,^{+0.014}_{-0.012\,{\textrm{SF}}}$ &
$2.359 \pm 0.130_{\textrm{exp}} \,^{+0.199}_{-0.170\,{\textrm{SF}}} \,^{+0.173}_{-0.133\,{\textrm{theory}}}$ & 
$4.507 \pm 0.124_{\textrm{exp}} \,^{+0.226}_{-0.204\,{\textrm{SF}}} \,^{+0.337}_{-0.275\,{\textrm{theory}}}$ & 
$12.36 \,^{+0.89}_{-0.83\,{\textrm{SF}}} \,^{+1.66}_{-1.66\,{\textrm{theory}}}$ \\ \\

\multicolumn{4}{c}{$X_c \ell \nu$, $m_c$ constraint fit of SF parameters with $\mu_i=1.5 \gev$, BLNP$_2$} \\ 
\rule[-0mm]{0mm}{4mm}
$0.376 \pm 0.021_{\textrm{exp}} \,^{+0.011}_{-0.010\,{\textrm{SF}}}$ & 
$2.110 \pm 0.117_{\textrm{exp}} \,^{+0.158}_{-0.143\,{\textrm{SF}}} \,^{+0.128}_{-0.087\,{\textrm{theory}}}$ & 
$4.356 \pm 0.120_{\textrm{exp}} \,^{+0.198}_{-0.190\,{\textrm{SF}}} \,^{+0.317}_{-0.265\,{\textrm{theory}}}$ & 
$12.55 \,^{+0.92}_{-0.85\,{\textrm{SF}}} \,^{+1.68}_{-1.64\,{\textrm{theory}}}$ \\ \\

\multicolumn{4}{c}{$X_c \ell \nu$, $X_s\gamma$ constraint fit of SF parameters with $\mu_i=2.0 \gev$, BLNP$_3$} \\
\rule[-0mm]{0mm}{4mm}
$0.389 \pm 0.022_{\textrm{exp}} \,^{+0.015}_{-0.013\,{\textrm{SF}}}$ & 
$2.244 \pm 0.124_{\textrm{exp}} \,^{+0.215}_{-0.183\,{\textrm{SF}}} \,^{+0.152}_{-0.117\,{\textrm{theory}}}$ & 
$4.367 \pm 0.121_{\textrm{exp}} \,^{+0.270}_{-0.248\,{\textrm{SF}}} \,^{+0.313}_{-0.257\,{\textrm{theory}}}$ & 
$12.91 \,^{+1.25}_{-1.17\,{\textrm{SF}}} \,^{+1.67}_{-1.67\,{\textrm{theory}}}$ \\ \\

\multicolumn{4}{c}{$X_c \ell \nu$, $X_s\gamma$ constraint fit of SF parameters with $\mu_i=1.5 \gev$, BLNP$_4$} \\
\rule[-0mm]{0mm}{4mm}
$0.370 \pm 0.020_{\textrm{exp}} \,^{+0.012}_{-0.010\,{\textrm{SF}}}$ & 
$2.013 \pm 0.111_{\textrm{exp}} \,^{+0.179}_{-0.153\,{\textrm{SF}}} \,^{+0.112}_{-0.075\,{\textrm{theory}}}$ & 
$4.225 \pm 0.116_{\textrm{exp}} \,^{+0.259}_{-0.239\,{\textrm{SF}}} \,^{+0.296}_{-0.250\,{\textrm{theory}}}$ & 
$13.10 \,^{+1.30}_{-1.20\,{\textrm{SF}}} \,^{+1.70}_{-1.66\,{\textrm{theory}}}$ \\ \\

\end{tabular}
\end{ruledtabular}
}
\end{table*}

The results for ${\cal B}(B\to X_u e \nu)$ and $|V_{ub}|$ are presented for the wide momentum range, $p_e=0.8-2.7~\gevc$, 
in Table~\ref{t:vub}, and for the narrower range, $p_e=2.1-2.7~\gevc$, in Table~\ref{t:vub_20}.
In these tables, the first uncertainty represents the combined statistical and systematic experimental uncertainties 
of the partial BF measurement, the second refers to the uncertainty in the determination of the shape
function parameters used by the DN, BLNP, and GGOU, and the third is due to the uncertainties
of the QCD calculations. 

For GGOU, we present results for two sets of SF parameters in the kinetic scheme, one based on fits to moments 
of lepton energy and hadron mass distributions from $B\to X_c \ell \nu$ decays and further constrained by the $c$-quark mass 
(GGOU$_1$), the other based on including the moments of the photon spectrum in $B \to X_s \gamma$ decays (GGOU$_2$). 

For BLNP, we present results for four sets of SF parameters in the SF scheme,  
based on fits of the moments of the lepton energy and hadron mass distributions in $B\to X_c \ell \nu$ decays: 
two with a constraint on the charm quark mass (BLNP$_1$, BLNP$_2$), and the other two on including the moments 
of the photon spectrum in $B \to X_s \gamma$ decays (BLNP$_3$, BLNP$_4$). 
For each pair of results, we choose two values for the scale parameter, 
$\mu_i=2.0 \gev$ and $\mu_i=1.5 \gev$. The results with the smaller scale parameter have the lower SF uncertainties. 

The resulting total BFs and $|V_{ub}|$ for DN, GGOU and DGE agree well within their uncertainties, while the 
BF results for BLNP are between about $25\%$ and $60\%$, and the values of $|V_{ub}|$ are about $8\%-20\%$ higher than 
for the other three QCD calculations. The BFs and the values of $|V_{ub}|$ that are extracted for the momentum 
range with $p_{\textrm{min}}=0.8$\gevc exceed those for $p_{\textrm{min}}=2.1$\gevc on average 
by $\sim 2\%$ and $\sim 1\%$, respectively.

To quantify the dependence of the total BF (and also $|V_{ub}|$) on the SF parameters we have introduced a simple relation, 

\begin{equation}
{\cal B}\times 10^{3}  = c_0 + c_1 \times \frac{x_1-x_1^0}{x_1^0} + c_2 \times \frac{x_2-x_2^0}{x_2^0}
\end{equation}   

\noindent The parameters $c_i$ and the default SF parameters values $x_i^0$ are given in Table ~\ref{t:SF_vary}. 

\begin{table}[tbp]
\caption{
Simple ansatz describing the dependence of the total branching fraction ${\cal B} (B \to X_u e \nu)$ and $|V_{ub}|$ 
on the shape function parameters $x_1$ and $x_2$, {\it i.e.},
$\overline{\Lambda}^{\textrm{SF}}$ and $\lambda_1^{\textrm{SF}}$ for DN, and $\m_b$ and $\mu^{2}_{\pi}$ for BLNP and GGOU.
}
\label{t:SF_vary}
{\small
\begin{ruledtabular}
\begin{tabular}{llllll}
${\cal B} \times 10^3$ & &         &        &          &         \\
QCD prediction  & $x^0_1$& $x^0_2$ &  $c_0$ &  $ c_1 $ & $c_2 $  \\ \hline
DN		& 0.49   & -0.24   &  1.494 & +1.498   & -0.072  \\
BLNP$_{1,3}$	& 4.561  & 0.149   &  2.418 & -34.608  & -0.252  \\
GGOU$_{1,2}$        & 4.560  & 0.453   & 1.665  & -13.714  & -0.314  \\ \hline            
\\
$|V_{ub}| \times 10^3$ & &     &        &          &          \\ 
QCD prediction  & $x^0_1$& $x^0_2$ &  $c_0$ &  $ c_1 $ & $c_2 $   \\ \hline
DN		& 0.49   & -0.24   &  3.794 & +1.949   & -0.109   \\
BLNP$_{1,3}$	& 4.561  & 0.149   &  4.563 & -43.621  & -0.178   \\
GGOU$_{1,2}$        & 4.560  & 0.453   &  3.959 & -25.024  & -0.357   \\             

\end{tabular}
\end{ruledtabular}
}
\end{table}

\section{Conclusions}

\begin{figure}[htbp]
\begin{center}
\includegraphics[height=5.cm]{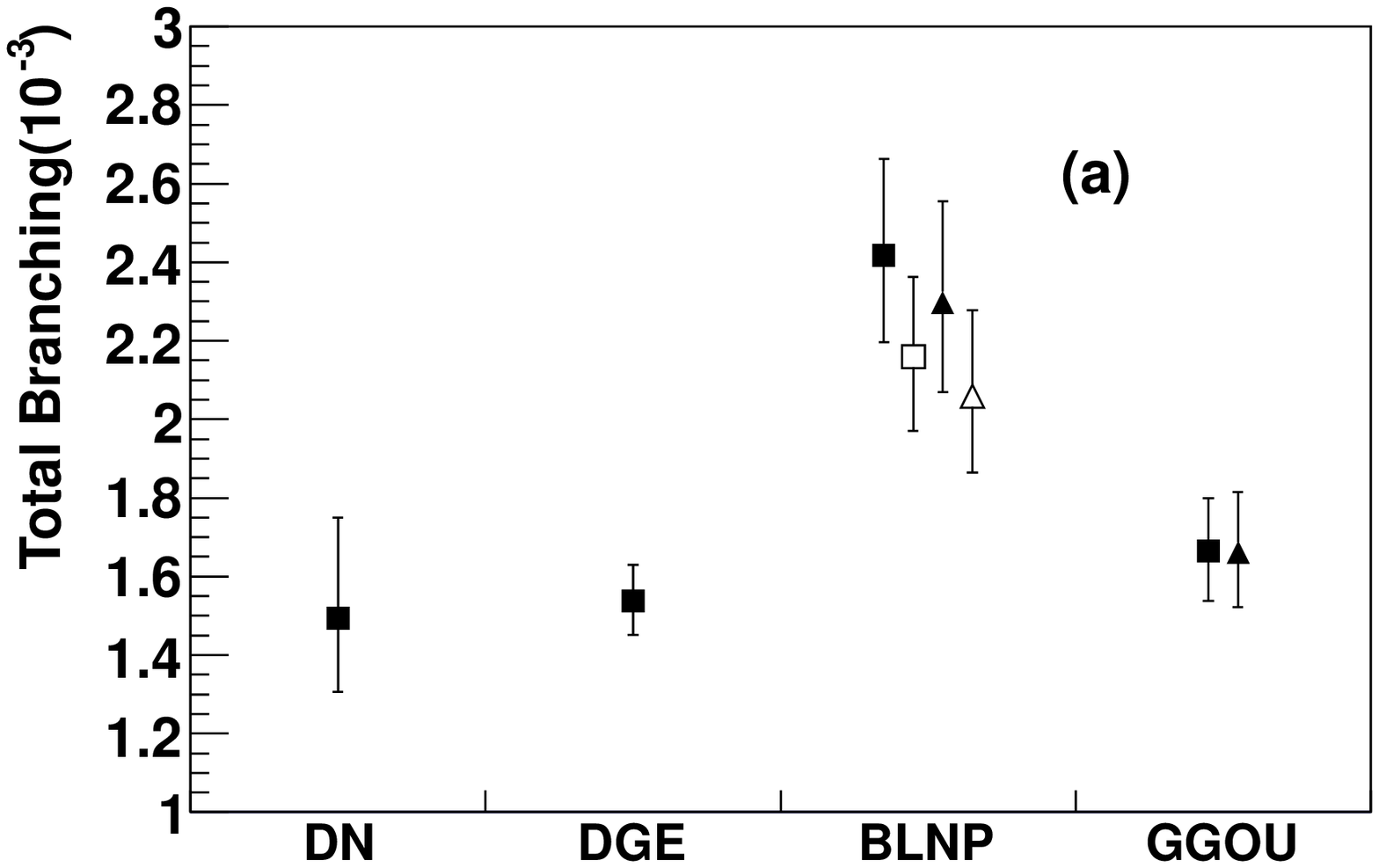}
\includegraphics[height=5.cm]{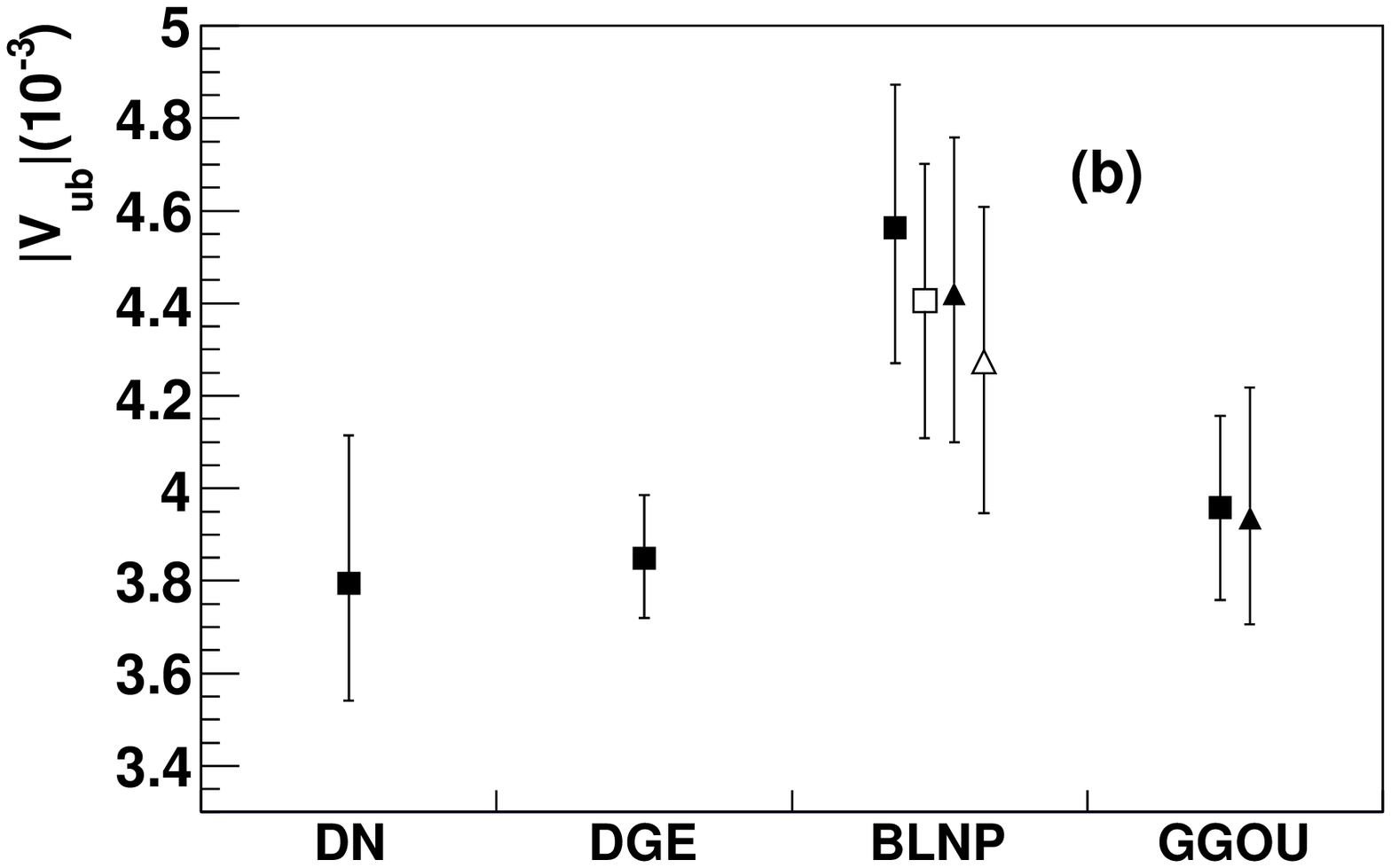}
\caption{
Results for (a) ${\cal B}(B \to X_u e \nu)$, and (b) $|V_{ub}|$ 
based on the momentum range $p_e=0.8-2.7$~GeV/c for different QCD calculations
and SF parametrizations (Table~\ref{t:vub}). For BLNP and GGOU: 
solid squares $- X_c \ell \nu$ moment fits with $m_c$ constraint (BLNP$_1$ and GGOU$_1$), 
solid triangles $- X_c \ell \nu$ and $X_s\gamma$ moment fits (BLNP$_3$ and GGOU$_2$), 
open squares or triangles $-$ the same constraints for translation of SF parameters from ``kinetic''
to ``shape-function'' scheme with $\mu_i$=1.5~GeV (BLNP$_2$ and BLNP$_4$).
}
\label{fig:vub_res}
\end{center}
\end{figure}

Based on the total \babar\ data sample, we have measured the inclusive electron spectrum and  BF,
averaged over $B^{\pm}$ and $B^0$ mesons, 
${\cal B}(B\rightarrow X e\nu) = (10.34 \pm 0.04_{\textrm{stat}} \pm 0.26_{\textrm{syst}})$\%.
From a fit to this observed inclusive spectrum, we have extracted the spectrum and partial BFs for the CKM suppressed 
$B\to X_u e \nu$ decays in the momentum range $0.8 < p_e < 2.7 \gevc$. This fit requires as input knowledge 
of the shapes of the signal and all background contributions, many of them derived from measurements.
The most challenging input is the prediction of the shape of the $B \to X_u e \nu$ spectrum, for which we rely 
on predictions of four sets of QCD calculations that involve estimates of the perturbative and nonperturbative 
hadronic corrections. Specifically, we have used the earlier calculations by De Fazio and Neubert~\cite{dFN}
and Kagan and Neubert~\cite{kagan_neubert}, and the more comprehensive approaches
by Lange, Neubert and Paz~\cite{neubert_extract}, Gambino, Giordano, Ossola and Uraltsev~\cite{ggou1,ggou2}, 
and Andersen and Gardi~\cite{gardi,dge1,dge3}. 
These calculations also enter the determination of $|V_{ub}|$. The resulting values of $|V_{ub}|$ and the total 
uncertainties are shown in Fig.~\ref{fig:vub_res}. The measurements based on DN, DGE, and GGOU agree well within their 
uncertainties, while the BLNP calculations result in significantly higher values for both the total BF and $|V_{ub}|$. 
The differences of measured $|V_{ub}|$ values and the world averages \cite{hfag14} are 0.3$\sigma$ (BLNP), 
1.9$\sigma$ (GGOU), 2.5$\sigma$ (DGE).

These results are in very good agreement with the earlier \babar\ measurements~\cite{babarVub} of the
inclusive lepton spectrum at the \FourS\ resonance, which showed similar differences between results based 
on the DN and BLNP predictions.

As in earlier measurements based on the lepton momentum spectrum, the uncertainties on the $B \to X_u e \nu$ lepton spectrum 
and the extraction of $|V_{ub}|$ are dominated by the current knowledge of the shapes of signal and background spectra, 
in particular in the theoretical predictions of the spectrum, both in terms of perturbative and nonperturbative corrections 
to the predicted rate.

\section*{ACKNOWLEDGMENTS}
\label{sec:acknowledgments}
We thank 
S.\ W.\ Bosch, B.\ O.\ Lange, M.\ Neubert,  and G.\ Paz,
and similarly P.\ Gambino, P.\ Giordano, G.\ Ossola, and N.\ Uraltsev,
for providing the software to implement their respective calculations. 
We are grateful for the 
extraordinary contributions of our \pep2\ colleagues in
achieving the excellent luminosity and machine conditions
that have made this work possible.
The success of this project also relies critically on the 
expertise and dedication of the computing organizations that 
support \babar.
The collaborating institutions wish to thank 
SLAC for its support and the kind hospitality extended to them. 
This work is supported by the
U.S. Department of Energy
and National Science Foundation, the
Natural Sciences and Engineering Research Council (Canada),
the Commissariat \`a l'Energie Atomique and
Institut National de Physique Nucl\'eaire et de Physique des Particules
(France), the
Bundesministerium f\"ur Bildung und Forschung and
Deutsche Forschungsgemeinschaft
(Germany), the
Istituto Nazionale di Fisica Nucleare (Italy),
the Foundation for Fundamental Research on Matter (The Netherlands),
the Research Council of Norway, the
Ministry of Education and Science of the Russian Federation, 
Ministerio de Econom\'{\i}a y Competitividad (Spain), and the
Science and Technology Facilities Council (United Kingdom).
Individuals have received support from 
the Marie-Curie IEF program (European Union) and the A. P. Sloan Foundation (USA).

\clearpage

\begin{widetext}
\begin{verbatim}

                                         Supplementary Material


The differential branching fraction (GGOU) for charmless 
semileptonic B decays as a function of the electron momentum (in the Upsilon(4S) 
rest frame) after background subtraction and corrections for bremsstrahlung and
final state radiation. The errors indicate the statistical errors on the background 
subtraction, including the uncertainties of the fit parameters.

p(GeV/c)      Delta B(10**{-3})/(50 MeV/c) 

0.80-0.85     0.0471 +- 0.0170
0.85-0.90     0.0275 +- 0.0140
0.90-0.95     0.0182 +- 0.0145
0.95-1.00     0.0090 +- 0.0127
1.00-1.05     0.0076 +- 0.0112
1.05-1.10     0.0320 +- 0.0103
1.10-1.15     0.0220 +- 0.0110
1.15-1.20     0.0447 +- 0.0105
1.20-1.25     0.0296 +- 0.0115
1.25-1.30     0.0496 +- 0.0110
1.30-1.35     0.0448 +- 0.0116
1.35-1.40     0.0557 +- 0.0111
1.40-1.45     0.0515 +- 0.0107
1.45-1.50     0.0665 +- 0.0107
1.50-1.55     0.0620 +- 0.0122
1.55-1.60     0.0510 +- 0.0119
1.60-1.65     0.0463 +- 0.0114
1.65-1.70     0.0449 +- 0.0100
1.70-1.75     0.0564 +- 0.0104
1.75-1.80     0.0552 +- 0.0110
1.80-1.85     0.0653 +- 0.0112
1.85-1.90     0.0664 +- 0.0103
1.90-1.95     0.0639 +- 0.0092
1.95-2.00     0.0676 +- 0.0081
2.00-2.05     0.0602 +- 0.0072
2.05-2.10     0.0675 +- 0.0064
2.10-2.15     0.0514 +- 0.0055
2.15-2.20     0.0521 +- 0.0041
2.20-2.25     0.0450 +- 0.0031
2.25-2.30     0.0414 +- 0.0023
2.30-2.35     0.0366 +- 0.0019
2.35-2.40     0.0344 +- 0.0017
2.40-2.45     0.0266 +- 0.0016
2.45-2.50     0.0222 +- 0.0016
2.50-2.55     0.0146 +- 0.0016
2.55-2.60     0.0091 +- 0.0016
2.60-2.65     0.0041 +- 0.0016
2.65-2.70     0.0040 +- 0.0015

Correlation matrix for statistical contributions is the following: 

p(GeV/c) 0.80-0.85  0.85-0.90  0.90-0.95  0.95-1.00  1.00-1.05  1.05-1.10  1.10-1.15  
1.15-1.20  1.20-1.25  1.25-1.30  1.30-1.35  1.35-1.40  1.40-1.45  1.45-1.50  1.50-1.55  
1.55-1.60  1.60-1.65  1.65-1.70  1.70-1.75  1.75-1.80  1.80-1.85  1.85-1.90  1.90-1.95  
1.95-2.00  2.00-2.05  2.05-2.10  2.10-2.15  2.15-2.20  2.20-2.25  2.25-2.30  2.30-2.35  
2.35-2.40  2.40-2.45  2.45-2.50  2.50-2.55  2.55-2.60  2.60-2.65  2.65-2.70    

0.80-0.85     1.0000  0.2887  0.1314  0.1027  -0.0057  -0.0021  -0.0288  0.0122  
0.0054  -0.0664  -0.0997  -0.1284  -0.1223  -0.1601  -0.1025  -0.0423  -0.0453  -0.1029  
-0.0667  0.0346  0.0212  -0.0046  0.0186  0.0349  0.0566  0.0686  0.0485  0.0308  
-0.0209  0.0178  -0.0244  -0.0083  -0.0121  -0.0083  0.0095  -0.0031  -0.0264  -0.0140  

0.85-0.90     0.2887  1.0000  0.3872  0.2255  0.0539  0.0336  0.0251  -0.0886  
-0.1193  -0.1503  -0.1770  -0.1753  -0.1086  -0.0395  -0.0250  -0.0777  -0.0615  -0.0866  
-0.1381  -0.1138  0.0045  -0.0024  0.0664  0.0633  -0.0249  0.0506  0.0212  0.0456  
-0.0449  0.1222  -0.0481  0.0065  0.0422  0.0166  -0.0395  0.0827  -0.0476  0.0941  

0.90-0.95     0.1314  0.3872  1.0000  0.4980  0.1404  0.1015  0.0294  -0.1461  
-0.2341  -0.2161  -0.2260  -0.1894  -0.1644  -0.0941  -0.0615  -0.0471  -0.0389  -0.0227  
-0.2051  -0.2035  -0.1089  -0.0541  0.0425  0.0030  0.0069  -0.0036  0.0133  0.0408  
-0.0385  0.0122  -0.0459  0.0719  -0.0243  0.0528  -0.0377  0.0984  -0.0416  0.0940  

0.95-1.00     0.1027  0.2255  0.4980  1.0000  0.2884  0.1141  -0.0381  -0.1223  
-0.1812  -0.0574  -0.0943  -0.1014  -0.1503  -0.0967  -0.0217  -0.0148  0.0040  -0.0316  
-0.1713  -0.2460  -0.1186  -0.0553  -0.0363  -0.0812  0.0362  -0.0170  -0.0141  0.0764  
-0.0123  0.0541  -0.0155  0.0775  0.0151  0.0350  0.0236  0.0514  0.0094  0.0601  

1.00-1.05     -0.0057  0.0539  0.1404  0.2884  1.0000  0.1229  -0.0540  -0.0758  
-0.0470  -0.1133  -0.0840  -0.0412  -0.0161  -0.0745  -0.0353  0.0307  0.0435  -0.0166  
-0.1473  -0.1842  -0.0882  -0.0981  -0.1204  -0.1442  -0.1148  -0.0595  -0.0123  0.0009  
0.0035  0.0206  0.0098  0.0300  0.0065  0.0094  -0.0320  0.0380  0.0367  0.0008  

1.05-1.10     -0.0021  0.0336  0.1015  0.1141  0.1229  1.0000  0.2971  0.1059  
-0.0262  0.0009  0.0050  -0.0183  -0.1227  -0.1060  -0.0959  -0.1126  -0.0710  -0.0114  
-0.1497  -0.1499  -0.0441  -0.0632  -0.0570  -0.0792  -0.0956  0.0050  0.0083  0.0435  
-0.0088  0.0077  0.0505  -0.0151  -0.0085  0.0312  -0.0247  0.0256  -0.0182  0.0462  

1.10-1.15     -0.0288  0.0251  0.0294  -0.0381  -0.0540  0.2971  1.0000  0.2324  
-0.0343  -0.0237  0.0452  0.0796  -0.0387  -0.0645  -0.1170  -0.1652  -0.1519  -0.0189  
-0.0694  -0.1264  -0.0690  -0.0544  -0.0231  -0.0179  0.0220  0.0437  0.0147  0.0164  
-0.0221  0.0185  0.0352  0.0193  -0.0307  0.0731  0.0053  -0.0112  -0.0116  0.0386  

1.15-1.20     0.0122  -0.0886  -0.1461  -0.1223  -0.0758  0.1059  0.2324  1.0000  
0.2893  0.0596  0.0272  0.0388  -0.0094  -0.0042  -0.0397  -0.1472  -0.1003  -0.0469  
-0.0371  -0.0561  -0.0814  -0.0683  -0.0205  0.0167  0.0172  0.0602  0.1063  0.0366  
0.0356  0.0758  0.0158  0.0110  -0.0352  0.0064  0.0529  -0.0639  0.0308  -0.0397  

1.20-1.25     0.0054  -0.1193  -0.2341  -0.1812  -0.0470  -0.0262  -0.0343  0.2893  
1.0000  0.1731  0.0294  -0.0268  -0.0501  -0.0401  0.0106  -0.0421  -0.0421  -0.0366  
-0.0402  0.0380  0.0506  0.0284  -0.0424  -0.0125  -0.0255  0.0157  0.0661  0.0357  
0.0931  0.0529  0.0565  0.0004  0.0047  -0.0026  0.0204  -0.0387  0.0278  -0.0570  

1.25-1.30     -0.0664  -0.1503  -0.2161  -0.0574  -0.1133  0.0009  -0.0237  0.0596  
0.1731  1.0000  0.4072  0.1927  0.0660  -0.0597  -0.0166  -0.0551  -0.0598  -0.0732  
-0.0586  -0.1287  -0.0914  -0.1138  -0.1145  -0.0793  -0.0298  -0.0336  0.0030  0.0006  
0.0422  0.0291  0.0065  -0.0722  0.0440  -0.0369  0.0722  -0.0490  0.0642  -0.0479  

1.30-1.35     -0.0997  -0.1770  -0.2260  -0.0943  -0.0840  0.0050  0.0452  0.0272  
0.0294  0.4072  1.0000  0.4353  0.1608  -0.0110  -0.0524  -0.1568  -0.0360  0.0402  
-0.0166  -0.1517  -0.2344  -0.2475  -0.0816  -0.0478  -0.0240  0.0133  0.0148  0.0751  
0.0653  0.0362  0.0475  -0.0530  0.0571  -0.0522  0.0656  -0.0257  0.0495  -0.0238  

1.35-1.40     -0.1284  -0.1753  -0.1894  -0.1014  -0.0412  -0.0183  0.0796  0.0388  
-0.0268  0.1927  0.4353  1.0000  0.2634  0.0773  -0.0139  -0.0565  -0.0397  0.0201  
-0.0377  -0.1427  -0.1397  -0.1597  -0.1047  -0.0598  0.0174  0.0601  0.0549  0.1184  
0.0531  0.0455  0.0503  0.0014  0.0751  -0.0546  0.0441  -0.0540  0.0392  -0.0039  

1.40-1.45     -0.1223  -0.1086  -0.1644  -0.1503  -0.0161  -0.1227  -0.0387  -0.0094  
-0.0501  0.0660  0.1608  0.2634  1.0000  0.2304  0.0818  0.0331  -0.0954  -0.0157  
-0.0272  -0.0351  -0.1209  -0.1252  -0.1635  -0.0475  0.0031  0.0565  0.0648  0.1228  
0.0752  0.0394  0.0286  0.0369  0.0418  -0.0068  0.0059  -0.0044  0.0512  -0.0254  

1.45-1.50     -0.1601  -0.0395  -0.0941  -0.0967  -0.0745  -0.1060  -0.0645  -0.0042  
-0.0401  -0.0597  -0.0110  0.0773  0.2304  1.0000  0.3575  0.2435  0.1535  0.0909  
0.0160  -0.0942  -0.1972  -0.1839  -0.1483  -0.0866  0.0481  0.0763  0.0931  0.1207  
0.1577  0.0873  0.0176  0.0434  0.0609  0.0067  0.0043  0.0516  0.0172  0.0115  

1.50-1.55     -0.1025  -0.0250  -0.0615  -0.0217  -0.0353  -0.0959  -0.1170  -0.0397  
0.0106  -0.0166  -0.0524  -0.0139  0.0818  0.3575  1.0000  0.4719  0.3019  0.0542  
-0.1196  -0.1531  -0.1880  -0.1334  -0.1163  -0.0785  0.0285  -0.0098  0.0188  0.0950  
0.1280  0.0905  0.0258  0.1117  0.0207  0.0442  0.0502  0.0134  0.0553  0.0348  

1.55-1.60     -0.0423  -0.0777  -0.0471  -0.0148  0.0307  -0.1126  -0.1652  -0.1472  
-0.0421  -0.0551  -0.1568  -0.0565  0.0331  0.2435  0.4719  1.0000  0.3841  0.0769  
-0.0896  -0.0805  -0.0925  -0.0901  -0.1146  -0.0719  0.0633  0.0380  0.0651  0.0893  
0.1587  0.0665  0.0229  0.0594  0.0685  0.0100  0.0201  0.0184  0.0364  0.0555  

1.60-1.65     -0.0453  -0.0615  -0.0389  0.0040  0.0435  -0.0710  -0.1519  -0.1003  
-0.0421  -0.0598  -0.0360  -0.0397  -0.0954  0.1535  0.3019  0.3841  1.0000  0.2247  
0.0303  -0.0788  -0.1348  -0.2102  -0.1781  -0.0408  0.1021  0.0597  0.0595  0.1067  
0.1093  0.0960  0.0802  0.0264  0.0278  0.0226  0.0010  0.0544  0.0165  0.0352  

1.65-1.70     -0.1029  -0.0866  -0.0227  -0.0316  -0.0166  -0.0114  -0.0189  -0.0469  
-0.0366  -0.0732  0.0402  0.0201  -0.0157  0.0909  0.0542  0.0769  0.2247  1.0000  
0.1938  0.0688  -0.0641  -0.1071  -0.0727  0.0369  0.0721  0.1169  0.1118  0.1188  
0.1079  0.0512  0.0672  0.0781  0.0074  0.0902  0.0222  0.0130  0.0336  0.0461  

1.70-1.75     -0.0667  -0.1381  -0.2051  -0.1713  -0.1473  -0.1497  -0.0694  -0.0371  
-0.0402  -0.0586  -0.0166  -0.0377  -0.0272  0.0160  -0.1196  -0.0896  0.0303  0.1938  
1.0000  0.3961  0.1305  0.0599  0.0874  0.0693  0.0908  0.1007  0.1210  -0.0055  
0.0459  -0.0168  0.0905  0.0059  0.0371  0.0118  0.0381  0.0777  -0.0021  0.0567  

1.75-1.80     0.0346  -0.1138  -0.2035  -0.2460  -0.1842  -0.1499  -0.1264  -0.0561  
0.0380  -0.1287  -0.1517  -0.1427  -0.0351  -0.0942  -0.1531  -0.0805  -0.0788  0.0688  
0.3961  1.0000  0.2892  0.2574  0.2306  0.1742  0.0981  0.0643  0.0247  -0.0407  
0.0300  0.0079  0.0245  0.0523  0.0534  0.0117  0.0773  -0.0012  0.0439  0.0301  

1.80-1.85     0.0212  0.0045  -0.1089  -0.1186  -0.0882  -0.0441  -0.0690  -0.0814  
0.0506  -0.0914  -0.2344  -0.1397  -0.1209  -0.1972  -0.1880  -0.0925  -0.1348  -0.0641  
0.1305  0.2892  1.0000  0.4800  0.2490  0.1530  0.0929  0.0913  0.0085  -0.0525  
-0.0354  -0.0018  0.0289  0.0686  0.0872  0.0203  0.0656  0.0477  0.0195  0.0359  

1.85-1.90     -0.0046  -0.0024  -0.0541  -0.0553  -0.0981  -0.0632  -0.0544  -0.0683  
0.0284  -0.1138  -0.2475  -0.1597  -0.1252  -0.1839  -0.1334  -0.0901  -0.2102  -0.1071  
0.0599  0.2574  0.4800  1.0000  0.3658  0.2014  0.1093  0.0060  -0.0271  -0.0214  
-0.0311  0.0104  0.0209  0.0757  0.0637  -0.0225  0.0244  0.0330  -0.0437  0.0620  

1.90-1.95     0.0186  0.0664  0.0425  -0.0363  -0.1204  -0.0570  -0.0231  -0.0205  
-0.0424  -0.1145  -0.0816  -0.1047  -0.1635  -0.1483  -0.1163  -0.1146  -0.1781  -0.0727  
0.0874  0.2306  0.2490  0.3658  1.0000  0.2847  0.2391  0.1143  0.1613  0.1621  
0.0881  0.0606  0.0496  0.0958  -0.0003  0.0231  0.0588  0.0306  -0.0180  0.0890  

1.95-2.00     0.0349  0.0633  0.0030  -0.0812  -0.1442  -0.0792  -0.0179  0.0167  
-0.0125  -0.0793  -0.0478  -0.0598  -0.0475  -0.0866  -0.0785  -0.0719  -0.0408  0.0369  
0.0693  0.1742  0.1530  0.2014  0.2847  1.0000  0.2321  0.3285  0.2592  0.2007  
0.1602  0.1452  0.0556  0.0503  0.0472  0.0556  -0.0086  0.0916  -0.0076  0.0982  

2.00-2.05     0.0566  -0.0249  0.0069  0.0362  -0.1148  -0.0956  0.0220  0.0172  
-0.0255  -0.0298  -0.0240  0.0174  0.0031  0.0481  0.0285  0.0633  0.1021  0.0721  
0.0908  0.0981  0.0929  0.1093  0.2391  0.2321  1.0000  0.3015  0.3927  0.3175  
0.2526  0.1953  0.1329  0.0581  0.0314  0.0505  0.0680  0.0259  0.0349  0.0735  

2.05-2.10     0.0686  0.0506  -0.0036  -0.0170  -0.0595  0.0050  0.0437  0.0602  
0.0157  -0.0336  0.0133  0.0601  0.0565  0.0763  -0.0098  0.0380  0.0597  0.1169  
0.1007  0.0643  0.0913  0.0060  0.1143  0.3285  0.3015  1.0000  0.3264  0.4134  
0.3254  0.2170  0.1046  0.0668  0.0559  0.0806  -0.0082  0.0875  0.0570  0.0676  

2.10-2.15     0.0485  0.0212  0.0133  -0.0141  -0.0123  0.0083  0.0147  0.1063  
0.0661  0.0030  0.0148  0.0549  0.0648  0.0931  0.0188  0.0651  0.0595  0.1118  
0.1210  0.0247  0.0085  -0.0271  0.1613  0.2592  0.3927  0.3264  1.0000  0.2714  
0.4015  0.2346  0.1308  0.0670  0.0513  0.0332  0.0236  0.0708  -0.0027  0.0756  

2.15-2.20     0.0308  0.0456  0.0408  0.0764  0.0009  0.0435  0.0164  0.0366  
0.0357  0.0006  0.0751  0.1184  0.1228  0.1207  0.0950  0.0893  0.1067  0.1188  
-0.0055  -0.0407  -0.0525  -0.0214  0.1621  0.2007  0.3175  0.4134  0.2714  1.0000  
0.1182  0.3326  0.1186  0.0705  0.0654  0.0639  0.0291  0.0139  0.0566  0.0964  

2.20-2.25     -0.0209  -0.0449  -0.0385  -0.0123  0.0035  -0.0088  -0.0221  0.0356  
0.0931  0.0422  0.0653  0.0531  0.0752  0.1577  0.1280  0.1587  0.1093  0.1079  
0.0459  0.0300  -0.0354  -0.0311  0.0881  0.1602  0.2526  0.3254  0.4015  0.1182  
1.0000  -0.0593  0.2122  0.0438  0.1097  0.0295  0.0827  0.0597  0.0468  0.0534  

2.25-2.30     0.0178  0.1222  0.0122  0.0541  0.0206  0.0077  0.0185  0.0758  
0.0529  0.0291  0.0362  0.0455  0.0394  0.0873  0.0905  0.0665  0.0960  0.0512  
-0.0168  0.0079  -0.0018  0.0104  0.0606  0.1452  0.1953  0.2170  0.2346  0.3326  
-0.0593  1.0000  -0.1891  0.1440  -0.0089  0.0925  0.0690  0.0834  0.0342  0.0842  

2.30-2.35     -0.0244  -0.0481  -0.0459  -0.0155  0.0098  0.0505  0.0352  0.0158  
0.0565  0.0065  0.0475  0.0503  0.0286  0.0176  0.0258  0.0229  0.0802  0.0672  
0.0905  0.0245  0.0289  0.0209  0.0496  0.0556  0.1329  0.1046  0.1308  0.1186  
0.2122  -0.1891  1.0000  -0.3216  0.2390  0.0501  0.0343  0.0526  -0.0078  0.1358  

2.35-2.40     -0.0083  0.0065  0.0719  0.0775  0.0300  -0.0151  0.0193  0.0110  
0.0004  -0.0722  -0.0530  0.0014  0.0369  0.0434  0.1117  0.0594  0.0264  0.0781  
0.0059  0.0523  0.0686  0.0757  0.0958  0.0503  0.0581  0.0668  0.0670  0.0705  
0.0438  0.1440  -0.3216  1.0000  -0.3655  0.1948  0.0407  0.0706  0.0593  0.0804  

2.40-2.45     -0.0121  0.0422  -0.0243  0.0151  0.0065  -0.0085  -0.0307  -0.0352  
0.0047  0.0440  0.0571  0.0751  0.0418  0.0609  0.0207  0.0685  0.0278  0.0074  
0.0371  0.0534  0.0872  0.0637  -0.0003  0.0472  0.0314  0.0559  0.0513  0.0654  
0.1097  -0.0089  0.2390  -0.3655  1.0000  -0.3778  0.1917  0.0702  -0.0008  0.1397  

2.45-2.50     -0.0083  0.0166  0.0528  0.0350  0.0094  0.0312  0.0731  0.0064  
-0.0026  -0.0369  -0.0522  -0.0546  -0.0068  0.0067  0.0442  0.0100  0.0226  0.0902  
0.0118  0.0117  0.0203  -0.0225  0.0231  0.0556  0.0505  0.0806  0.0332  0.0639  
0.0295  0.0925  0.0501  0.1948  -0.3778  1.0000  -0.3806  0.1933  0.0341  0.0278  

2.50-2.55     0.0095  -0.0395  -0.0377  0.0236  -0.0320  -0.0247  0.0053  0.0529  
0.0204  0.0722  0.0656  0.0441  0.0059  0.0043  0.0502  0.0201  0.0010  0.0222  
0.0381  0.0773  0.0656  0.0244  0.0588  -0.0086  0.0680  -0.0082  0.0236  0.0291  
0.0827  0.0690  0.0343  0.0407  0.1917  -0.3806  1.0000  -0.4335  0.1970  0.0433  

2.55-2.60     -0.0031  0.0827  0.0984  0.0514  0.0380  0.0256  -0.0112  -0.0639  
-0.0387  -0.0490  -0.0257  -0.0540  -0.0044  0.0516  0.0134  0.0184  0.0544  0.0130  
0.0777  -0.0012  0.0477  0.0330  0.0306  0.0916  0.0259  0.0875  0.0708  0.0139  
0.0597  0.0834  0.0526  0.0706  0.0702  0.1933  -0.4335  1.0000  -0.4425  0.2553  

2.60-2.65     -0.0264  -0.0476  -0.0416  0.0094  0.0367  -0.0182  -0.0116  0.0308  
0.0278  0.0642  0.0495  0.0392  0.0512  0.0172  0.0553  0.0364  0.0165  0.0336  
-0.0021  0.0439  0.0195  -0.0437  -0.0180  -0.0076  0.0349  0.0570  -0.0027  0.0566  
0.0468  0.0342  -0.0078  0.0593  -0.0008  0.0341  0.1970  -0.4425  1.0000  -0.4198  

2.65-2.70     -0.0140  0.0941  0.0940  0.0601  0.0008  0.0462  0.0386  -0.0397  
-0.0570  -0.0479  -0.0238  -0.0039  -0.0254  0.0115  0.0348  0.0555  0.0352  0.0461  
0.0567  0.0301  0.0359  0.0620  0.0890  0.0982  0.0735  0.0676  0.0756  0.0964  
0.0534  0.0842  0.1358  0.0804  0.1397  0.0278  0.0433  0.2553  -0.4198  1.0000

The differential branching fraction (GGOU) for semileptonic 
B decays as a function of the electron momentum (in the Upsilon(4S)
rest frame) after background subtraction and corrections for bremsstrahlung and
final state radiation. The errors indicate the statistical errors on the background
subtraction, including the uncertainties of the fit parameters.

p(GeV/c)      Delta B(10**{-3})/(50 MeV/c)

0.80-0.85     2.1762 +- 0.0184
0.85-0.90     2.2868 +- 0.0153
0.90-0.95     2.4719 +- 0.0154
0.95-1.00     2.6582 +- 0.0135
1.00-1.05     2.8527 +- 0.0120
1.05-1.10     3.0683 +- 0.0108
1.10-1.15     3.2503 +- 0.0110
1.15-1.20     3.4596 +- 0.0101
1.20-1.25     3.6255 +- 0.0109
1.25-1.30     3.8184 +- 0.0101
1.30-1.35     3.9735 +- 0.0105
1.35-1.40     4.1359 +- 0.0098
1.40-1.45     4.2622 +- 0.0092
1.45-1.50     4.3878 +- 0.0091
1.50-1.55     4.4713 +- 0.0106
1.55-1.60     4.5138 +- 0.0104
1.60-1.65     4.5219 +- 0.0099
1.65-1.70     4.4863 +- 0.0086
1.70-1.75     4.3998 +- 0.0090
1.75-1.80     4.2331 +- 0.0092
1.80-1.85     4.0039 +- 0.0090
1.85-1.90     3.6911 +- 0.0080
1.90-1.95     3.2900 +- 0.0074
1.95-2.00     2.8298 +- 0.0065
2.00-2.05     2.3039 +- 0.0053
2.05-2.10     1.7660 +- 0.0046
2.10-2.15     1.2257 +- 0.0038
2.15-2.20     0.7860 +- 0.0030
2.20-2.25     0.4481 +- 0.0025
2.25-2.30     0.2207 +- 0.0020
2.30-2.35     0.0921 +- 0.0017
2.35-2.40     0.0430 +- 0.0016
2.40-2.45     0.0274 +- 0.0016
2.45-2.50     0.0222 +- 0.0016
2.50-2.55     0.0146 +- 0.0016
2.55-2.60     0.0091 +- 0.0016
2.60-2.65     0.0041 +- 0.0016
2.65-2.70     0.0040 +- 0.0015

Correlation matrix for statistical contributions is the following:

p(GeV/c) 0.80-0.85  0.85-0.90  0.90-0.95  0.95-1.00  1.00-1.05  1.05-1.10  1.10-1.15
1.15-1.20  1.20-1.25  1.25-1.30  1.30-1.35  1.35-1.40  1.40-1.45  1.45-1.50  1.50-1.55
1.55-1.60  1.60-1.65  1.65-1.70  1.70-1.75  1.75-1.80  1.80-1.85  1.85-1.90  1.90-1.95
1.95-2.00  2.00-2.05  2.05-2.10  2.10-2.15  2.15-2.20  2.20-2.25  2.25-2.30  2.30-2.35
2.35-2.40  2.40-2.45  2.45-2.50  2.50-2.55  2.55-2.60  2.60-2.65  2.65-2.70

0.80-0.85     1.0000  0.5555  0.3934  0.3957  0.3226  0.3387  0.2857  0.3352  
0.2908  0.2050  0.1365  0.1003  0.1033  0.0754  0.0736  0.1056  0.0849  0.0359  
0.0254  0.0765  0.0472  -0.0009  0.0303  0.0234  0.0262  0.0496  0.0290  0.0169  
-0.0498  0.0207  -0.0126  -0.0002  -0.0146  0.0009  -0.0068  0.0159  -0.0382  0.0052  

0.85-0.90     0.5555  1.0000  0.5753  0.4764  0.3499  0.3470  0.3085  0.2412  
0.1657  0.1132  0.0454  0.0300  0.0868  0.1531  0.1197  0.0599  0.0506  0.0233  
-0.0637  -0.0822  0.0058  -0.0221  0.0531  0.0360  -0.0725  0.0178  -0.0113  0.0157  
-0.0883  0.1009  -0.0427  0.0072  0.0276  0.0231  -0.0456  0.0863  -0.0564  0.0937  

0.90-0.95     0.3934  0.5753  1.0000  0.6476  0.3690  0.3455  0.2616  0.1356  
0.0152  0.0025  -0.0475  -0.0364  -0.0224  0.0508  0.0467  0.0441  0.0406  0.0502  
-0.1444  -0.1764  -0.0962  -0.0550  0.0541  -0.0046  -0.0224  -0.0244  -0.0054  0.0253  
-0.0682  0.0170  -0.0412  0.0683  -0.0268  0.0543  -0.0456  0.1035  -0.0523  0.0975  

0.95-1.00     0.3957  0.4764  0.6476  1.0000  0.4959  0.3559  0.2106  0.1471  
0.0523  0.1227  0.0520  0.0227  -0.0332  0.0191  0.0588  0.0573  0.0573  0.0255  
-0.1185  -0.2122  -0.0999  -0.0567  -0.0255  -0.0966  -0.0078  -0.0570  -0.0655  0.0291  
-0.0718  0.0353  -0.0265  0.0682  0.0027  0.0406  0.0021  0.0651  -0.0122  0.0685  

1.00-1.05     0.3226  0.3499  0.3690  0.4959  1.0000  0.3526  0.1703  0.1564  
0.1375  0.0515  0.0337  0.0441  0.0621  0.0171  0.0323  0.0974  0.0928  0.0375  
-0.1034  -0.1598  -0.0707  -0.0817  -0.0719  -0.1109  -0.0990  -0.0252  0.0228  0.0150  
-0.0139  0.0266  -0.0051  0.0242  -0.0062  0.0192  -0.0491  0.0549  0.0073  0.0197  

1.05-1.10     0.3387  0.3470  0.3455  0.3559  0.3526  1.0000  0.4665  0.2896  
0.1439  0.1316  0.0902  0.0434  -0.0598  -0.0512  -0.0549  -0.0625  -0.0327  0.0327  
-0.0946  -0.1046  0.0124  -0.0087  0.0193  -0.0335  -0.0795  0.0214  0.0146  0.0231  
-0.0581  -0.0106  0.0206  -0.0176  -0.0263  0.0396  -0.0460  0.0431  -0.0454  0.0584  

1.10-1.15     0.2857  0.3085  0.2616  0.2106  0.1703  0.4665  1.0000  0.3958  
0.1038  0.0955  0.1202  0.1308  0.0088  -0.0263  -0.0914  -0.1370  -0.1365  0.0019  
-0.0397  -0.1115  -0.0455  -0.0283  0.0295  0.0111  0.0406  0.0675  0.0248  -0.0089  
-0.0757  -0.0033  0.0053  0.0092  -0.0492  0.0762  -0.0190  0.0067  -0.0397  0.0509  

1.15-1.20     0.3352  0.2412  0.1356  0.1471  0.1564  0.2896  0.3958  1.0000  
0.4103  0.1565  0.0870  0.0671  0.0213  0.0222  -0.0299  -0.1244  -0.0972  -0.0303  
0.0015  -0.0231  -0.0500  -0.0335  0.0369  0.0440  0.0225  0.0694  0.1028  -0.0155  
-0.0400  0.0334  -0.0301  -0.0076  -0.0574  0.0162  0.0168  -0.0422  -0.0057  -0.0234  

1.20-1.25     0.2908  0.1657  0.0152  0.0523  0.1375  0.1439  0.1038  0.4103  
1.0000  0.2623  0.0645  -0.0298  -0.0435  -0.0396  0.0115  -0.0211  -0.0472  -0.0406  
-0.0226  0.0520  0.0835  0.0575  -0.0075  0.0116  -0.0280  0.0199  0.0688  -0.0146  
0.0244  0.0067  -0.0017  -0.0232  -0.0190  0.0003  -0.0078  -0.0257  -0.0056  -0.0488  

1.25-1.30     0.2050  0.1132  0.0025  0.1227  0.0515  0.1316  0.0955  0.1565  
0.2623  1.0000  0.4519  0.1861  0.0521  -0.0992  -0.0566  -0.0799  -0.0970  -0.1007  
-0.0522  -0.1151  -0.0521  -0.0696  -0.0540  -0.0430  -0.0104  -0.0306  0.0009  -0.0442  
-0.0178  0.0017  -0.0314  -0.0897  0.0153  -0.0329  0.0368  -0.0351  0.0274  -0.0425  

1.30-1.35     0.1365  0.0454  -0.0475  0.0520  0.0337  0.0902  0.1202  0.0870  
0.0645  0.4519  1.0000  0.4496  0.1145  -0.1121  -0.1508  -0.2494  -0.1183  -0.0244  
-0.0217  -0.1401  -0.2222  -0.2261  -0.0270  -0.0314  -0.0464  -0.0258  -0.0376  0.0050  
-0.0169  -0.0080  0.0114  -0.0728  0.0300  -0.0540  0.0326  -0.0200  0.0162  -0.0241  

1.35-1.40     0.1003  0.0300  -0.0364  0.0227  0.0441  0.0434  0.1308  0.0671  
-0.0298  0.1861  0.4496  1.0000  0.2252  -0.0532  -0.1527  -0.1875  -0.1662  -0.0958  
-0.0799  -0.1616  -0.1441  -0.1560  -0.0835  -0.0761  -0.0330  0.0062  -0.0124  0.0372  
-0.0537  -0.0158  0.0074  -0.0224  0.0467  -0.0585  0.0093  -0.0507  0.0036  -0.0062  

1.40-1.45     0.1033  0.0868  -0.0224  -0.0332  0.0621  -0.0598  0.0088  0.0213  
-0.0435  0.0521  0.1145  0.2252  1.0000  0.1760  -0.0240  -0.0678  -0.2154  -0.1302  
-0.0424  -0.0060  -0.0850  -0.0701  -0.1132  -0.0269  -0.0046  0.0441  0.0442  0.0680  
-0.0079  -0.0085  -0.0176  0.0118  0.0091  -0.0109  -0.0346  0.0001  0.0114  -0.0315  

1.45-1.50     0.0754  0.1531  0.0508  0.0191  0.0171  -0.0512  -0.0263  0.0222  
-0.0396  -0.0992  -0.1121  -0.0532  0.1760  1.0000  0.3121  0.1525  0.0461  -0.0193  
0.0089  -0.0449  -0.1486  -0.1159  -0.1024  -0.1106  -0.0312  -0.0425  -0.0401  -0.0294  
0.0205  0.0171  -0.0365  0.0230  0.0227  0.0012  -0.0385  0.0560  -0.0229  0.0014  

1.50-1.55     0.0736  0.1197  0.0467  0.0588  0.0323  -0.0549  -0.0914  -0.0299  
0.0115  -0.0566  -0.1508  -0.1527  -0.0240  0.3121  1.0000  0.4648  0.2316  -0.0683  
-0.1699  -0.1424  -0.1573  -0.0826  -0.0884  -0.1175  -0.0674  -0.1536  -0.1334  -0.0455  
0.0087  0.0329  -0.0190  0.0959  -0.0121  0.0345  0.0147  0.0119  0.0220  0.0237  

1.55-1.60     0.1056  0.0599  0.0441  0.0573  0.0974  -0.0625  -0.1370  -0.1244  
-0.0211  -0.0799  -0.2494  -0.1875  -0.0678  0.1525  0.4648  1.0000  0.3533  -0.0543  
-0.1590  -0.0650  -0.0376  -0.0321  -0.0991  -0.1128  -0.0275  -0.0965  -0.0631  -0.0298  
0.0630  0.0323  0.0056  0.0501  0.0362  -0.0015  -0.0151  0.0159  0.0026  0.0422  

1.60-1.65     0.0849  0.0506  0.0406  0.0573  0.0928  -0.0327  -0.1365  -0.0972  
-0.0472  -0.0970  -0.1183  -0.1662  -0.2154  0.0461  0.2316  0.3533  1.0000  0.1866  
-0.0149  -0.0675  -0.1129  -0.2100  -0.1929  -0.0841  0.0185  -0.0680  -0.0668  -0.0002  
0.0053  0.0496  0.0611  0.0054  -0.0055  0.0112  -0.0361  0.0493  -0.0169  0.0246  

1.65-1.70     0.0359  0.0233  0.0502  0.0255  0.0375  0.0327  0.0019  -0.0303  
-0.0406  -0.1007  -0.0244  -0.0958  -0.1302  -0.0193  -0.0683  -0.0543  0.1866  1.0000  
0.1895  0.0443  -0.0862  -0.1628  -0.1362  -0.0583  -0.0837  -0.0296  -0.0180  -0.0036  
-0.0178  -0.0059  0.0396  0.0495  -0.0345  0.0797  -0.0206  0.0049  -0.0029  0.0243  

1.70-1.75     0.0254  -0.0637  -0.1444  -0.1185  -0.1034  -0.0946  -0.0397  0.0015  
-0.0226  -0.0522  -0.0217  -0.0799  -0.0424  0.0089  -0.1699  -0.1590  -0.0149  0.1895  
1.0000  0.3901  0.0230  -0.0827  -0.0106  -0.0130  0.0119  0.0583  0.1228  -0.0244  
0.0101  -0.0371  0.0673  -0.0267  -0.0017  0.0051  0.0043  0.0577  -0.0285  0.0344  

1.75-1.80     0.0765  -0.0822  -0.1764  -0.2122  -0.1598  -0.1046  -0.1115  -0.0231  
0.0520  -0.1151  -0.1401  -0.1616  -0.0060  -0.0449  -0.1424  -0.0650  -0.0675  0.0443  
0.3901  1.0000  0.1680  0.0916  0.1165  0.1085  0.0652  0.0728  0.0723  0.0018  
0.0463  0.0098  -0.0053  0.0074  0.0292  0.0036  0.0540  -0.0252  0.0250  0.0102  

1.80-1.85     0.0472  0.0058  -0.0962  -0.0999  -0.0707  0.0124  -0.0455  -0.0500  
0.0835  -0.0521  -0.2222  -0.1441  -0.0850  -0.1486  -0.1573  -0.0376  -0.1129  -0.0862  
0.0230  0.1680  1.0000  0.3786  0.0938  0.0463  0.0374  0.0982  0.0411  -0.0221  
-0.0414  -0.0091  -0.0160  0.0256  0.0678  0.0150  0.0477  0.0235  0.0020  0.0187  

1.85-1.90     -0.0009  -0.0221  -0.0550  -0.0567  -0.0817  -0.0087  -0.0283  -0.0335  
0.0575  -0.0696  -0.2261  -0.1560  -0.0701  -0.1159  -0.0826  -0.0321  -0.2100  -0.1628  
-0.0827  0.0916  0.3786  1.0000  0.2667  0.0867  0.0142  -0.0579  -0.0469  -0.0074  
-0.0511  -0.0092  -0.0373  0.0308  0.0396  -0.0367  -0.0007  0.0021  -0.0676  0.0398  

1.90-1.95     0.0303  0.0531  0.0541  -0.0255  -0.0719  0.0193  0.0295  0.0369  
-0.0075  -0.0540  -0.0270  -0.0835  -0.1132  -0.1024  -0.0884  -0.0991  -0.1929  -0.1362  
-0.0106  0.1165  0.0938  0.2667  1.0000  0.1773  0.0616  -0.0835  0.0216  0.0750  
-0.0230  -0.0225  -0.0347  0.0420  -0.0376  0.0045  0.0315  -0.0071  -0.0424  0.0595  

1.95-2.00     0.0234  0.0360  -0.0046  -0.0966  -0.1109  -0.0335  0.0111  0.0440  
0.0116  -0.0430  -0.0314  -0.0761  -0.0269  -0.1106  -0.1175  -0.1128  -0.0841  -0.0583  
-0.0130  0.1085  0.0463  0.0867  0.1773  1.0000  0.0200  0.1213  0.0472  0.0062  
-0.0198  0.0247  -0.0290  -0.0063  0.0083  0.0389  -0.0426  0.0609  -0.0294  0.0659  

2.00-2.05     0.0262  -0.0725  -0.0224  -0.0078  -0.0990  -0.0795  0.0406  0.0225  
-0.0280  -0.0104  -0.0464  -0.0330  -0.0046  -0.0312  -0.0674  -0.0275  0.0185  -0.0837  
0.0119  0.0652  0.0374  0.0142  0.0616  0.0200  1.0000  -0.0128  0.0975  0.0169  
0.0007  0.0285  0.0552  0.0099  -0.0154  0.0319  0.0413  -0.0144  0.0194  0.0324  

2.05-2.10     0.0496  0.0178  -0.0244  -0.0570  -0.0252  0.0214  0.0675  0.0694  
0.0199  -0.0306  -0.0258  0.0062  0.0441  -0.0425  -0.1536  -0.0965  -0.0680  -0.0296  
0.0583  0.0728  0.0982  -0.0579  -0.0835  0.1213  -0.0128  1.0000  -0.0202  0.1137  
0.0645  0.0403  0.0159  0.0269  0.0214  0.0705  -0.0428  0.0620  0.0536  0.0255  

2.10-2.15     0.0290  -0.0113  -0.0054  -0.0655  0.0228  0.0146  0.0248  0.1028  
0.0688  0.0009  -0.0376  -0.0124  0.0442  -0.0401  -0.1334  -0.0631  -0.0668  -0.0180  
0.1228  0.0723  0.0411  -0.0469  0.0216  0.0472  0.0975  -0.0202  1.0000  -0.1395  
0.1578  0.0342  0.0300  0.0254  0.0188  0.0108  -0.0030  0.0466  -0.0213  0.0354  

2.15-2.20     0.0169  0.0157  0.0253  0.0291  0.0150  0.0231  -0.0089  -0.0155  
-0.0146  -0.0442  0.0050  0.0372  0.0680  -0.0294  -0.0455  -0.0298  -0.0002  -0.0036  
-0.0244  0.0018  -0.0221  -0.0074  0.0750  0.0062  0.0169  0.1137  -0.1395  1.0000  
-0.2312  0.1727  0.0124  0.0345  0.0438  0.0557  0.0111  -0.0152  0.0539  0.0758  

2.20-2.25     -0.0498  -0.0883  -0.0682  -0.0718  -0.0139  -0.0581  -0.0757  -0.0400  
0.0244  -0.0178  -0.0169  -0.0537  -0.0079  0.0205  0.0087  0.0630  0.0053  -0.0178  
0.0101  0.0463  -0.0414  -0.0511  -0.0230  -0.0198  0.0007  0.0645  0.1578  -0.2312  
1.0000  -0.2762  0.1571  0.0103  0.1053  0.0157  0.0746  0.0414  0.0419  0.0306  

2.25-2.30     0.0207  0.1009  0.0170  0.0353  0.0266  -0.0106  -0.0033  0.0334  
0.0067  0.0017  -0.0080  -0.0158  -0.0085  0.0171  0.0329  0.0323  0.0496  -0.0059  
-0.0371  0.0098  -0.0091  -0.0092  -0.0225  0.0247  0.0285  0.0403  0.0342  0.1727  
-0.2762  1.0000  -0.3044  0.1337  -0.0291  0.0925  0.0647  0.0765  0.0335  0.0698  

2.30-2.35     -0.0126  -0.0427  -0.0412  -0.0265  -0.0051  0.0206  0.0053  -0.0301  
-0.0017  -0.0314  0.0114  0.0074  -0.0176  -0.0365  -0.0190  0.0056  0.0611  0.0396  
0.0673  -0.0053  -0.0160  -0.0373  -0.0347  -0.0290  0.0552  0.0159  0.0300  0.0124  
0.1571  -0.3044  1.0000  -0.3498  0.2442  0.0523  0.0326  0.0507  -0.0104  0.1362  

2.35-2.40     -0.0002  0.0072  0.0683  0.0682  0.0242  -0.0176  0.0092  -0.0076  
-0.0232  -0.0897  -0.0728  -0.0224  0.0118  0.0230  0.0959  0.0501  0.0054  0.0495  
-0.0267  0.0074  0.0256  0.0308  0.0420  -0.0063  0.0099  0.0269  0.0254  0.0345  
0.0103  0.1337  -0.3498  1.0000  -0.3574  0.2001  0.0404  0.0715  0.0593  0.0841  

2.40-2.45     -0.0146  0.0276  -0.0268  0.0027  -0.0062  -0.0263  -0.0492  -0.0574  
-0.0190  0.0153  0.0300  0.0467  0.0091  0.0227  -0.0121  0.0362  -0.0055  -0.0345  
-0.0017  0.0292  0.0678  0.0396  -0.0376  0.0083  -0.0154  0.0214  0.0188  0.0438  
0.1053  -0.0291  0.2442  -0.3574  1.0000  -0.3759  0.1931  0.0714  -0.0001  0.1412  

2.45-2.50     0.0009  0.0231  0.0543  0.0406  0.0192  0.0396  0.0762  0.0162  
0.0003  -0.0329  -0.0540  -0.0585  -0.0109  0.0012  0.0345  -0.0015  0.0112  0.0797  
0.0051  0.0036  0.0150  -0.0367  0.0045  0.0389  0.0319  0.0705  0.0108  0.0557  
0.0157  0.0925  0.0523  0.2001  -0.3759  1.0000  -0.3805  0.1935  0.0341  0.0281  

2.50-2.55     -0.0068  -0.0456  -0.0456  0.0021  -0.0491  -0.0460  -0.0190  0.0168  
-0.0078  0.0368  0.0326  0.0093  -0.0346  -0.0385  0.0147  -0.0151  -0.0361  -0.0206  
0.0043  0.0540  0.0477  -0.0007  0.0315  -0.0426  0.0413  -0.0428  -0.0030  0.0111  
0.0746  0.0647  0.0326  0.0404  0.1931  -0.3805  1.0000  -0.4334  0.1970  0.0433  

2.55-2.60     0.0159  0.0863  0.1035  0.0651  0.0549  0.0431  0.0067  -0.0422  
-0.0257  -0.0351  -0.0200  -0.0507  0.0001  0.0560  0.0119  0.0159  0.0493  0.0049  
0.0577  -0.0252  0.0235  0.0021  -0.0071  0.0609  -0.0144  0.0620  0.0466  -0.0152  
0.0414  0.0765  0.0507  0.0715  0.0714  0.1935  -0.4334  1.0000  -0.4425  0.2554  

2.60-2.65     -0.0382  -0.0564  -0.0523  -0.0122  0.0073  -0.0454  -0.0397  -0.0057  
-0.0056  0.0274  0.0162  0.0036  0.0114  -0.0229  0.0220  0.0026  -0.0169  -0.0029  
-0.0285  0.0250  0.0020  -0.0676  -0.0424  -0.0294  0.0194  0.0536  -0.0213  0.0539  
0.0419  0.0335  -0.0104  0.0593  -0.0001  0.0341  0.1970  -0.4425  1.0000  -0.4198  

2.65-2.70     0.0052  0.0937  0.0975  0.0685  0.0197  0.0584  0.0509  -0.0234  
-0.0488  -0.0425  -0.0241  -0.0062  -0.0315  0.0014  0.0237  0.0422  0.0246  0.0243  
0.0344  0.0102  0.0187  0.0398  0.0595  0.0659  0.0324  0.0255  0.0354  0.0758  
0.0306  0.0698  0.1362  0.0841  0.1412  0.0281  0.0433  0.2554  -0.4198  1.0000

\end{verbatim}

\end{widetext}

\end{document}